\documentclass[12pt,a4paper]{article}
\usepackage[latin9]{inputenc}
\usepackage{a4wide}
\usepackage{amssymb}
\usepackage{slashed}
\usepackage{ifpdf}
\ifpdf
\usepackage[pdftex]{graphicx}
\usepackage[pdftex,unicode,implicit]{hyperref}
\hypersetup{%
  pdftitle    = {The supersymmetric configurations of gauged N=2,d=4 
supergravity coupled to vector supermultiplets},
  pdfkeywords = {N=2 Supersymmetry, special Kähler geometry, BPS solutions},
  pdfauthor   = {M. Hübscher, P. Meessen, T. Ortín, S. Vaulà},
  pdfcreator  = {pdf\LaTeXe\ with package \flqq hyperref\frqq},
  pdfproducer = {pdf\LaTeXe\ with package \flqq hyperref\frqq},
  pdfpagemode = None,  
  pdffitwindow= true,  
  unicode     = true,
  plainpages  = false,
  colorlinks  = true,  
  citecolor   = blue,
  urlcolor    = red,
  linkcolor   = black
}
\newcommand{\hepth}[1]{({\tt \href{http://www.arXiv.org/abs/hep-th/#1}{hep-th/#1}})}

\newcommand{\arxiv}[1]{{\tt \href{http://www.arXiv.org/abs/#1}{arXiv:#1}}}
\else
  \usepackage[dvips]{graphicx}
  \usepackage[unicode,implicit]{hyperref}
  \newcommand{\hepth}[1]{{\tt hep-th/#1}}

  \newcommand{\arxiv}[1]{{\tt arXiv:#1}}
\fi
\makeatletter
\@addtoreset{equation}{section}
\makeatother

\begin{document}
\begin{flushright}
\small
IFT-UAM/CSIC-08-32\\
\texttt{arXiv:0806.1477}\\
June  $9^{\rm th}$, $2008$
\normalsize
\end{flushright}
\begin{center}
\vspace{2cm}
{\LARGE\bf $N=2$ Einstein-Yang-Mills's BPS solutions}
\vspace{2cm}

{\sl\large Mechthild H\"ubscher}
\footnote{E-mail: {\tt Mechthild.Huebscher@uam.es}},
{\sl\large Patrick Meessen}
\footnote{E-mail: {\tt Patrick.Meessen@uam.es}},
{\sl\large Tom{\'a}s Ort\'{\i}n}
\footnote{E-mail: {\tt Tomas.Ortin@uam.es}}
{\sl\large and Silvia Vaul\`a}
\footnote{E-mail: {\tt Silvia.Vaula@uam.es}}

\vspace{1cm}

{\it Instituto de F\'{\i}sica Te\'orica UAM/CSIC\\
Facultad de Ciencias C-XVI,  C.U.~Cantoblanco,  E-28049-Madrid, Spain}\\

\vspace{2cm}


{\bf Abstract}

\end{center}

\begin{quotation}\small
  We find the general form of all the supersymmetric configurations and
  solutions of $N=2,d=4$ Einstein-Yang-Mills theories. In the timelike case,
  which we study in great detail, giving many examples, the solutions to the
  full supergravity equations can be constructed from known flat spacetime
  solutions of the Bogomol'nyi equations. This allows the regular
  supersymmetric embedding in supergravity of regular monopole solutions ('t
  Hooft-Poyakov's, Weinberg's, Wilkinson and Bais's) but also embeddings of
  irregular solutions to the Bogomol'nyi equations which turn out to be
  regular black holes with different forms of non-Abelian hair once the
  non-triviality of the spacetime metric is taken into account. The attractor
  mechanism is realized in a gauge-covariant way.

  In the null case we determine the general equations that supersymmetric
  configurations and solutions must satisfy but we do not find relevant new
  supersymmetric solutions.

\end{quotation}

\newpage
\pagestyle{plain}

\tableofcontents

\newpage

\section*{Introduction}

Supersymmetric solutions of supergravity theories are playing a curcial r\^ole
in may of the developments that Superstring Theory has seen in the last few
years. The knowledge of all the possible solutions can lead to new interesting
models from which we can learn more about the possible vacua of the theory,
their potential holographic relations with  CFTs etc. Achieving a complete
characterization and classification of all the supersymmetric solutions of
supergravity theories is, thus, an important goal with may potential spin-offs.
 
Most of the work done so far in this subject has been focussed on
higher-dimensional ungauged theories. The 4-dimensional theories are equally
interesting, though, since they admit solutions such as the much-studied
families of charged extreme black holes found in Refs.~\cite{Gibbons:1982ih}
in ungauged $N=4,d=4$ supergravity and in
Refs.~\cite{Ferrara:1995ih,Behrndt:1997ny} in $N=2,d=4$ ungauged supergravity
coupled to vector multiplets.

The systematic study and classification of supersymmetric solutions of
4-dimensional supergravities was pioneered by Tod 25 years ago in
Ref.~\cite{Tod:1983pm}, in which he completeley solved the problem in pure,
ungauged, $N=2,d=4$ supergravity. Apart from another work on $N=4,d=4$
supergravity \cite{Tod:1995jf}, the subject was not reanimated until quite
recently: the problem was solved for pure, gauged $N=2,d=4$ supergravity
in Refs.~\cite{Caldarelli:2003pb,Cacciatori:2004rt,Cacciatori:2007vn}, for
ungauged $N=2,d=4$ supergravity coupled to vector supermultiplets in
Ref.~\cite{Meessen:2006tu}, and for the same theory with a $U(1)$ gauging in
Ref.~\cite{Cacciatori:2008ek}. Ungauged $N=2,d=4$ supergravity coupled to
vector supermultiplets and hypermultiplets was dealt with in
Ref.~\cite{Huebscher:2006mr}. Finally, the problem was solved for pure,
ungauged $N=4,d=4$ supergravity in \cite{Bellorin:2005zc} and for
matter-coupled $N=1,d=4$ supergravity in Ref.~\cite{Ortin:2008wj} for the
ungauged case (without superpotential but with non-trivial kinetic matrix) and
in Ref.~\cite{Gran:2008vx} for the gauged case with superpotential but without
kinetic matrix.

The cases considered so far (the above list) only include non-Abelian gauge
groups in the $N=1,d=4$ case, which does not admit supersymmetric
black-hole-type nor static monopole-like solutions. They can only exist in
$N>1,d=4$ theories. We are, therefore, led to consider $N>1,d=4$ theories with
non-Abelian gaugings. Some interesting non-Abelian monopole solutions are
known in gauged $N=4,d=4$ supergravity (namely, the Chamseddine-Volkov
monopole \cite{Chamseddine:1997nm}) and similar solutions must exist in
$N=2,d=4$ theories with non-Abelian gaugings, of which there is a much wider
variety.

In this paper we are going to study the classification of the supersymmetric
solutions of $N=2,d=4$ supergravity coupled to vector multiplets with
non-Abelian gaugins of the special-K\"ahler manifold (that we will call
$N=2,d=4$ super-Einstein-Yang-Mills (SEYM) theories for short) with the aim of
finding non-Abelian generalizations of the known supersymmetric extreme black
holes of the ungauged theories \cite{Behrndt:1997ny} and supersymmetric
embeddings of YM monopoles in supergravity. We will present the full
classification of the general solutions and the explicit construction of
several examples of the kind of solutions we were searching for: non-Abelian
monopoles and black-holes\footnote{ A few examples have been published in
  Refs.~\cite{Huebscher:2007hj,Meessen:2008kb}.  }.  We will actually give a
recipe (see Section~\ref{sec-recipe}) which allows the (not always regular)
embedding into $N=2,d=4$ supergravity of virtually any solution to the
Bogomol'nyi equation \cite{kn:Bog}.

While the existence of the monopoles was expected due to the existence of the
globally regular Chamseddine-Volkov \cite{Chamseddine:1997nm} and Harvey-Liu
\cite{Harvey:1991jr} monopole solutions, the existence of regular extreme
black-holes with non-trivial non-Abelian hair is a bit more surprising given
the existence of a \textit{non-Abelian baldness theorem}
\cite{Galtsov:1989ip}\footnote{
  See the review
  paper~\cite{Volkov:1998cc} for further on this subject.
}  
that states that
all the regular black-hole solutions of the $SU(2)$ Einstein-Yang-Mills theory
with colour charges are actually embeddings of solutions with Abelian
charges. The truly non-Abelian solutions of the EYM theory (the Bartnik-McKinnon
particle \cite{Bartnik:1988am} and its black hole generalizations 
\cite{Bizon:1990sr}), which are known only
numerically, do not have any asymptotic gauge charges. By contrast, some of
our solutions, which are fully analytical, do have genuinely non-Abelian
charges at infinity. Some of our solution also have non-Abelian hair that does
not result into any gauge charges at infinity. It is evident that the
non-Abelian baldness theorems do not apply to $N=2,d=4$ SEYM theories, which
have a different matter content, one in which the scalars play a prominent r\^ole.

One of the most interesting aspects of the supersymmetric black holes of
ungauged $N=2,d=4$ supergravity is the existence of the \textit{attractor
mechanism} for the values of the scalars
\cite{Ferrara:1995ih,Strominger:1996kf}: independently of their asymptotic
values, the values of the scalars on the event horizon are fully determined by
the conserved charges. As a result, the Bekenstein-Hawking entropy only depends
on conserved charges which is, by itself, a strong indication that it admits a
microscopic interpretation. It is of utmost interest, then, to study if and
how the attractor mechanism works for the supersymmetric non-Abelian black
holes in these theories. Our answer will be positive in a properly generalized
way.

The plan of this article is as follows: in Section~\ref{sec:GaugedN2} we will
review gauged $N=2$ $d=4$ supergravity without hypermultiplets (to which
we shall refer as $N=2,d=4$ SEYM), 
leaving information about isometries in Special Geometry and their
implementation in supergravity for the Appendix~\ref{sec-gauging}. In
Section~\ref{sec-setup} we shall discuss the generic characteristics of the
supersymmetric solutions, such as the Killing Spinor Equations and their
implications for the equations of motion. In Section~\ref{sec-timelike}, we
shall characterize the solutions in the timelike case obtaining the minimal
set of equations that need to be solved in order to have supersymmetric
solutions to $N=2$ $d=4$ SEYM. Section~\ref{sec-recipe} contains the
step-by-step procedure to construct supersymmetric solutions of the theory
starting with a solution of the Bogomol'nyi equations on $\mathbb{R}^{3}$ and
which we will use in Section~\ref{sec:EYMsols} to construct and study
different examples of solutions belonging to this class. These solutions
split up into globally regular monopoles and black holes. Appendices~\ref{sec:ST2n}
and~\ref{sec:bais} contain some complementary information needed for Sec.~(\ref{sec:EYMsols}). 
In Section~\ref{sec-null} we solve the null case. A discussion of our results and
our conclusions are contained in Section~\ref{sec:concl}.


\section{Gauged $N=2,d=4$ supergravity coupled to vector supermultiplets}
\label{sec:GaugedN2}

In this section we shall describe the action, equations of motion and supersymmetry
transformation rules of gauged $N=2,d=4$ supergravity coupled to vector
multiplets. In order to make this description brief, we only discuss the
differences with the ungauged case, which is described in detail in
Ref.~\cite{Huebscher:2006mr}. Some definitions and formulae related to the
gauging of holomorphic isometries of special K\"ahler manifolds are contained
in Appendix~\ref{sec-gauging}. We also refer the reader to
Ref.~\cite{Andrianopoli:1996cm}, the review Ref.~\cite{kn:toinereview}, and
the original works Refs.~\cite{deWit:1984pk,deWit:1984px} for more information.

The action restricted to the bosonic fields of these theories is

\begin{equation}
\label{eq:action}
\begin{array}{rcl}
 S & = & {\displaystyle\int} d^{4}x \sqrt{|g|}
\left[R +2\mathcal{G}_{ij^{*}}\mathfrak{D}_{\mu}Z^{i}
\mathfrak{D}^{\mu}Z^{*\, j^{*}}
+2\Im{\rm m}\mathcal{N}_{\Lambda\Sigma} 
F^{\Lambda\, \mu\nu}F^{\Sigma}{}_{\mu\nu}
 \right. \\
& & \\
& & \left. 
\hspace{2cm}
-2\Re{\rm e}\mathcal{N}_{\Lambda\Sigma}  
F^{\Lambda\, \mu\nu}{}^{\star}F^{\Sigma}{}_{\mu\nu}
-V(Z,Z^{*})
\right]\, ,
\end{array}
\end{equation}

\noindent
where the potential $V(Z,Z^{*})$,  is given by

\begin{equation}
V(Z,Z^{*})= 2\mathcal{G}_{ij^{*}}W^i W^*{}^{j^*}\, ,
\end{equation}

\noindent
where

\begin{equation}
W^{i}\; \equiv\; {\textstyle\frac{1}{2}}g\mathcal{L}^{*\, \Lambda}k_{\Lambda}{}^{i}\, .  
\end{equation}

\noindent
In these expressions $g$ is the gauge coupling constant, the $k_{\Lambda}{}^{i}(Z)$ are holomorphic
Killing vectors of $\mathcal{G}_{ij^{*}}$ and $\mathfrak{D}$ the gauge
covariant derivative (also K\"ahler-covariant when acting on fields of
non-trivial K\"ahler weight) and is defined in Appendix~\ref{sec-gauging}.

This is not the most general gauged $N=2,d=4$ supergravity: if the
$\mathfrak{sp}(2\bar{n})$ matrices $\mathcal{S}_{\Lambda}$ that provide a
representation of the Lie algebra of the gauge group $G_{V}$, see
Eq.~(\ref{eq:SLiealgebra}), are written in the form

\begin{equation}
\mathcal{S}_{\Lambda} =
\left(
  \begin{array}{cc}
a_{\Lambda}{}^{\Omega}{}_{\Sigma} & b_{\Lambda}{}^{\Omega\Sigma} \\
& \\
c_{\Lambda\Omega\Sigma} & d_{\Lambda\Omega}{}^{\Sigma} \\
\end{array}
\right)\, , 
\end{equation}

\noindent
we are then considering only the cases in which $b=0$, so that only symmetries
of the action are gauged, and $c=0$. This last restriction is only made for
the sake of simplicity as theories in which symmetries with $c\neq 0$ are
gauged have complicated Chern-Simons terms.

Within this restricted class of theories, then, we can use Eqs.~(\ref{eq:LP0}) and
(\ref{eq:LKfP}) to rewrite the potential as

\begin{equation}
V(Z,Z^{*}) =
{\textstyle\frac{1}{2}}g^{2}f^{*\Lambda\, i}f^{\Sigma}{}_{i} 
\mathcal{P}_{\Lambda}\mathcal{P}_{\Sigma}
 =-{\textstyle\frac{1}{4}}g^{2} (\Im{\rm m}\mathcal{N})^{-1|\Lambda\Sigma}
\mathcal{P}_{\Lambda}\mathcal{P}_{\Sigma}\, .
\end{equation}

\noindent
Then, since $\Im{\rm m}\mathcal{N}_{\Lambda\Sigma}$ is negative definite and
the momentum map is real, the potential is positive semi-definite
$V(Z,Z^{*})\geq 0$.  For constant values of the scalars $V(Z,Z^{*})$ behaves as a
non-negative cosmological constant $\Lambda=V(Z,Z^{*})/2$ which leads to Minkowski
($\Lambda=0$) or $dS$ ($\Lambda > 0$) vacua. The latter cannot be maximally
supersymmetric, however.

For convenience, we denote the bosonic equations of motion by

\begin{equation}
\mathcal{E}_{a}{}^{\mu}\equiv 
-\frac{1}{2\sqrt{|g|}}\frac{\delta S}{\delta e^{a}{}_{\mu}}\, ,
\hspace{.5cm}
\mathcal{E}^{i} \equiv -\frac{\mathcal{G}^{ij^{*}}}{2\sqrt{|g|}}
\frac{\delta S}{\delta Z^{*j^{*}}}\, ,
\hspace{.5cm}
\mathcal{E}_{\Lambda}{}^{\mu}\equiv 
\frac{1}{8\sqrt{|g|}}\frac{\delta S}{\delta A^{\Lambda}{}_{\mu}}\, .
\end{equation}

\noindent
and the Bianchi identities for the vector field strengths by

\begin{equation}
\label{eq:Bianchiidentities}
\mathcal{B}^{\Lambda\, \mu} \equiv \mathfrak{D}_{\nu}\star F^{\Lambda\,
  \nu\mu}\, ,
\,\,\,\,\, 
\star\mathcal{B}^{\Lambda}\equiv -\mathfrak{D}F^{\Lambda}\, .
\end{equation}

Then, using the action Eq.~(\ref{eq:action}), we find

\begin{eqnarray}
\mathcal{E}_{\mu\nu} & = & 
G_{\mu\nu}
+2\mathcal{G}_{ij^{*}}[\mathfrak{D}_{\mu}Z^{i} \mathfrak{D}_{\nu}Z^{*\, j^{*}}
-{\textstyle\frac{1}{2}}g_{\mu\nu}
\mathfrak{D}_{\rho}Z^{i}\mathfrak{D}^{\rho}Z^{*\, j^{*}}]\nonumber \\
& & \nonumber \\
& & 
+8\Im {\rm m}\mathcal{N}_{\Lambda\Sigma}
F^{\Lambda\, +}{}_{\mu}{}^{\rho}F^{\Sigma\, -}{}_{\nu\rho}
+{\textstyle\frac{1}{2}}g_{\mu\nu}V(Z,Z^{*})\, ,
\label{eq:Emn}\\
& & \nonumber \\
\mathcal{E}_{\Lambda}{}^{\mu} & = &
\mathfrak{D}_{\nu} \star F_{\Lambda}{}^{\nu\mu}
+{\textstyle\frac{1}{2}}g\Re{\rm e}
(k_{\Lambda\, i^{*}}\mathfrak{D}^{\mu} Z^{*}{}^{i^{*}})\, ,
\label{eq:EmL}\\
& & \nonumber \\
\mathcal{E}^{i} &=& 
\mathfrak{D}^{2}Z^{i} 
+\partial^{i}
\tilde{F}_{\Lambda}{}^{\mu\nu}\star F^{\Lambda}{}_{\mu\nu}
+{\textstyle\frac{1}{2}} \partial^{i}V(Z,Z^{*})\, .
\label{eq:Ei2}
\end{eqnarray}

In differential-form notation, the Maxwell equation takes the form

\begin{equation}
\label{eq:EmLdiff}
-\star \hat{\mathcal{E}}_{\Lambda} =\mathfrak{D}F_{\Lambda}
-
{\textstyle\frac{1}{2}}g\star\Re{\rm e}\,(k^{*}_{\Lambda\, i}\mathfrak{D}Z^{i}) \; .
\end{equation}

For vanishing fermions, the supersymmetry transformation rules of the
fermions are

\begin{eqnarray}
\delta_{\epsilon}\psi_{I\, \mu} & = & 
\mathfrak{D}_{\mu}\epsilon_{I} 
+\epsilon_{IJ}T^{+}{}_{\mu\nu}\gamma^{\nu}\epsilon^{J}\, ,
\label{eq:gravisusyrule}\\
& & \nonumber \\
\delta_{\epsilon}\lambda^{Ii} & = & 
i\not\!\!\mathfrak{D} Z^{i}\epsilon^{I} 
+\epsilon^{IJ}[\not\!G^{i\, +} +W^{i}]\epsilon_{J}\, .
\label{eq:gaugsusyrule}
\end{eqnarray}

\noindent
$\mathfrak{D}_{\mu}\epsilon_{I}$ is given in Eq.~(\ref{eq:dei}).

The supersymmetry transformations of the bosons are the same as in the
ungauged case

\begin{eqnarray}
  \delta_{\epsilon} e^{a}{}_{\mu} & = & 
-{\textstyle\frac{i}{4}} (\bar{\psi}_{I\, \mu}\gamma^{a}\epsilon^{I}
+\bar{\psi}^{I}{}_{\mu}\gamma^{a}\epsilon_{I})\, ,
\label{eq:susytranse}\\
& & \nonumber \\ 
  \delta_{\epsilon} A^{\Lambda}{}_{\mu} & = & 
{\textstyle\frac{1}{4}}
(\mathcal{L}^{\Lambda\, *}
\epsilon^{IJ}\bar{\psi}_{I\, \mu}\epsilon_{J}
+
\mathcal{L}^{\Lambda}
\epsilon_{IJ}\bar{\psi}^{I}{}_{\mu}\epsilon^{J}) \nonumber \\
& & \nonumber \\
& & 
+
{\textstyle\frac{i}{8}}(f^{\Lambda}{}_{i}\epsilon_{IJ}
\bar{\lambda}^{Ii}\gamma_{\mu}
\epsilon^{J}
+
f^{\Lambda *}{}_{i^{*}}\epsilon^{IJ}
\bar{\lambda}_{I}{}^{i^{*}}\gamma_{\mu}\epsilon_{J})\, ,
\label{eq:susytransA}\\
& & \nonumber \\
  \delta_{\epsilon} Z^{i} & = & 
{\textstyle\frac{1}{4}} \bar{\lambda}^{Ii}\epsilon_{I}\, .
\label{eq:susytransZ}
\end{eqnarray}


\section{Supersymmetric configurations: general setup}
\label{sec-setup}

Our first goal is to find all the bosonic field configurations
$\{g_{\mu\nu},F^{\Lambda}{}_{\mu\nu}, Z^{i}\}$ for which the Killing
spinor equations (KSEs):

\begin{eqnarray}
\delta_{\epsilon}\psi_{I\, \mu} & = & 
\mathfrak{D}_{\mu}\epsilon_{I} 
+\epsilon_{IJ}T^{+}{}_{\mu\nu}\gamma^{\nu}\epsilon^{J} =0\, ,
\label{eq:KSE1}\\
& & \nonumber \\
\delta_{\epsilon}\lambda^{Ii} & = & 
i\!\not\!\!\mathfrak{D} Z^{i}\epsilon^{I} 
+\epsilon^{IJ}[\not\!G^{i\, +}+W^{i}]\epsilon_{J}=0\, ,
\label{eq:KSE2}
\end{eqnarray}

\noindent
admit at least one solution. 

Our second goal will be to identify among all the supersymmetric field
configurations those that satisfy all the equations of motion (including
the Bianchi identities).

Let us initiate the analysis of the KSEs by studying their integrability
conditions.


\subsection{Killing Spinor Identities (KSIs)}
\label{sec-KSIs}

The off-shell equations of motion of the bosonic fields of bosonic
supersymmetric configurations satisfy certain relations known as
(\textit{Killing spinor identities}, KSIs)
\cite{Kallosh:1993wx,Bellorin:2005hy}. If we assume that the Bianchi
identities are always identically satisfied everywhere, the KSIs only depend
on the supersymmetry transformation rules of the bosonic fields. These are
identical for the gauged and ungauged theories, implying that their KSIs are
also identical. If we do not assume that the Bianchi identities are
identically satisfied everywhere, then they also occur in the KSIs, which now
have to be found via the integrability conditions of the KSEs. In the ungauged
case they occur in symplectic-invariant combinations, as one would expect, and
take the form \cite{Meessen:2006tu}

\begin{eqnarray}
\mathcal{E}_{a}{}^{\mu}\gamma^{a}\epsilon^{I}
-4i\epsilon^{IJ}
\langle\, \mathcal{E}^{\mu} \mid \mathcal{V}\, \rangle \epsilon_{J}
& = & 0\, , \label{eq:ksi1}\\
& & \nonumber \\
\mathcal{E}^{i}\epsilon^{I} 
-2i\epsilon^{IJ}\langle\, \not\!\mathcal{E} \mid \mathcal{U}^{*i}\, \rangle 
\epsilon_{J} & = & 0\, ,
\label{eq:ksi2}
\end{eqnarray}

\noindent
where

\begin{equation}
\mathcal{E}^{a}\equiv
\left(
  \begin{array}{rcl}
  \mathcal{B}^{\Lambda\, a} \\ \mathcal{E}_{\Lambda}{}^{a}  
  \end{array}
\right)\, .  
\end{equation}

We have checked through explicit computation that these relations remain valid in
the non-Abelian gauged case at hand.

Taking products of these expressions with Killing spinors and gamma matrices,
one can derive KSIs involving the bosonic equations and tensors constructed as
bilinears of the commuting Killing spinors.\footnote{See the appendix in
  Ref.~\cite{Bellorin:2005zc} for the definitions and properties of these
  bilinears.}
In the case in which the bilinear $V^{\mu}\equiv
i\bar{\epsilon}^{I}\gamma^{\mu}\epsilon_{I}$ is a timelike vector
(referred to as the \textit{timelike case}), one obtains \cite{Bellorin:2006xr}
 the following identities 
({\em w.r.t.\/} an orthonormal
frame with $e_{0}{}^{\mu}\equiv V^{\mu}/|V|$) 

\begin{eqnarray}
\label{eq:ksi3}
\mathcal{E}^{ab} & = & \eta^{a}{}_{0}  \eta^{b}{}_{0}\mathcal{E}^{00}\, ,\\
& & \nonumber \\
\label{eq:ksi4}
\langle\, \mathcal{V}/X \mid \, \mathcal{E}^{a} \, \rangle  & = & 
{\textstyle\frac{1}{4}}|X|^{-1} \mathcal{E}^{00}\delta^{a}{}_{0}\, ,\\
& & \nonumber \\
\label{eq:ksi5}
\langle\, \mathcal{U}^{*}_{i^{*}}\mid \, \mathcal{E}^{a} \, \rangle  & = & 
{\textstyle\frac{1}{2}} e^{-i\alpha}\mathcal{E}_{i^{*}}\delta^{a}{}_{0}\, ,
\end{eqnarray}

\noindent
where $X\equiv {\textstyle\frac{1}{2}}
\varepsilon_{IJ}\bar{\epsilon}^{I}\epsilon^{J}$ and is non-zero in the
timelike case.

As discussed in Ref.~\cite{Bellorin:2006xr}, these identities contain a great
deal of physical information.  In this paper we shall exploit only one fact,
namely the fact that if the Maxwell equation and the Bianchi identity are
satisfied for a supersymmetric configuration, then so are the rest of the
equations of motion.  The strategy to be followed is, therefore, to first
identify the supersymmetric configurations and impose the Maxwell equations
and the Bianchi identities. This will lead to some differential equations that
need be solved in order to construct a supersymmetric solution.

In the case in which $V^{\mu}$ is a null vector (the \textit{null case}),
renaming it as $l^{\mu}$ for reasons of clarity,  one gets

\begin{eqnarray}
(\mathcal{E}_{\mu\nu}
-{\textstyle\frac{1}{2}}g_{\mu\nu}\mathcal{E}^{\rho}{}_{\rho}) l^{\nu}
= 
(\mathcal{E}_{\mu\nu}
-{\textstyle\frac{1}{2}}g_{\mu\nu}\mathcal{E}^{\rho}{}_{\rho}) m^{\nu}
& = & 0\, ,
\label{eq:ksinull0}\\
& & \nonumber \\
\mathcal{E}_{\mu\nu} l^{\nu}  
=
\mathcal{E}_{\mu\nu} m^{\nu}
& = & 0\, ,
\label{eq:ksinull1}\\
& & \nonumber \\
\langle\,  \mathcal{V} \mid\, \mathcal{E}^{\mu}\, \rangle & = & 0\, ,
\label{eq:ksinull2}\\
& & \nonumber \\
\langle\, \mathcal{U}^{*}_{i^{*}}\mid \, \mathcal{E}^{\mu} \, \rangle\,  
l_{\mu}
= 
\langle\, \mathcal{U}^{*}_{i^{*}}\mid \, \mathcal{E}^{\mu} \, \rangle\, 
 m^{*}_{\mu}
& = & 0\, ,
\label{eq:ksinull3}\\
& & \nonumber \\
\mathcal{E}^{i} & = & 0\, ,
\label{eq:ksinull4}
\end{eqnarray}

\noindent
where $l,n,m,m^{*}$ is a null tetrad constructed with the Killing spinor
$\epsilon^{I}$ and an auxiliary spinor $\eta$ as explained in
Ref.~\cite{Meessen:2006tu}. 

These identities imply that the only independent equations of motion that one
has to check on supersymmetric configurations are $\mathcal{E}_{\mu\nu}
n^{\mu} n^{\nu}$ and $\langle\, \mathcal{U}^{*}_{i^{*}}\mid \,
\mathcal{E}_{\mu} \, \rangle\, n^{\mu}$. As before, these are the equations
that need to be imposed in order for a supersymmetric configuration to be 
a supersymmetric solution.


\subsection{Killing equations for the bilinears}
\label{sec-bilinearkse}

In order to find the most general background admitting a solution to the KSEs,
Eqs.~(\ref{eq:KSE1}) and (\ref{eq:KSE2}), we shall assume that the background
admits one Killing spinor. Using this assumption we will derive consistency
conditions that the background must satisfy, after which we will prove that
these necessary conditions are also sufficient.

It is convenient to work with spinor bilinears, and consequently we start by
deriving equations for these bilinears by contracting the KSEs with gamma
matrices and Killing spinors.

{}From the gravitino supersymmetry transformation rule
Eq.~(\ref{eq:gravisusyrule}) we get the independent equations

\begin{eqnarray}
\label{eq:TV}
\mathfrak{D}_{\mu}X & = & -i T^{+}{}_{\mu\nu}V^{\nu}  \, ,
\\
& & \nonumber \\
\mathfrak{D}_{\mu} V^{I}{}_{J\, \nu} & = & 
i \delta^{I}{}_{J} [XT^{*-}{}_{\mu\nu} -X^{*}T^{+}{}_{\mu\nu}]
\\
& & \nonumber \\
& & 
-i[\epsilon^{IK}T^{*-}{}_{\mu\rho}\Phi_{KJ}{}^{\rho}{}_{\nu}
-\epsilon_{JK}T^{+}{}_{\mu\rho}\Phi^{KI}{}^{\rho}{}_{\nu}]\, ,
\end{eqnarray}

\noindent
which have the same functional form as their equivalents in the ungauged case.
Hence, as in the ungauged case, $V^{\mu}$ is a Killing vector and the 1-form
$\hat{V}\equiv V_{\mu}dx^{\mu}$ satisfies the equation

\begin{equation}
\label{eq:dV}
d\hat{V} = 4i [XT^{*-} -X^{*}T^{+}]\, .
\end{equation}

\noindent
The remaining 3 independent 1-forms $\hat{V}^{x}\equiv
\frac{1}{\sqrt{2}} V^{I}{}_{J\, \mu}\sigma^{x\, J}{}_{I} dx^{\mu}$
($x=1,2,3$ and the $\sigma^{x}$ are the Pauli matrices) are exact,
{\em i.e.}

\begin{equation}
d\hat{V}^{x} =0\, .
\end{equation}

{}From the gauginos' supersymmetry transformation rules,
Eqs.~(\ref{eq:gaugsusyrule}), we obtain 

\begin{eqnarray}
V^{I}{}_{K}{}^{\mu}\mathfrak{D}_{\mu}Z^{i} +\epsilon^{IJ}\Phi_{KJ}{}^{\mu\nu}
G^{i\, +}{}_{\mu\nu} +W^{i}\epsilon^{IJ}M_{KJ}& = & 0\, ,\\
& & \nonumber \\
iM^{KI}\mathfrak{D}_{\mu}Z^{i} +i\Phi^{KI}{}_{\mu}{}^{\nu}\mathfrak{D}_{\nu}Z^{i}
-4i\epsilon^{IJ}V^{K}{}_{J}{}^{\nu}G^{i\, +}{}_{\mu\nu}
-i W^{i}\epsilon^{IJ}V^{K}{}_{J\;\mu} & = & 0\, .
\end{eqnarray}

\noindent
The trace of the first equation gives

\begin{equation}
  \label{eq:VdZ}
V^{\mu}\mathfrak{D}_{\mu}Z^{i}+2X W^{i}=0\, ,  
\end{equation}

\noindent
while the antisymmetric part of the second equation gives

\begin{equation}
\label{eq:GV}
2 X^{*}\mathfrak{D}_{\mu}Z^{i}+4 G^{i\, +}{}_{\mu\nu}V^{\nu} 
+W^{i} V_{\mu}=0\, .  
\end{equation}

\noindent
The well-known special geometry completeness relation implies that 

\begin{equation}
\label{eq:completeness}
F^{\Lambda\, +}=
i\mathcal{L}^{*\, \Lambda}T^{+} +2f^{\Lambda}{}_{i}G^{i\, +}\, ,  
\end{equation}

\noindent
which allows us to combine Eqs.~(\ref{eq:TV}) and (\ref{eq:GV}), 
as to obtain

\begin{equation}
\label{eq:FV}
\begin{array}{rcl}
V^{\nu}F^{\Lambda\, +}{}_{\nu\mu} 
& = &
i\mathcal{L}^{*\, \Lambda}V^{\nu}T^{+}{}_{\nu\mu} 
+2f^{\Lambda}{}_{i}V^{\nu}G^{i\, +}{}_{\nu\mu}
\\
& & \\
& = &
\mathcal{L}^{*\, \Lambda}\mathfrak{D}_{\mu}X
+X^{*}\mathfrak{D}_{\mu} \mathcal{L}^{\Lambda} 
+{\textstyle\frac{1}{2}}W^{i} V_{\mu}\, .\\
\end{array}
\end{equation}

Multiplying this equation by $V^{\mu}$ and using Eq.~(\ref{eq:VdZ}), we find

\begin{equation}
\label{eq:VdX}
V^{\mu}\mathfrak{D}_{\mu}X=0\, .
\end{equation}

At this point in the investigation, it is convenient to take into account the
norm of the Killing vector $V^{\mu}$: we shall investigate the timelike case
in Section~\ref{sec-timelike} and the null case in Section~\ref{sec-null}.


\section{The timelike case}
\label{sec-timelike}


\subsection{The vector field strengths}
\label{sec-vectorfieldstrengths}

As is well-known, the contraction of the (anti-) self-dual part of a 2-form
with a non-null vector, such as $V^{\mu}$ in the current timelike case, completely
determines the 2-form, {\em i.e.\/}

\begin{equation}
\label{eq:decomposition2}
C^{\Lambda\, +}{}_{\mu}\equiv V^{\nu}
F^{\Lambda\, +}{}_{\nu\mu}\,\,\, \Rightarrow\,\,\,
F^{\Lambda\, +}=V^{-2}[\hat{V}\wedge \hat{C}^{\Lambda\, +}
+ i\,\star\!(\hat{V}\wedge \hat{C}^{\Lambda\, +})]\, .
\end{equation}

\noindent
As $C^{\Lambda\, +}{}_{\mu}$ is given by Eq.~(\ref{eq:FV}), the vector
field strengths are written in terms of the scalars $Z^{i}$, $X$ and the vector
$V$.  Observe that the component of $C^{\Lambda\, +}{}_{\mu}$ proportional to
$V^{\mu}$ is projected out in this formula: this implies that the field
strengths have the same functional form as in the ungauged case.  The
covariant derivatives that appear in the r.h.s., however, contain explicitly
the vector potentials.

The next item on the list is the determination of the spacetime metric:


\subsection{The metric}
\label{sec-metric}

As in the ungauged case we define a time coordinate $t$ by

\begin{equation}
V^{\mu}\partial_{\mu}\equiv\sqrt{2} \partial_{t}\, .
\end{equation}

\noindent
Unlike the ungauged case, however, the scalars in a supersymmetric
configuration need not automatically be time-independent: with respect to the
chosen $t$-coordinate Eq.~(\ref{eq:VdZ}) takes the form

\begin{equation}
\label{eq:effe1}
\partial_{t}Z^{i} +gA^{\Lambda}{}_{t}k_{\Lambda}{}^{i}
+\sqrt{2}X W^{i} = 
\partial_{t}Z^{i} +g(A^{\Lambda}{}_{t}
+{\textstyle\frac{1}{\sqrt{2}}}X\mathcal{L}^{*\Lambda}) k_{\Lambda}{}^{i}
=0\, .
\end{equation}

\noindent
It is convenient to choose a $G_{V}$ gauge in which the complex fields
$Z^{i}$ are time-independent, and one accomplishing just that is

\begin{equation}
\label{eq:gaugechoice}
A^{\Lambda}{}_{t}\; =\;
-\sqrt{2}\Re{\rm e}\, (X\mathcal{L}^{*\Lambda})\; =\;
-\sqrt{2}|X|^{2}\Re{\rm e}\,
(\mathcal{L}^{*\Lambda}/X^{*})\, .
\end{equation}

\noindent
This gauge choice reduces Eq.~(\ref{eq:effe1}) to

\begin{equation}
\partial_{t}Z^{i} 
-{\textstyle\frac{1}{\sqrt{2}}}gX^{*}\mathcal{L}^{\Lambda} k_{\Lambda}{}^{i}
\; =\; \partial_{t}Z^{i} 
\; =\; 0\, ,
\end{equation}

\noindent
on account of Eq.~(\ref{eq:LK0}). It should be pointed out that this gauge choice
is identical to the expression for $A_{t}$ obtained in ungauged case in 
Refs.~\cite{Meessen:2006tu,Huebscher:2006mr}.
{}Further, using the above $t$-independence and gauge choice in Eq.~(\ref{eq:VdX}),
we can derive

\begin{equation}
\begin{array}{rcl}
\partial_{t}X +i\mathcal{Q}_{t}X
+igA^{\Lambda}{}_{t}\mathcal{P}_{\Lambda} & = &  
\partial_{t}X 
+{\textstyle\frac{1}{2}}
(\partial_{t}Z^{i}\partial_{i}\mathcal{K}
-\mathrm{c.c})X
+igA^{\Lambda}{}_{t}\mathcal{P}_{\Lambda}X 
\\
& & \\
& = & 
\partial_{t}X 
-\sqrt{2}ig|X|^{2}\Re{\rm e}\,
(\mathcal{L}^{*\Lambda}/X^{*})
\mathcal{P}_{\Lambda}X 
\\
& & \\
& = & 
\partial_{t}X =0\, ,
\\
\end{array}
\end{equation}

\noindent
where we made use of Eq.~(\ref{eq:LP0}) and the reality of $\mathcal{P}_{\Lambda}$.
Thus, with the standard coordinate choice and the gauge choice
(\ref{eq:gaugechoice}) the scalars $Z^{i}$ and $X$ are time-independent.

Using the exactness of the 1-forms $\hat{V}^{x}$ to define spacelike
coordinates $x^{x}$ by

\begin{equation}
\hat{V}^{x} \;\equiv\; dx^{x}\, ,
\end{equation}

\noindent
the metric takes on the form

\begin{equation}
\label{eq:metric}
ds^{2} \; =\; 2|X|^{2}(dt+\hat{\omega})^{2} -\frac{1}{2|X|^{2}}dx^{x}dx^{x}
\hspace{1cm}
(x,y=1,2,3)\; ,
\end{equation}

\noindent
where $\hat{\omega}=\omega_{\underline{i}}dx^{i}$ is a time-independent 1-form.
This 1-form is determined by the following condition

\begin{equation}
\label{eq:do}
d\hat{\omega} = {\textstyle\frac{i}{2\sqrt{2}}}\star 
\left[\hat{V}\wedge \frac{X\mathfrak{D}X^{*}
-X^{*}\mathfrak{D}X}{|X|^{4}}\right] 
\end{equation}

\noindent 
Observe that this equation has, apart from a different definition of the
covariant derivative, the same functional form as in the ungauged case; before
we start rewriting the above result in order to get to the desired result,
however, we would like to point out that due to the stationary character of
the metric, the resulting covariant derivatives on the transverse
$\mathbb{R}^{3}$ contain a piece proportional to $\omega_{\underline{x}}$.
The end-effect of this pull-back is that we introduce a new connection on
$\mathbb{R}^{3}$, denoted by $\tilde{\mathfrak{D}}_{\underline{x}}$, which is
formally the same as $\mathfrak{D}_{\underline{x}}$ but for a redefinition of the
gauge field, {\em i.e.\/}

\begin{equation}
\label{eq:TildeDer}
\tilde{A}^{\Lambda}{}_{\underline{x}} \; =\; 
A^{\Lambda}{}_{\underline{x}}- \omega_{\underline{x}}\ A^{\Lambda}{}_{t} \; .
\end{equation}

In order to compare the results in this article with the ones found in
\cite{Meessen:2006tu}, we introduce the real symplectic sections $\mathcal{I}$
and $\mathcal{R}$ defined by

\begin{equation}
\label{eq:realsections}
\mathcal{R}\equiv \Re{\rm e}(\mathcal{V}/X)\, ,
\hspace{1.5cm}
\mathcal{I}\equiv \Im{\rm m}(\mathcal{V}/X)\, .
\end{equation}

\noindent
$\mathcal{V}$ is the symplectic section defining special geometry and 
thence satisfies

\begin{equation}
\label{eq:SGDefFund}
\mathcal{V} = 
\left( \!
\begin{array}{c}
\mathcal{L}^{\Lambda}\\
\mathcal{M}_{\Sigma}\\
\end{array}
\!\right)\, ,
\hspace{1cm}
\langle \mathcal{V}\mid\mathcal{V}^{*}\rangle 
 \equiv  
\mathcal{L}^{*\, \Lambda}\mathcal{M}_{\Lambda} 
-\mathcal{L}^{\Lambda}\mathcal{M}^{*}_{\Lambda}
= -i\, .
\end{equation}

\noindent
This then implies that our gauge choice can be expressed in the form 

\begin{equation}
\label{eq:gaugechoice2}
A^{\Lambda}{}_{t}  =    
-\sqrt{2}|X|^{2}\mathcal{R}^{\Lambda}\, ,
\end{equation}

\noindent
and that the metric function $|X|$ can be written as

\begin{equation}
\label{eq:Gtt}
\frac{1}{2|X|^{2}}  =  \langle\, \mathcal{R}\mid\mathcal{I} \,\rangle\, ,
\end{equation}

\noindent
Similar to the ungauged case, we can then rewrite Eq.~(\ref{eq:do}) as

\begin{equation}
\label{eq:oidi}
(d\hat{\omega})_{xy} =2 \epsilon_{xyz}
\langle\,\mathcal{I}\mid \tilde{\mathfrak{D}}_{z}\mathcal{I}\, \rangle\, ,  
\end{equation}

\noindent
whose integrability condition reads

\begin{equation}
\label{eq:IntConOmega}
\langle\,\mathcal{I}\mid \tilde{\mathfrak{D}}_{x}\tilde{\mathfrak{D}}_{x}
\mathcal{I}\, \rangle =0\, ,    
\end{equation}

\noindent
and we shall see that, apart from possible singularities
\cite{Denef:2000nb,Bellorin:2006xr}, the integrability condition is
identically satisfied for supersymmetric solutions.


\subsection{Solving the Killing spinor equations}
\label{sec-solvingKSEtimelike}

In the previous sections we have found that timelike supersymmetric
configurations have a metric and vector field strengths given by
Eqs.~(\ref{eq:metric},\ref{eq:FV}) and (\ref{eq:decomposition2}) in terms of
the scalars $X,Z^{i}$. It is easy to see that all configurations of this form
admit spinors $\epsilon_{I}$ that satisfy the Killing spinor equations
(\ref{eq:KSE1},\ref{eq:KSE2}).  The Killing spinors have exactly the same form
as in the ungauged case \cite{Meessen:2006tu}

\begin{equation}
\epsilon_{I}=X^{1/2}\epsilon_{I\, 0}\, ,
\hspace{1cm}
\partial_{\mu}\epsilon_{I\, 0}=0\, ,
\hspace{1cm}  
\epsilon_{I\, 0} +i\gamma_{0}
  \epsilon_{IJ}\epsilon^{J}{}_{0}=0\, .  
\end{equation}

We conclude that we have identified all the supersymmetric configurations of
the theory.


\subsection{Equations of motion}
\label{sec-eoms}

The results of Section~\ref{sec-KSIs} imply that in order to have a classical 
solution, we only need to impose the
Maxwell equations and Bianchi identities on the supersymmetric configurations.
In this section, then, we will discuss the differential equations arrising
from the applying the Maxwell and Bianchi equations on the supersymmetric
configurations obtained thus far.

As we mentioned in Section~\ref{sec-vectorfieldstrengths} the field strengths
of supersymmetric configurations take the same form as in the ungauged case
\cite{Meessen:2006tu} with the K\"ahler-covariant derivatives replaced by
K\"ahler- and $G_{V}$-covariant derivatives. Therefore, the symplectic vector
of field strengths and dual field strengths takes the form

\begin{equation}
\label{eq:F}
F=\frac{1}{2|X|^{2}}
\left\{\hat{V}\wedge \mathfrak{D}(|X|^{2}\mathcal{R}) 
-\star [\hat{V}\wedge 
\Im{\rm m}(\mathcal{V}^{*}\mathfrak{D}X
+X^{*}\mathfrak{D}\mathcal{V})]
\right\}\, .  
\end{equation}

\noindent
Operating in the first term we can rewrite it in the form

\begin{equation}
F=-{\textstyle\frac{1}{2}}
\left\{ \mathfrak{D}(\mathcal{R}\hat{V}) 
-2\sqrt{2}|X|^{2}\mathcal{R}d\hat{\omega} 
+\star \left[\hat{V}\wedge 
\frac{\Im{\rm m}(\mathcal{V}^{*}\mathfrak{D}X
+X^{*}\mathfrak{D}\mathcal{V})}{|X|^{2}}\right]
\right\}\, ,  
\end{equation}

\noindent
and using the equation of 1-form $\hat{\omega}$, Eq.~(\ref{eq:do}), which is also
identical to that of the ungauged case with the same substitution of covariant
derivatives,  we arrive at

\begin{equation}\label{eq:Fsuper}
F=-{\textstyle\frac{1}{2}}
\left\{ \mathfrak{D}(\mathcal{R}\hat{V}) 
+\star (\hat{V}\wedge \mathfrak{D} \mathcal{I})
\right\}\, .  
\end{equation}

In what follows we shall use the following Vierbein ($e^{0},e^{x}$) and the
corresponding directional derivatives ($\theta_{0},\theta_{a}$), normalized as
$e^{a}(\theta_{b})=\delta^{a}{}_{b}$, that are given by

\begin{equation}
\label{eq:Maxwell3}
\begin{array}{lclclcl}
e^{0} 
& = & 
\sqrt{2}|X|\left( dt\ +\ \omega\right)\, , 
&\hspace{.5cm}&
\theta_{0} & = & \textstyle{1\over \sqrt{2}}|X|^{-1}\ \partial_{t} \; ,
\\
& & & & & & \\
e^{x} 
& = & \textstyle{1\over \sqrt{2}}|X|^{-1}\ dx^{x}\, , 
&  &
\theta_{x} & =& \sqrt{2}|X|
\left( \partial_{\underline{x}} \ -\ \omega_{\underline{x}}\partial_{t}\right) \; .
\end{array}
\end{equation}

\noindent
With respect to this basis we 

\begin{equation}
\label{eq:Maxwell4}
V^{\mu}\partial_{\mu} \; =\; 
2|X|\ \theta_{0}\, , \hspace{1cm}
  \hat{V} \; =\; 2|X|\ e^{0} \; ,
\end{equation}

\noindent
and the gauge fixing (\ref{eq:gaugechoice}) and the constraint (\ref{eq:effe1})
read

\begin{equation}
\label{eq:Maxwell5}
A^{\Lambda}{}_{0} \; = \; 
-|X|\ \mathcal{R}^{\Lambda}\, ,
\hspace{1cm}
X^{*}\ \mathfrak{D}_{0}Z^{i} \; =\; -|X|\ W^{i} \; .
\end{equation}

\noindent
The equation that the spacelike components of the field strengths
$F^{\Lambda}{}_{\underline{x}\underline{y}}$ satisfy can be rewritten in the form

\begin{equation}
\label{eq:Maxwell6}
\tilde{F}^{\Lambda}{}_{\underline{x}\underline{y}} 
\; =\; 
-{\textstyle\frac{1}{\sqrt{2}}}\epsilon_{xyz} 
\tilde{\mathfrak{D}}_{\underline{z}}\mathcal{I}^{\Lambda} \; ,
\end{equation}

\noindent 
where the tilde indicates that the gauge field that appears in this equation
is the combination $\tilde{A}^{\Lambda}{}_{\underline{x}}$ defined in
Eq.~(\ref{eq:TildeDer}). 

This equation is easily recognized as the well-known Bogomol'nyi equation
\cite{kn:Bog} for the connection $\tilde{A}^{\Lambda}{}_{\underline{x}}$ and
the real ``Higgs'' field $\mathcal{I}^{\Lambda}$ on $\mathbb{R}^{3}$. Its
integrability condition uses the Bianchi identity for the 3-dimensional
gauge connection $\tilde{A}^{\Lambda}{}_{\underline{x}}$ and, as it turns out,
is equivalent to the complete Bianchi identity for the 4-dimensional gauge
connection $A^{\Lambda}{}_{\mu}$. It takes the form

\begin{equation}
\label{eq:Maxwell9}
\tilde{\mathfrak{D}}_{\underline{x}}\tilde{\mathfrak{D}}_{\underline{x}}\ 
\mathcal{I}^{\Lambda} 
\; =\; 
0 \; .
\end{equation}

Taking the Maxwell equation in form notation
Eq.~(\ref{eq:EmLdiff}) and  using heavily the formulae in
Appendix~\ref{sec-gauging} we find that all the components are satisfied
(as implied by the KSIs) except for one which leads to the equation

\begin{equation}
\label{eq:Maxwell16}
\tilde{\mathfrak{D}}_{\underline{x}}
\tilde{\mathfrak{D}}_{\underline{x}}\mathcal{I}_{\Lambda} 
\; =\; 
\textstyle{1\over 2}g^{2}\ 
\left[
f_{\Lambda (\Sigma}{}^{\Gamma}f_{\Delta )\Gamma}{}^{\Omega}\ 
\mathcal{I}^{\Sigma}\mathcal{I}^{\Delta}
\right]\; \mathcal{I}_{\Omega} \; .
\end{equation}

Plugging the above equation and the Bianchi identity (\ref{eq:Maxwell9}) into the integrability
condition for $\omega$, Eq.~(\ref{eq:IntConOmega}), leads to

\begin{equation}
\label{eq:intautomatic}
\langle\, \mathcal{I}\, |\, \tilde{\mathfrak{D}}_{\underline{x}}
\tilde{\mathfrak{D}}_{\underline{x}}\mathcal{I}\, \rangle
 \; =\; 
-\mathcal{I}^{\Lambda}\tilde{\mathfrak{D}}_{\underline{x}}
\tilde{\mathfrak{D}}_{\underline{x}}\mathcal{I}_{\Lambda}
\; =\;
-\textstyle{1\over 2}g^{2}f_{\Lambda (\Sigma}{}^{\Gamma}f_{\Delta )\Gamma}{}^{\Omega}\ 
\mathcal{I}^{\Lambda}\mathcal{I}^{\Sigma}\mathcal{I}^{\Delta}\ \mathcal{I}_{\Omega}
 \; =\; 
0 \; ,
\end{equation}
which is, ignoring possible singularities, therefore identically satisfied.


\subsection{Construction of supersymmetric solutions of  $N=2,d=4$ SEYM}
\label{sec-recipe}

According to the KSIs, the supersymmetric configurations that satisfy the pair
of Eqs.~(\ref{eq:Maxwell9}) and (\ref{eq:Maxwell16}), or, equivalently, the
pair of Eqs.~(\ref{eq:Maxwell6}) and (\ref{eq:Maxwell16}) solve all the
equations of motion of the theory. This implies that one can give a
step-by-step prescription to construct supersymmetric solutions of any
$N=2,d=4$ SEYM starting from any solution of the YM-Higgs Bogomol'nyi
equations on $\mathbb{R}^{3}$:

\begin{enumerate}
\item Take a solution
  $\tilde{A}^{\Lambda}{}_{\underline{x}}$, $\mathcal{I}^{\Lambda}$ to the equations

\begin{displaymath}
\tilde{F}^{\Lambda}{}_{\underline{x}\underline{y}} 
\; =\; 
-{\textstyle\frac{1}{\sqrt{2}}}\epsilon_{xyz} 
\tilde{\mathfrak{D}}_{\underline{z}}\mathcal{I}^{\Lambda} \; .
\end{displaymath}

As we have stressed repeatedly, these equations are nothing but YM-Higgs
Bogomol'nyi equations on $\mathbb{R}^{3}$ and there are plenty of solutions
available in the literature. However, since in most cases the authors' goal is
to obtain regular monopole solutions on $\mathbb{R}^{3}$, there are many
solutions to the same equations that have been discarded because they present
singularities. We know, however, that in the Abelian case, the singularities
might be hidden by an event horizon\footnote{More precisely they turn out to be
  coordinate singularities in the full spacetime and correspond, not to a
  singular point, but to an event horizon.}. Therefore, we will not require the
solutions to the Bogomol'nyi equations to be globally regular on
$\mathbb{R}^{3}$.

\item Given the solution
  $\tilde{A}^{\Lambda}{}_{\underline{x}}$, $\mathcal{I}^{\Lambda}$,
  Eq.~(\ref{eq:Maxwell16}), which we write here again for the sake of
  clarity (as we will do with other relevant equations):

\begin{displaymath}
\tilde{\mathfrak{D}}_{\underline{x}}
\tilde{\mathfrak{D}}_{\underline{x}}\ \mathcal{I}_{\Lambda} 
\; =\; 
\textstyle{1\over 2}g^{2}\ 
\left[
f_{\Lambda (\Sigma}{}^{\Gamma}f_{\Delta )\Gamma}{}^{\Omega}\ 
\mathcal{I}^{\Sigma}\mathcal{I}^{\Delta}
\right]\; \mathcal{I}_{\Omega} \; .
\end{displaymath}

\noindent
becomes a linear equation for the $\mathcal{I}_{\Lambda}$s alone which has to
be solved. For compact gauge groups a possible solution is

\begin{equation}
\label{eq:IJI}
\mathcal{I}_{\Lambda}= \mathcal{J} \mathcal{I}^{\Lambda}\, ,
\end{equation}

\noindent
for an arbitrary real constant $\mathcal{J}$ (the r.h.s.~of Eq.~(\ref{eq:Maxwell16})
vanishes for this Ansatz). 

\item The first two steps provide
  $\mathcal{I}=(\mathcal{I}^{\Lambda},\mathcal{I}_{\Lambda}) =\Im\mathrm{m}\left(\mathcal{V}/X\right)$. 
  The next step, then, is to obtain
  $\mathcal{R}=(\mathcal{R}^{\Lambda},\mathcal{R}_{\Lambda}) =\Re\mathrm{e}\left(\mathcal{V}/X\right)$ 
  as functions of $\mathcal{I}$ by solving the model-dependent
  \textit{stabilization equations}. The stabilization equations depend only on the specific
  model one is considering and does not depend on whether the model is gauged or not.

\item Given $\mathcal{R}$ and $\mathcal{I}$, one can compute the metric
  function $|X|$ using Eq.~(\ref{eq:Gtt})

\begin{displaymath}
\frac{1}{2|X|^{2}}  =  \langle\, \mathcal{R}\mid\mathcal{I} \,\rangle\, ;
\end{displaymath}

\noindent
the $n$ physical complex scalars $Z^{i}$ by

\begin{equation}
Z^{i}\,\equiv\, \frac{\mathcal{L}^{i}}{\mathcal{L}^{0}}
\ =\  
\frac{\mathcal{L}^{i}/X}{\mathcal{L}^{0}/X}
\ =\
\frac{\mathcal{R}^{i}+i\mathcal{I}^{i}}{\mathcal{R}^{0}+i\mathcal{I}^{0}}\, ,
\end{equation}

\noindent
and the metric 1-form $\hat{\omega}$ using Eq.~(\ref{eq:oidi})

\begin{displaymath}
(d\hat{\omega})_{\underline{x}\underline{y}} =2 \epsilon_{xyz}
\langle\,\mathcal{I}\mid \tilde{\mathfrak{D}}_{\underline{z}}\mathcal{I}\, \rangle\, .  
\end{displaymath}

This last equation can always be solved locally, as according to
Eq.~(\ref{eq:intautomatic}) its integrability equation
is solved automatically, at least locally: 
Since the solutions to the covariant Laplace
equations are usually local (they generically have singularities), the
integrability condition may fail to be satisfied everywhere, as discussed for example in
Refs.~\cite{Denef:2000nb,Bates:2003vx,Bellorin:2006xr}, leading to
singularities in the metric. The solution Eq.~(\ref{eq:IJI}), however, always leads to
exactly vanishing $\hat{\omega}$, whence to static solutions.

$|X|$ and $\hat{\omega}$ completely determine the metric of the supersymmetric
solutions, given in Eq.~(\ref{eq:metric})

\begin{displaymath}
ds^{2} \; =\; 2|X|^{2}(dt+\hat{\omega})^{2} -\frac{1}{2|X|^{2}}dx^{x}dx^{x}
\hspace{1cm}
(x,y=1,2,3)\; .
\end{displaymath}

\item Once $\mathcal{I},\mathcal{R},|X|$ and $\hat{\omega}$ have been
  determined, the 4-dimensional gauge potential can be found from
  Eq.~(\ref{eq:gaugechoice2})

\begin{displaymath}
A^{\Lambda}{}_{t}  =    
-\sqrt{2}|X|^{2}\mathcal{R}^{\Lambda}\, ,
\end{displaymath}

\noindent
and from the definition of $\tilde{A}^{\Lambda}{}_{\underline{x}}$ Eq.~(\ref{eq:TildeDer})

\begin{displaymath}
A^{\Lambda}{}_{\underline{x}}
=
\tilde{A}^{\Lambda}{}_{\underline{x}} +\omega_{\underline{x}}\ A^{\Lambda}{}_{t}\; .
\end{displaymath}

The procedure we have followed ensures that this is the gauge potential whose
field strength is given in Eq.~(\ref{eq:Fsuper}).

\end{enumerate}

In the next section we are going to construct, following this procedure,
several solutions.


\section{Monopoles and hairy black holes}
\label{sec:EYMsols}

As we have seen, the starting point in the construction of $N=2,d=4$ SEYM
supersymmetric solutions is the Bogomol'nyi equation on $\mathbb{R}^{3}$.
Of course, the most interesting solutions to the
Bogomol'nyi equations are the monopoles that can be characterised by saying
that they are finite energy solutions that are everywhere regular. The fact
that the gauge fields are regular does, however, not imply that the full
supergravity solution is regular. Indeed, the metric and the physical scalar
fields are built out of the ``Higgs field'', {\em i.e.\/} $\mathcal{I}$, and the
precise relations are model dependent and requires knowing the solutions to
the stabilization equation.

As the Higgs field in a monopole asymptotes to a non-trivial constant
configuration, it asymptotically breaks the gauge group through the Higgs
effect.  In fact, as we are dealing with supergravity and supersymmetry
preserving solutions, monopoles in our setting would have to implement the
super-Higgs effect as for example discussed in
Refs.~\cite{Andrianopoli:2002vq}. If we were to insist on an asymptotic
supersymmetric effective action, we would be forced to introduce
hypermultiplets in order to fill out massive supermultiplets, but this point
will not be pursued in this article.

The Bogomol'nyi equations admit more than just regular solutions, and we shall
give families of solutions, labelled by a continuous parameter $s>0$, having
the same asymptotic behaviour as the monopole solutions. As they are singular
on $\mathbb{R}^{3}$, however, we will use them to construct metrics describing
the regions outside regular black holes: as will be shown, the members of a
given family lead to black holes that are not distinguished by their
asymptotic data, such as the moduli or the asymptotic mass, nor by their
entropy and as such illustrate the non-applicability of the no-hair theorem to
supersymmetric EYM theories. Furthermore, in all examples considered, the
attractor mechanisms is at work, meaning that the physical scalars at the
horizon and the entropy depend only on the asymptotic charges and not on the
moduli nor on the parameter $s$.

The plan of this section is as follows: in section (\ref{sec:SO3EYM}) we
shall repeat briefly the embedding of the spherically symmetric solutions to
the $SO(3)$ Bogomol'nyi equations in the $\overline{\mathbb{CP}}^{3}$ models.
In all but one of these solutions, the asymptotic gauge symmetry breaking is
maximal, {\em i.e.\/} the $SO(3)$ gauge symmetry is broken down to $U(1)$.  In
section (\ref{sec:SO5wein}), we will investigate the embedding of solutions
that manifest a non-maximal asymptotic symmetry breaking: for this we take
E. Weinberg's spherically symmetric $SO(5)$-monopole \cite{Weinberg:1982jh}
embedded into $\overline{\mathbb{CP}}^{10}$.  This monopole breaks the $SO(5)$
down to $U(2)$ and has the added characteristic that, unlike the 't
Hooft-Polyakov monopole, the Higgs field does not vanish at the origin.

An interesting question is whether one can embed monopoles also into more
complicated models. This question will be investigated in
Section~\ref{sec:Magic}, where we consider gauged ``Magic'' supergravities.


\subsection{Spherically symmetric solutions in  $SO(3)$ gauged 
$\overline{\mathbb{CP}}^{3}$}
\label{sec:SO3EYM}

Before discussing the solutions we need to make some comments on the model:
the model we shall consider in this and the next section is the so-called
$\overline{\mathbb{CP}}^{n}$ model.\footnote{ The solutions in this and the
  next section can also be embedded into the $\mathcal{ST}$-models, with
  similar conclusions. Contrary to Ref.~\cite{Huebscher:2007hj}, however, we
  have chosen not to deal with this model explicitly, and refer the reader to
  Appendix~\ref{sec:ST2n} for more details.} In this model the metric on
the scalar manifold is that of the symmetric space $SU(1,n)/U(n)$ and the
prepotential is given by

\begin{equation}
\label{eq:CP1}
\mathcal{F} \; =\; 
\textstyle{1\over 4i}\ \eta_{\Lambda\Sigma}\ \mathcal{X}^{\Lambda}\
\mathcal{X}^{\Sigma}\, , 
\hspace{.5cm}
\eta \ =\ \mathrm{diag}\left(\ +\ ,\ [-]^{n}\ \right) \; ,
\end{equation}

\noindent
which is manifestly $SO(1,n)$ invariant.

The K\"ahler potential is straightforwardly derived by fixing
$\mathcal{X}^{0}=1$ and introducing the notation $\mathcal{X}^{i}\ =\ Z^{i}$;
this results in

\begin{equation}
\label{eq:CP2}
e^{-\mathcal{K}} \; =\; 
|\mathcal{X}^{0}|^{2} \ -\ \sum_{i=1}^{n}\ |\mathcal{X}^{i}|^{2}
\; =\; 1\ -\ \sum_{i=1}^{n}\ |Z^{i}|^{2} \equiv 
 1\ -\ |Z|^{2} \; .
\end{equation}




Observe that this expression for the K\"ahler potential implies that the $Z$'s
are constrained by $0 \leq |Z|^{2} < 1$.

As the model is quadratic, the stabilization equations are easily solved and leads to

\begin{equation}
\label{eq:CP3}
\mathcal{R}_{\Lambda} \; =\; 
 \textstyle{1\over 2}\eta_{\Lambda\Sigma}\ \mathcal{I}^{\Sigma} \;\;\;\; ,\;\;\;\;
\mathcal{R}^{\Lambda} \; =\; -2\eta^{\Lambda\Sigma}\ \mathcal{I}_{\Sigma} \; .
\end{equation}

\noindent
With this solution to the stabilization equation, we can express the metrical 
factor, Eq.~(\ref{eq:Gtt}), in terms of the $\mathcal{I}$ as

\begin{equation}
\label{eq:CP4}
\frac{1}{2|X|^{2}} \; =\; 
\textstyle{1\over 2}\ \eta_{\Lambda\Sigma}\ \mathcal{I}^{\Lambda}\mathcal{I}^{\Sigma}
+2\eta^{\Lambda\Sigma}\ \mathcal{I}_{\Lambda}\mathcal{I}_{\Sigma} 
\; =\; 
\textstyle{1\over 2}\ \eta_{\Lambda\Sigma}\ \mathcal{I}^{\Lambda}\mathcal{I}^{\Sigma} \; ,
\end{equation}

\noindent
where in that last step we used the fact that in this article we shall
consider only purely magnetic solutions, so that
$\mathcal{I}_{\Lambda}=0$. The fact that we choose to consider magnetic
embeddings only, implies be means of Eq.~(\ref{eq:oidi}) that we will be dealing
with static solutions.

In order to finish the discussion of the model, we must discuss the possible
gauge groups that can occur in the $\overline{\mathbb{CP}}^{n}$-models: as we
saw at the beginning of this section, these models have a manifest $SO(1,n)$
symmetry, under which the $\mathcal{X}$'s transform as a vector. Furthermore,
as we are mostly interested in monopole-like solutions, we shall restrict our
attention to compact simple groups, which, as implied by Eq.~(\ref{eq:2PreP}),
must be subgroups of $SO(n)$.  In fact, Eq.~(\ref{eq:2PreP}) and
Eq.~(\ref{eq:GGchoice}) make the stronger statement that given a gauge algebra
$\mathfrak{g}$, the action of $\mathfrak{g}$ on the $\mathcal{X}$'s must be
such that only singlets and the adjoint representation appear. For the
$\overline{\mathbb{CP}}^{n}$-models there is no problem whatsoever as we can
choose $n$ to be large enough as to accomodate any Lie algebra.  Indeed, as is
well-known any compact simple Lie algebra $\mathfrak{g}$ is a subalgebra of
$\mathfrak{so}(\mathrm{dim}(\mathfrak{g}))$ and the branching of the latter's
vector representation is exactly the adjoint representation of $\mathfrak{g}$.

The simplest possibility, namely the $SO(3)$-gauged model on
$\overline{\mathbb{CP}}^{3}$, will be used in the remainder of this section,
and the $SO(5)$-gauged $\overline{\mathbb{CP}}^{10}$ model will be used in
section (\ref{sec:SO5wein}).  The $SO(4)$- and the $SU(3)$-gauged models will
not be treated, but solutions to these models can be created with great ease
using the information in this section and Appendix~\ref{sec:bais}.

As we are restricting ourselves to purely magnetic solutions, which are
automatically static, the construction of explicit supergravity solutions goes
through the explicit solutions to the $SO(3)$ Bogomol'nyi equation
(\ref{eq:Maxwell6}). Having applications to the attractor mechanism in mind,
and being fully aware of the fact
that this class consists of only the tip of the iceberg of solutions, we shall
restrict ourselves to spherically symmetric solutions to the Bogomol'nyi
equations.

Working in gauge theories opens up the possibility of compensating the
spacetime rotations with gauge transformations, and in the case of an $SO(3)$
gauge group this means that the gauge connection and the Higgs field,
$\mathcal{I}$, after a suitable gauge fixing, takes on the form (See {\em
  e.g.\/} \cite{Goddard:1977da})

\begin{equation}
\label{eq:SO3Ansatz}
A^{i}{}_{m} \; =\; -\varepsilon_{mn}{}^{i}\ x^{n}\ P(r) \;\;\; ,\;\;\;
  \mathcal{I}^{i} \; =\; -\sqrt{2}\ x^{i}\ H(r) \; .
\end{equation}

\noindent
Substituting this Ansatz into the Bogomol'nyi equation we find that $H$ and
$P$ must satisfy

\begin{eqnarray}
\label{eq:SO3eq1}
r\partial_{r}\left( H+P\right) & =& gr^{2}\ P\ \left( H+P\right) \; ,\\
& & \nonumber \\
\label{eq:SO3eq2}
r\partial_{r} P \ +\ 2P & =& H\ \left( 1+gr^{2}P\right) \; .
\end{eqnarray}

\noindent
All the solutions to the above equations were found in
Ref.~\cite{Protogenov:1977tq} and all but one of them contain
singularities. Furthermore, not all of them have the correct asymptotics to
lead to asymptotic flat spaces and only part of the ones that do can be used
to construct regular supergravity solutions
\cite{Huebscher:2007hj,Meessen:2008kb}. Here, by regular supergravity
solutions we mean that the solutions is either free of singularities, which is
what is meant by a globally regular solution, or has a singularity but, like
the black hole solutions in the Abelian theories, has the interpretation of
describing the physics outside the event horizon of a regular black hole. The
criterion for this last to occur is that the geometry near the singularity is
that of a Robinson-Bertotti/$aDS_{2}\times S^{2}$ spacetime, implying that the
black hole has a non-vanishing horizon area, whence also entropy.

The suitable solutions, then, break up into 3 classes:


\subsubsection*{(I)~'t~Hooft-Polyakov monopole}

This is the most famous  solution and reads

\begin{equation}
\label{eq:SO3tHP}
H \ =\ 
-\frac{\mu}{gr}\left[ \coth (\mu r) - \frac{1}{\mu r}\right] 
\ \equiv\ -\frac{\mu}{gr}\ \overline{H}(r) \;\; ,\;\;
P \; =\; 
-\frac{1}{gr^{2}}\left[ 1\ -\ \mu r\sinh^{-1}(\mu r)\right] \; ,
\end{equation}

\noindent
where $\mu$ is a positive constant.  The renowned regularity of the
't~Hooft-Polyakov monopole opens up the possibility of creating a globally
regular solution to the supergravity equations which is in fact trivial to
achieve: for the moment we have been ignoring $\mathcal{I}^{0}$, which, since
it is uncharged under the gauge group, is just a real, spherically symmetric
harmonic function we can parametrize as 

\begin{equation}
\label{eq:i0}
\mathcal{I}^{0} =\sqrt{2}( h +p/r)\, .
\end{equation}

\noindent
It is clear, however, that if we want to avoid singularities, we must
take $p=0$, so that the only free parameter is $h$.

Let us then discuss the regularity conditions imposed by the metric: as was
said before, the solutions are automatically static, so that if singularities
in the metric are to appear, they arise from the metrical factor
$|X|^{2}$. Plugging the solution for the Higgs field into the expression
(\ref{eq:CP4}), we find

\begin{equation}
  \label{eq:SOHPmet}
  \frac{1}{2|X|^{2}} \; =\; h^{2} \; -\; \frac{\mu^{2}}{g^{2}}\ \overline{H}^{2}(r) \; .
\end{equation}

As one can infer from its definition in Eq.~(\ref{eq:SO3tHP}), the function
$\overline{H}$ is a monotonic, positive semi-definite function on
$\mathbb{R}^{+}$ and vanishes only at $r=0$, where it behaves as
$\overline{H}\sim \mu r/3 +\mathcal{O}(r^{2})$; its behaviour for large $r$ is
given by $\overline{H}= 1 -1/(\mu r)$, which means that we should choose $h$
large enough in order to ensure the positivity of the metrical factor. A
convenient choice for $h$ is given by imposing that
asymptotically we recover the standard Minkowskian metric in spherical
coordinates: this condition gives $h^{2}=1+\mu^{2}g^{-2}$ from which we find
the final metrical factor and can then also calculate the asymptotic mass,
{\em i.e.\/}

\begin{equation}
  \label{eq:SOHPmass}
  \frac{1}{2|X|^{2}} \ =\ 1 \ +\ \frac{\mu^{2}}{g^{2}}\left[ 1\ -\ \overline{H}^{2}\right]
  \;\;\rightarrow\;\; 
  M \ =\ \frac{\mu}{g^{2}} \; .
\end{equation}

Written in this form, it is paramount that the metric is globally regular and
interpolates between two Minkowksi spaces, one at $r=0$ and one at $r=\infty$.

In order to show that the solution is a globally regular supergravity
solution, we should show that the physical scalars are regular. In the
$\overline{\mathbb{CP}}^{n}$-models the scalars are given by (introducing the
outward-pointing unit vector $\vec{n}=\vec{x}/r$)

\begin{equation}
\label{eq:SOHPscal}
Z^{i} \;\equiv\; 
\frac{\mathcal{R}^{i}+i\mathcal{I}^{i}}{\mathcal{R}^{0}+i\mathcal{I}^{0}}
\; =\;  \frac{\mathcal{I}^{i}}{\mathcal{I}^{0}}
\; =\; \frac{\mu}{gh}\ \overline{H}\ n^{i} \; ,
\end{equation}

\noindent
so that the regularity is obvious. The scalars also respect the bound $0\leq
|Z|^{2}<1$ as can be seen from the fact that the bound corresponds to the
positivity of the metrical factor. This regularity of the scalars and that of
the spacetime metric are related \cite{Bellorin:2006xr}.


\subsubsection*{(II)~Hairy black holes}

A generic class of singular solutions is indexed by a free parameter $s>0$,
called the {\em Protogenov hair}, and can be seen as a deformation of the
't~Hooft-Polyakov monopole, {\em i.e.\/}

\begin{equation}
\label{eq:SO3hairy}
H \ =\ 
-\frac{\mu}{gr}\left[\coth (\mu r +s)\ -\ \frac{1}{\mu r}\right]
  \ \equiv\ -\frac{\mu}{gr}\ \overline{H}_{s}(r) \;\; ,\;\;
P \; =\; 
-\frac{1}{gr^{2}}\left[ 1\ -\ \mu r\sinh^{-1}(\mu r+s)\right] \; .
\end{equation}

\noindent
The effect of introducing the parameter $s$ is to shift the singularity of the
cotangent from $r=0$ to $\mu r=-s$, {\em i.e.\/} outside the domain of $r$,
but leaving unchanged its asymptotic behaviour.\footnote{ One can consider the
  limiting solution for $s\rightarrow\infty$, the result of which was called a
  black hedgehog in Ref.~\cite{Huebscher:2007hj}. This solution has, apart
  from not containing hyperbolic functions, no special properties and will not
  be considered seperately.}  This not only means that the function
$\overline{H}_{s}$ vanishes at some $r_{s}>0$, but also that it becomes
singular at $r=0$, so that in order to build a regular solution we must have
$p\neq 0$. Using then the general Ansatz for $\mathcal{I}^{0}$,
Eq.~(\ref{eq:i0}), in order to calculate the metrical factor, we find in stead
of Eq.~(\ref{eq:SOHPmet})

\begin{equation}
\label{eq:HPPmet}
\frac{1}{2|X|^{2}} \; =\; \left( h + \frac{p}{r}\ \right)^{2}
\ -\ \frac{\mu^{2}}{g^{2}}\ \overline{H}_{s}^{2} \; .
\end{equation}

\noindent
As the asymptotic behaviour of $\overline{H}_{s}$ is the same as the one for 
the 't~Hooft-Polyakov monopole, the condition imposed by asymptotic flatness
still is $h^{2}= 1+ \mu^{2}g^{-2}$. Given this normalization, the asymptotic mass
is 

\begin{equation}
\label{eq:HPPmass}
M \; =\; hp \ +\ \frac{\mu}{g^{2}} \; , 
\end{equation}

\noindent
which should be positive for a physical solution. In this respect, we would
like to point out that the product $hp$ should be positive as otherwise the
metrical factor would become negative or zero, should it coincide with the
zero of $\overline{H}_{s}$, at a finite distance, ruining our interpretation
of the metric as describing the outside of a regular black hole.  This then
implies that the mass is automatically positive.  Finally, let us point out
that neither the mass nor the modulus $h$ depend on the Protogenov hair
parameter $s$.

The metrical factor is clearly singular at $r=0$, but given the interpretation
of the metric this is not a problem as long as the geometry near $r=0$, which
corresponds to the near horizon geometry, is that of an $aDS_{2}\times S^{2}$
space. This is the case if

\begin{equation}
\label{eq:HPPentropy}
S_{bh} \;\equiv\; \lim_{r\rightarrow 0}\ \frac{r^{2}}{2|X|^{2}}
\; =\; p^{2} \ -\ \frac{1}{g^{2}} \; , 
\end{equation}

\noindent
is positive and can thence be identified with the entropy of the black hole.

The scalars for this solution are given by

\begin{equation}
  \label{eq:HPPscal}
  Z^{i} \; =\; \frac{\mu}{g}\ \frac{r\overline{H}_{s}}{p\ +\ hr} \; n^{i} \; ,
\end{equation}

\noindent
whose asymptotic behaviour is the same as for the 't~Hooft-Polyakov monopole.
Its behaviour near the horizon, {\em i.e.\/} near $r=0$, is easily calculated
to be 

\begin{equation}
  \label{eq:HPPattr}
  \lim_{r\rightarrow 0}\ Z^{i} \; =\; -\frac{1}{gp}\ n^{i} \; ,
\end{equation}

\noindent
and does not depend on the moduli nor on the Protogenov hair, but only on the
asymptotic charges.  Observe, however, that since $\overline{H}_{s}=0$ at some
finite $r_{s}>0$, there is a 2-sphere outside the horizon at which the scalars
vanish, which is not a singularity for the scalars of this model.





\subsection*{(III)~Coloured black holes}

There is another particular solution to the $SO(3)$ Bogomol'nyi equation that
has all the necessary properties, and this solution is given by

\begin{equation}
\label{eq:SO3Colour}
H 
\; =\; 
-P 
\; =\; 
\frac{1}{gr^{2}}\left[\frac{1}{1+\lambda^{2}r}\right] \; .
\end{equation}

This solution has the same $r\rightarrow 0$ behaviour as the hairy solutions,
but is such that in the asymptotic regime it has no Higgs v.e.v. nor colour
charge.  Given the foregoing discussion, it is clear that this solution can be
used to build a regular black hole solution, and we can and will be brief.

The regularity of the metric goes once again through the judicious election of
$h$ and $p$: the normalization condition implies that $|h|=1$ which then also
implies that the asymptotic mass of the solution is $M=|p|$. It may seem
strange that the YM-configuration does not contribute to the mass, but it does
so, at least for a regular black hole solution, in an indirect fashion: the
condition for a regular horizon is clearly given by Eq.~(\ref{eq:HPPentropy}),
which implies that $|p|> 1/g$.  With these choices then, the scalars $Z$ are
regular for $r>0$ and at the horizon they behave as in Eq.~(\ref{eq:HPPattr}).


\subsection{Non-maximal symmetry breaking in $SO(5)$ gauged 
$\overline{\mathbb{CP}}^{10}$}
\label{sec:SO5wein}

In Ref.~\cite{Weinberg:1982jh}, E.~Weinberg presented an explicit solution for
a spherically symmetric monopole solution that breaks the parent $SO(5)$ gauge
group down to $U(2)$; in this section we will discuss the embedding of
this solution into supergravity and also generalize it to a family of hairy
black holes by introducing Protogenov hair\footnote{ In
  Ref.~\cite{Guo:2008ks} the general equations for a spherically symmetric
  solution to the $SO(5)$ Bogomol'nyi equations were derived. This opens up
  the possibility of analysing the system along the lines of
  Ref.~\cite{Protogenov:1977tq}, but for the moment this has not lead to
  anything new.}.

The starting point of the derivation of Weinberg's monopole is the explicit
embedding of an 't~Hooft-Polyakov monopole into an $\mathfrak{so}(3)$
subalgebra of $\mathfrak{so}(5)$. In order to make this embedding paramount we
take the generators of $\mathfrak{so}(5)$ to be $J_{i}$, $\overline{J}_{i}$
($i=1,2,3$) and $P_{a}$ ($a=1,\ldots ,4$). These generators satisfy the
following commutation relations

\begin{equation}
\label{eq:Wein5}
\begin{array}{lclclcl}
\left[ J_{i},J_{j}\right] 
& = & 
\varepsilon_{ijk}\ J_{k}\, , 
&\hspace{1cm}&
\left[ J_{i},P_{a}\right] 
& = & 
P_{c}\ \Sigma_{i}{}^{c}{}_{a} \; , 
\\
 & & & & & & \\
\left[ \bar{J}_{i},\bar{J}_{j}\right] 
& = & \varepsilon_{ijk}\ \bar{J}_{k} \, ,
& &
\left[ \bar{J}_{i},P_{a}\right] 
& = & 
P_{c}\ \overline{\Sigma}_{i}{}^{c}{}_{a} \; , 
\\
& & & & & & \\
\left[ J_{i},\bar{J}_{j}\right] 
& = & 
0\, , 
& & 
\left[ P_{a},P_{b}\right] 
& = & 
-2\ J_{i}\ \Sigma^{i}_{ab} 
- 2\ \bar{J}_{i}\ \overline{\Sigma}^{i}_{ab} \; ,
\end{array}
\end{equation}

\noindent
where we have introduced the 't~Hooft symbols $\Sigma_{i}^{ab}$ and
$\overline{\Sigma}_{i}^{ab}$.  The $\Sigma$ ({\em resp.\/}
$\overline{\Sigma}$) are self-dual ({\em resp.\/} anti-selfdual) 2-forms on
$\mathbb{R}^{4}$ and satisfy the following relations

\begin{equation}
\label{eq:Wein3}
\begin{array}{rclcrclclcl}
\left[ \Sigma_{i},\Sigma_{j}\right] 
& = & 
\varepsilon_{ijk}\Sigma_{k}\, , 
&\hspace{1cm}&
\left[ \overline{\Sigma}_{i},\overline{\Sigma}_{j}\right] 
& = & 
\varepsilon_{ijk}\overline{\Sigma}_{k}\, , 
&\hspace{1cm}&
\left[ \Sigma_{i},\overline{\Sigma}_{j}\right] 
& = & 
0 \; ,
\\
& & & & & & & & & & \\
\Sigma_{i}^{2} 
& = & 
-\textstyle{1\over 4}\ 1_{4}\, , 
& & 
\overline{\Sigma}_{i}^{2} 
& = & 
-\textstyle{1\over 4}\ 1_{4}\, , 
& & 
\Sigma_{iab}\overline{\Sigma}_{j}^{ab} 
& = & 
0 \; . 
\end{array}
\end{equation}

\noindent
We would like to stress that $\overline{\Sigma}$ is not the complex nor the
Hermitean conjugate of $\Sigma$.

Following Weinberg we make the following Ansatz for the $\mathfrak{so}(5)$-valued
connection and Higgs field, taking $T_{A}$ ($A=1,\ldots ,10$) to be the generators
of $\mathfrak{so}(5)$,

\begin{eqnarray}
\label{eq:Wein9a}
\mathtt{A}_{m} 
& \equiv& A^{A}{}_{m}\ T_{A} \; =\; 
-\varepsilon_{mj}{}^{i}n^{j}\left[ rP\ J_{i} \ +\ rB\ \bar{J}_{i}\right]
\ +\ M_{m}{}^{a}\ P_{a}\; ,
\\
& & \nonumber \\
\label{eq:Wein9b}
-\textstyle{1\over \sqrt{2}}\mathtt{I} 
& \equiv& 
-\textstyle{1\over\sqrt{2}}\mathcal{I}^{A}\ T_{A} \; =\; 
rH\ n^{i}J_{i} \ +\ rK\ n^{i}\bar{J}_{i}
\ +\ \Omega^{a}\ P_{a} \; ,
\end{eqnarray}

\noindent
where $P$, $B$, $H$ and $K$ are functions of $r$ only.  $M$ and $\Omega$ are
determined by the criterion that we have an 't~Hooft-Polyakov monopole in some
$\mathfrak{so}(3)$-subalgebra, which we take to be generated by the $J_{i}$.
One way of satisfying this criterion is by choosing

\begin{equation}
\label{eq:Wein10}
M_{m}{}^{a} \; =\; F\ \delta_{m}^{a} \;\;\; ,\;\;\;
\Omega^{a} \; =\; -F\ \delta^{a0}\; , 
\end{equation}

\noindent
which implies that the Bogomol'nyi equation in the $J_{i}$ sector reduce to
Eqs.~(\ref{eq:SO3eq1}) and (\ref{eq:SO3eq2}).

The analysis of the Bogomol'nyi equations in the remaining sectors impose the
constraint that $K=-B$ and the differential equations\footnote{ In order to go
  from Weinberg's notation \cite{Weinberg:1982jh} to ours one needs to change
  $A\rightarrow -rP$, $G\rightarrow -rB$, $H\rightarrow rH$, $K\rightarrow
  rK$, $e\rightarrow -g$ and also $F\rightarrow F/\sqrt{2}$.  }

\begin{eqnarray}
\label{eq:Wein12a}
2g\ F^{2} 
& = & 
rK^{\prime} \ +\ 2K \ +\ K(1-gr^{2}K) \; ,
\\
& &  \nonumber \\
\label{eq:Wein12b}
F^{\prime} 
& = & 
\textstyle{1\over 2}gr\ F\left[ 2P\ +\ H\ +\ K \right] \; .
\end{eqnarray}

The final ingredient, needed for the calculation of the metrical factor, consists 
of finding an expression for the $SO(5)$-invariant quantity
$\mathcal{I}^{A}\mathcal{I}^{A}$: this is

\begin{equation}
\label{eq:6}
\textstyle{1\over 2}\ \mathcal{I}^{A}\mathcal{I}^{A} \; =\; r^{2}H^{2} \ +\ r^{2}K^{2} \ +\ 2\ F^{2} \; .
\end{equation}

In conclusion, given a solution to
Eqs.~(\ref{eq:SO3eq1},\ref{eq:SO3eq2},\ref{eq:Wein12a}) and (\ref{eq:Wein12b})
we can discuss their embedding into the $SO(5)$-gauged
$\overline{\mathbb{CP}}^{10}$-model by means of Eq.~(\ref{eq:6}).


\subsubsection*{Weinberg's monopole in supergravity}
\label{sec:WmonSugra}

The explicit form of Weinberg's monopole is given by the solution in
Eq.~(\ref{eq:SO3tHP}) and

\begin{eqnarray}
\label{eq:2Wmon}
K(r) 
& = & -P(r)\ L(r;a) \; \equiv\; \frac{\mu}{gr}\ \overline{K}\; , 
\\
& & \nonumber \\
F(r) 
& = & 
\frac{\mu}{2g\ \cosh\left(\mu r/2\right) }\ 
L^{1/2}(r;a) \; \equiv\; \frac{\mu}{g}\ \overline{F} \; ,
\end{eqnarray}

\noindent
where the profile function $L$, given by

\begin{equation}
\label{eq:3Wmon}
L(r;a) \; =\; 
\left[  1\ +\ {\mu r\over 2a}
\coth\left(\mu r/2 \right)\ \right]^{-1} \; ,
\end{equation}

\noindent
depends on a positive parameter $a$ called the \textit{cloud parameter}.  The
cloud parameter $a$ is a measure for the extention of the region in which the
Higgs field in the $\overline{J}_{i}$- and the $P_{a}$-directions are active:
in fact when $a=0$ the profile functions vanishes identically and we are
dealing with an embedding of the 't Hooft-Polyakov monopole.  The maximal
extention is for $a\rightarrow\infty$ which then means that $L=1$.

As one can see from the definitions, $K$ and $F$ are positive semi-definite
functions that asymptote exponentially to zero.  This not only means that the
gauge symmetry is asymptotically broken to $U(2)$, but also that $K$ and $F$
will not contribute to the asymptotic mass, nor to the normalization
condition.  Unlike the 't~Hooft-Polyakov monopole or the degenerate
Wilkinson-Bais $SU(3)$-monopole (\ref{eq:WB11}), however, the regularity of
the solution does {\em not} imply that the Higgs field vanishes at $r=0$!  In fact,
near $r=0$ one finds that

\begin{equation}
\label{eq:WM12}
\overline{F} \; \sim\; \textstyle{1\over 2}\ 
\sqrt{\ {\displaystyle\frac{a}{1+a}} \ }\ +\ldots \;\;\; ,\;\;\;
  \overline{K} \; \sim\; 
{\displaystyle\frac{\mu a}{3! (a+1)}}\ r \ +\ldots \; .
\end{equation}

\noindent
It is this behaviour that may pose a problem for creating a globally regular
solution and is the reason for including it in this article.

Using Eqs.~(\ref{eq:CP4}) and (\ref{eq:6}) and choosing as in
Sec.~(\ref{sec:SO3EYM}) $p=0$, we can write the metrical factor as

\begin{equation}
\label{eq:WM13}
\frac{1}{2|X|^{2}}\; =\; 1 \ +\ \frac{\mu^{2}}{g^{2}}\left[
1\ -\ \overline{H}^{2}\ -\ \overline{K}^{2}\ -\ 2\overline{F}^{2}
\right] \; ,
\end{equation}

\noindent 
where we already used the normalization condition $h^{2}= 1+\mu^{2}g^{-2}$.
As mentioned above, $\overline{K}$ and $\overline{F}$ asymptote exponentially
to zero and cannot contribute to the mass, which is the one for the
't~Hooft-Polyakov monopole, {\em i.e.\/} $M=\mu g^{-2}$.

\begin{figure}[t]
  \centering
  \label{fig:weinMon}
    \includegraphics[width=8cm]{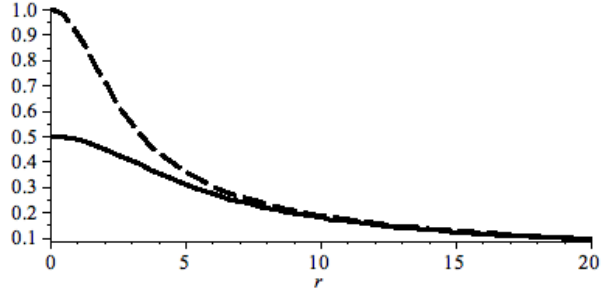}
    \caption{A plot of
      $1-\overline{H}^{2}-\overline{K}^{2}-2\overline{F}^{2}$: the dashed line
      corresponds to $a=0$ and the solid line corresponds to the maximal cloud
      extention, {\em i.e.\/} $L=1$.}
\end{figure}

Let us then investigate the behaviour of (\ref{eq:WM13}) at $r=0$: a simple
substitution shows that

\begin{equation}
  \left. \frac{1}{2|X|^{2}}\right|_{r=0} \; =\; 1\; +\; \frac{\mu^{2}}{g^{2}}\
            \frac{2a+1}{2(a+1)} \; ,
\end{equation}

\noindent
which is always positive so that the non-zero value of the Higgs field at the
origin is no obstruction to the construction of a globally regular
supergravity solution. The remaining question as far as the global regularity of the
solution is concerned, is whether there are values of $r$ for which the metrical factor
(\ref{eq:WM13}) becomes negative. This however never happens as one can see from
Fig.~(1) 
which shows a plot of
$1-\overline{H}^{2}-\overline{K}^{2}-2\overline{F}^{2}$ for the values of
$a=0$ and $a=\infty$.


\subsubsection*{Another hairy black hole}

The introduction of Protogenov hair, {\em i.e.} a real and positive parameter
$s$, in Weinberg's monopole solution is trivial and leads to the following
solution

\begin{eqnarray}
\label{eq:WH1a}
L_{s}(r;a) 
& = &  
{\displaystyle \left[  1\ +\ {\mu r\over 2a}
                         \coth\left({\mu r +s\over 2} \right)\ 
                 \right]^{-1} } \; , 
\\
& & \nonumber \\
\label{eq:WH1b}
F 
& = & 
\frac{\mu}{g}\ \overline{F}_{s} 
\ =\ \frac{\mu}{2g\cosh\left(\textstyle{\mu r +s\over 2}\right)}\ L_{s}^{1/2}
\, , 
\\
& & \nonumber \\
\label{eq:WH1c}
K 
& = & 
\frac{\mu}{gr}\ \overline{K}_{s}
\ =\ \frac{\mu}{gr}\ \left[
\frac{1}{\mu r} \ -\ \frac{1}{\sinh (\mu r +s)}
\right]\ L_{s} \; .
\end{eqnarray}


\noindent
supplemented by the expression for $H$ and $P$ given in
Eq.~(\ref{eq:SO3hairy}).  As far as the limiting cases of this family is
concerned, it is clear that Weinberg's monopole is obtained in the limit
$s\rightarrow 0$; in the limit $s\rightarrow\infty$ we find that $F\rightarrow
0$ and the solution splits up into the direct sum of an $SO(3)$ black
hedgehog, {\em i.e.\/} an $s\rightarrow\infty$ limit of (\ref{eq:SO3hairy}),
and an $SO(3)$ coloured black hole, Eq.~(\ref{eq:SO3Colour}).

As in the case of the hairy $SO(3)$ black holes, the introduction of the hair
parameter $s$ preserves the asymptotic behaviour of Weinberg's monopole and
the solution is regular for $r>0$. This immediately implies that the
normalization condition for $h$ once again reads $h^{2}=1+\mu^{2}g^{-2}$ and
that the asymptotic mass of this solution is given by Eq.~(\ref{eq:HPPmass}),
which is positive with the usual proviso that $hp>0$.

As in the case of the hairy black holes in the $SO(3)$-gauged
$\overline{\mathbb{CP}}^{3}$-models, the regularity of the metric imposes the
constraint that the entropy

\begin{equation}
  \label{eq:3}
  S_{bh} \; =\; p^{2} \; -\; \frac{2}{g^{2}} \; ,
\end{equation}

\noindent
be positive. This positivity of the entropy also ensures that the physical
scalars stay in their domain of definition at $r=0$. Indeed, the physical
scalars can be compactly written as

\begin{equation}
\label{eq:28}
\mathtt{Z} \; =\; Z^{A}\ T_{A} 
            \; =\;  \frac{\mu}{g}\left[
                          \frac{r\overline{H}_{s}}{p+hr}\ n^{i}J_{i}
                         \ -\ 
                          \frac{r\overline{K}_{s}}{p+hr}\ n^{i}\overline{J}_{i}
                         \ +\
                           \frac{r\overline{F}_{s}}{p+hr}\ P_{0}
                      \right] \; ,
\end{equation}

\noindent
which are therefore regular for $r>0$. Their value at $r=0$ is

\begin{equation}
  \label{eq:2}
  \left. \mathtt{Z}\right|_{r=0} \; =\; -\frac{1}{gp}\ n^{i}\left( J_{i} \ +\ \overline{J}_{i}\right) \; ,
\end{equation}

\noindent
which, as in the case of the $SO(3)$ solution, depend only on the asymptotic charges.


\subsection{Non-Abelian solutions in Magic models}
\label{sec:Magic}

\begin{table}
  \centering
  {\small
   \begin{tabular}{|c|c||c|c|c|c|c|c|}
    \hline
         $\mathbf{A}$ 
       & $\mathsf{G}$  
       & $\mathsf{H}$ 
       & $\mathsf{G}\circ\mathcal{V}$
       & $\mathsf{H}\circ\mathcal{X}^{0}$
       & $\mathsf{H}\circ\mathcal{X}^{i}$ 
       & $\mathsf{I}_{3}(\mathcal{X}^{i})$ 
       & $\mathrm{max}(G)$ 
       \\[.1cm] \hline\hline
         $\mathbb{R}$
       & $Sp(3;\mathbb{R})$ 
       & $U(3)$ 
       & $\mathbf{14}^{\prime}$
       & $\mathbf{1}_{-3}$
       & $\mathbf{6}_{-1}$ 
       & $\mathrm{det}(\mathcal{X})$
       & 
       \\ \hline
         $\mathbb{C}$
       & $SU(3,3)$
       & $S[U(3)\otimes U(3)]$ 
       & $\mathbf{20}$
       & $(\mathbf{1},\mathbf{1})_{-3}$
       & $(\mathbf{3},\overline{\mathbf{3}})_{-1}$
       & $\mathrm{det}(\mathcal{X})$ 
       & $SU(3)_{diag}$
       \\ \hline
         $\mathbb{Q}$
       & $SO^{*}(12)$ 
       & $U(6)$
       & $\mathbf{32}^{\prime}$ 
       & $\mathbf{1}_{-3}$
       & $\mathbf{15}_{-1}$ 
       & $\mathrm{Pf}\left(\mathcal{X}\right)$
       & $SU(4)$
       \\ \hline
         $\mathbb{O}$
       & $E_{7(-25)}$
       & $E_{6}\otimes SO(2)$
       & $\mathbf{56}$ 
       & $\mathbf{1}_{3}$
       & $\mathbf{27}_{1}$
       & $\mathrm{Tr}\left( [\Omega\mathcal{X}]^{3}\right) /3!$
       & 
       \\ \hline
  \end{tabular}
  }
  \caption{List of characteristics of Symmetric Special Geometries; 
    all the names of the representations are the ones used by Slansky 
\cite{Slansky:1981yr}. The meaning of the different columns
is explained in the main text.}
  \label{tab:SSG}
\end{table}

In this section we would like to discuss the embeddings of monopole solutions
into the gauged Magic supergravity theories. We want to show that it is not
always possible to construct, given a prepotential for a theory, a globally
regular solution based on a given monopole solution.  We would like to stress
that this holds for a given prepotential, as the choice of symplectic section
for a given gauged model is physical due to the breakdown of
symplectic invariance.

To start looking for ways to embed monopoles into gauged magic supergravities,
we must discuss first the possible gaugings of the magic models, which boils
down to a group theory problem whose outcome is given in Table~\ref{tab:SSG},
which we are going to explain now.

The scalar manifolds of the magic models are based on symmetric coset spaces
$\mathsf{G}/\mathsf{H}$, which are given in the second and the third column in
the table. As the isometry-group of the scalar manifold, which for the magic
models is isomorphic to $\mathsf{G}$, acts on the symplectic section defining
the model (see Appendix~\ref{sec-gauging}), we should specify under what
representation of $\mathsf{G}$ it transforms; this representation is given in
the column denoted as $\mathsf{G}\circ\mathcal{V}$. The following 2 columns
determine how the isotropy subgroup $\mathsf{H}$ acts on the complex scalars
$Z^{i}=\mathcal{X}^{i}/\mathcal{X}^{0}$; the reason why this is important will
be discussed presently.

As we are interested in monopoles, we shall restrict ourselves to {\em compact
  gauge groups} $G$, which implies that $G\subseteq\mathsf{H}$.  Moreover, as
we restricted ourselves to a specific class of gaugings, {\em i.e.\/}~gaugings
that satisfy Eq.~(\ref{eq:GGchoice}), we should use a prepotential that is
$G$-invariant. Manifestly $\mathsf{H}$-invariant prepotentials for the magic
models were given in Ref.~\cite{Ferrara:2006yb}.  These prepotentials are of
the $STU$-type and have the form

\begin{equation}
\label{eq:MagPre}
\mathcal{F}\left(\mathcal{X}\right) \; =\; 
\frac{\mathsf{I}_{3}\left(\mathcal{X}^{i}\right)}{\mathcal{X}^{0}} \; ,
\end{equation}

\noindent
where $\mathsf{I}_{3}$ is a cubic $\mathsf{H}^{\prime}$-invariant\footnote{
  By $\mathsf{H}^{\prime}$ we mean $\mathsf{H}$ minus the $U(1)$-factors.
}, whose
value for the specific magic model can be found in the seventh column of
Table~\ref{tab:SSG}.

Another implication of our choice of possible gauge groups is that we can only
consider $G\subseteq\mathsf{H}$ for which the branching of the
$\mathsf{H}$-representation of the $\mathcal{X}^{i}$ to $G$-representations
contains only the adjoint representation and singlets. This is a very
restrictive property and the maximal possibilities we found are listed in the
last column of Table~\ref{tab:SSG}.

Having discussed the possible models, we must then start discussing the actual
embedding of the magnetic monopoles.  The first thing is to solve the
stabilization equation to find $\mathcal{R}$ in terms of $\mathcal{I}$.  This
is a complicated question but luckily a general solution exists and was found
by Bates and Denef \cite{Bates:2003vx}; this solution uses the fact that the
generic entropy functions for these models are known.  For our purposes,
however, the full machinery is not needed. Instead, we shall consider the
simpler setting of embedding a purely magnetic monopole in the matter sector
and only turn on an electric component for the graviphoton. This means that we
should solve the stabilization equations,

\begin{equation}
\label{eq:MagStEq}
\begin{array}{lclclcl}
0 
& = &
\Im{\rm m}\, \mathcal{L}^{0}
&
\hspace{.2cm},\hspace{.2cm}
&
\mathcal{I}_{0} 
& = & 
-\Im{\rm m}\,\left[ \mathsf{I}_{3}(\mathcal{L}^{i})/(\mathcal{L}^{0})^{2}\right] \; ,\\
& & & & & & \\
\mathcal{I}^{i}
 & = & 
\Im{\rm m}\,\mathcal{L}^{i}
& , & 
0 
& = & 
\Im{\rm m}\,\left[\partial_{i}\mathsf{I}_{3}(\mathcal{L}^{i})\ /
\mathcal{L}^{0}\right] \; ,
\end{array}
\end{equation}

\noindent
where we absorbed the function $X$ into the $\mathcal{L}$'s. This system
admits a solution

\begin{equation}
  \label{eq:MagStEqSol}
  \mathcal{R}^{i} \; =\; 0 \;\; ,\;\;
  \mathcal{R}^{0} \; =\; -\frac{\sqrt{\ \mathcal{I}_{0}\ \mathsf{I}_{3}(\mathcal{I}^{i})  \ }}{\mathcal{I}_{0}}
  \;\;\mbox{provided that}\;\; \mathcal{I}_{0}\ \mathsf{I}_{3}(\mathcal{I}^{i}) > 0 \; .
\end{equation}

\noindent
With this solution to the stabilization equation, it is then straightforward to use Eq.~(\ref{eq:Gtt})
to determine

\begin{equation}
\label{eq:MagMetFac}
\frac{1}{2|X|^{2}} \; =\; 
4\sqrt{\ \mathcal{I}_{0}\ \mathsf{I}_{3}(\mathcal{I}^{i})\ } \; .
\end{equation}


\subsubsection{The $\mathbb{C}$-magic model}

Let us then consider the $\mathbb{C}$-magic model, which allows an $SU(3)$
gauging. The reason why this is the case is easy to understand: as one can see
from Table~\ref{tab:SSG} the $\mathcal{L}$'s transform under $SU(3)\otimes
SU(3)$ as a $(\mathbf{1},\mathbf{1})\oplus (\mathbf{3},\overline{\mathbf{3}})$
representation. Choosing to gauge the diagonal $SU(3)$ means identifying the
left and the right $SU(3)$ actions so that w.r.t.~the diagonal action the
$\mathcal{L}$'s transform as $\mathbf{1}\oplus
\mathbf{3}\otimes\overline{\mathbf{3}} = \mathbf{1}\oplus
\mathbf{1}\oplus\mathbf{8}$, which is just what we wanted.

The spherically symmetric monopole solution to the $SU(3)$ Bogomol'nyi
equations were found by Wilkinson and Bais in Ref.~\cite{Wilkinson:1978zh},
and a discussion of these solutions is given in Appendix~\ref{sec:bais}.  In
order to discuss the embedding of the WB-monopole, we gather the components of
the symplectic vector $\mathcal{I}$ into a $3\times 3$ matrix,
$\mathcal{I}^{\mathbf{1}\oplus\mathbf{8}}$, and as this matrix behaves as the
sum of a singlet and the adjoint under the diagonal $SU(3)$, we must take it
to be

\begin{equation}
  \mathcal{I}^{\mathbf{1}\oplus\mathbf{8}} \; =\; 
     \textstyle{1\over \sqrt{2}}\left(
          \lambda\ \mathbf{I}_{3} \ -\ 2\Phi 
     \right) \; ,
\end{equation}

\noindent
where $\Phi$ is defined in Eq.~(\ref{eq:WB2a}) and 

\begin{equation}
\lambda = l + L/r\, ,
\end{equation}

\noindent
is a real and spherically symmetric harmonic function.  If we then also
conveniently redefine $\sqrt{2}\mathcal{I}_{0} \equiv H$, where 

\begin{equation}
H=h+ q/r\, ,
\end{equation}

\noindent
is another real harmonic function, we can express Eq.~(\ref{eq:MagMetFac}) as

\begin{equation}
\label{eq:MetFacSU33}
\frac{1}{2|X|^{2}} \; =\; \sqrt{\    
                                H\ (\lambda \ -\ \phi_{1})\ 
                                   (\lambda \ -\ \phi_{2}\ +\ \phi_{1})
                                   (\lambda \ +\ \phi_{2})
                            \ } \; .
\end{equation}

\noindent
Given the asymptotic behaviour of the WB solution, let us for clarity discuss the
non-degenerate solution whose asymptotic behaviour is given in
Eq.~(\ref{eq:WB10}), we can normalize the solution to be asymptotically
Minkowski by demanding that

\begin{equation}
  \label{eq:SU33AsFl}
  1\; =\; h\ \prod_{a=1}^{3} \left( l +\mu_{a}\right) \; .
\end{equation}

\noindent
Using this normalization, we can then extract the asymptotic mass which turns
out to be

\begin{equation}
  \label{eq:SU33mass}
  M \; =\; {\textstyle\frac{1}{4}}\left[ \frac{q}{h} 
                \ +\ L\ \sum_{i=1}^{3}\left( l+\mu_{i}\right)^{-1}
                \ +\ 2\frac{\mu_{3}-\mu_{1}}{(l+\mu_{1})(l+\mu_{3})}
           \right] \; ,
\end{equation}

\noindent
and must be ensured to be positive.

Let us then look for a globally regular embedding of the WB-monopole by tuning
the free parameters: as before, we shall take $q=L=0$ in order to avoid the
Coulomb singularities in the Abelian field strengths. The first obvious remark
is that $h$ is already fixed in terms of $l$ and the $\mu_{a}$ due to
Eq.~(\ref{eq:SU33AsFl}), so that we need to discuss the possible values for
$l$: a first constraint for $l$ comes from the positivity of the mass. Using
the facts that $\mu_{1} < 0$ and $\mu_{3}>0$, which follow from the constraint
and the chosen ordering, in the mass formula (\ref{eq:SU33mass}) we see that
this implies

\begin{equation}
  \label{eq:SU33Monmass}
  M \; =\; \frac{\mu_{3}\ -\ \mu_{1}}{2(l+\mu_{1})(l+\mu_{3})} \ >\ 0 \;\;\; \Longrightarrow\;\;\;
  l < -\mu_{3} \;\;\mbox{or}\;\;
  l > -\mu_{1} \; .
\end{equation}

\noindent
As we are interested in finding globally regular embeddings, we should discuss
the regularity of the metric at $r=0$: as the $\phi_{i}$'s vanish at the
origin we see that regularity implies that

\begin{equation}
  h\ l^{3} \; =\; \prod_{a}\left( 1 +\frac{\mu_{a}}{l}\right)^{-1} \; > \; 0\; .
\end{equation}

\noindent
It is not hard to see that the above holds for the 2 bounds on $l$ derived in
Eq.~(\ref{eq:SU33Monmass}).  At this point then, the real question is whether,
given the constraints on $h$ and $l$ derived above, there are values for $r$
other than $r= 0$ or $r=\infty$ for which the metrical factor in
Eq.~(\ref{eq:MetFacSU33}) vanishes; from the monotonicity of $\phi_{1}$ and
$\phi_{2}$ it is clear that if this is to happen, then this is because the
factor $\lambda -\phi_{2}+\phi_{1}$ vanishes. Seeing, then, that the
combination $\phi_{1}-\phi_{2}$ takes values between $-\mu_{3}$ and
$-\mu_{1}$, we see that Eq.~(\ref{eq:MetFacSU33}) never vanishes if

\begin{equation}
  \lambda \ >\ \mathrm{max}\left( |\mu_{1}|,|\mu_{3}|\right) \;\;\;\mbox{or}\;\;\;
  \lambda \ <\ -\mathrm{max}\left( |\mu_{1}|,|\mu_{3}|\right) \; .
\end{equation}

In order to finish the discussion of the regularity, we must have a look at
the physical scalars: for the above embedding they are schematically given by
$Z^{\mathbf{1}\oplus\mathbf{8}} = i\
\mathcal{I}^{\mathbf{1}\oplus\mathbf{8}}/\mathcal{R}^{0}$, where
$\mathcal{R}^{0}$ is given in Eq.~(\ref{eq:MagStEqSol}). The regularity then
follows straightforwardly from the regularity of monopole solution and the
metric.


\subsubsection{The $\mathbb{Q}$-magic model}

All the embeddings of YM monopoles discussed till now, share a common
ingredient, namely the occurrence of additional Abelian fields, whose
associated harmonic functions can be used to compensate for the vanishing of
the Higgs field at $r=0$. In the above example, this r\^{o}le is played by
$\lambda$ and $\mathcal{I}_{0}$ and in the $\overline{\mathbb{CP}}^{n}$ and
$\mathcal{ST}[2,n]$-models by the graviphoton. In fact, a model in which no
such a compensator exists is the $\mathbb{Q}$-magic model.

As displayed in Table~\ref{tab:SSG}, the $\mathcal{X}$ in the matter sector
lie in the $\mathbf{15}$ of $SU(6)$, which corresponds to holomorphic
2-forms. As $SU(6)$ admits an $SO(6)\sim SU(4)$ as a singular subgroup for
which the relevant branching is $\mathbf{15}\rightarrow\mathbf{15}$, we can
try to embed an $SU(4)$ WB monopole \cite{Wilkinson:1978zh}. This monopole is
given, as in the $SU(3)$ case, by 3 functions $\phi_{i}$ ($i=1,2,3$) and their
embedding into the $\mathbb{Q}$-model has $\mathsf{I}_{3}(\mathcal{I}) =
\mathrm{Pf}(\mathcal{X})= \phi_{1}\phi_{2}\phi_{3}$.  The asymptotic behaviour
can of course be compensated for by choosing $\mathcal{I}_{0}$ judiciously,
but the real problem lies at $r=0$. At the origin the $\phi_{i}$ vanish as
$\phi_{1}\sim r^{3}$, $\phi_{2}\sim r^{4}$ and $\phi_{3}\sim r^{3}$
\cite{Wilkinson:1978zh}, which means that at the origin we have
$\mathsf{I}_{3}(\mathcal{I}) \sim r^{7}+\ldots$ The only freedom we then have
is to use the harmonic function $\mathcal{I}_{0}$, but it is straightforward
to see that this is of no use whatsoever, meaning that the resulting spacetime, as well as
the physical scalars, are singular at $r=0$.


\subsubsection*{Growing hair on the $SU(3)$ WB-monopole}

Let us then end this section, with a small discussion of the hairy black hole
version of the $SU(3)$-monopole. As is discussed in Appendix
(\ref{sec:WBhairy}), singular deformations of the $SU(3)$-monopole can be
found with great ease, and is determined by constants $\beta_{a}$ ($a=1,2,3$)
whose sum is zero. The hard part is to determine the values for the $\beta$'s
for which the metrical factor (\ref{eq:MetFacSU33}) does not vanish for
$r>0$. In fact, lacking general statements about the behaviour of the
$\phi$'s, or the $Q$'s, for general $\beta$, we shall restrict ourselves to
the minimal choice $\beta_{a} = s\mu_{a}$ for $s>0$. For this choice of $\beta$'s,
seeing as we are only shifting the position of where the $Q$'s vanish from
$r=0$ to $r=-s$, the $Q$ are monotonic, positive definite functions on
$\mathbb{R}^{+}$. If we then rewrite the $\phi$'s as

\begin{equation}
  \label{eq:SU33kutje}
  \phi_{i}(r) \ =\ -\partial_{r}\log (Q_{i})  +\frac{2}{r}
              \ =\ -\partial_{r}\log (Q_{i})  + \frac{2}{r+s}  +\frac{2s}{r(s+r)} 
              \ \equiv\ \varphi_{i}(r;s)  + \frac{2s}{r(s+r)} \; ,
\end{equation}

\noindent
where the $\varphi_{i}$ are regular and vanish only at $r=-s$; in fact, they
correspond to the monopole's Higgs field, and are therefore negative definite
on $\mathbb{R}^{+}$.  As pointed out in the appendix, the asymptotic behaviour
of the $\phi_{i}$'s remain the same as in the monopole case, so that also the
normalization condition (\ref{eq:SU33AsFl}) and the asymptotic mass of the
object (\ref{eq:SU33mass}) remain the same.

The negativity of the $\varphi_{i}$ brings us to the next point, namely the
absence of zeroes of the metrical factor at non-zero $r$. This is best
illustrated by having a look at the function $H$ in Eq.~(\ref{eq:MetFacSU33}):
it is clear that if $H$ is to have no zeroes for $r>0$, then $h$ and $q$ must
be either both positive or negative, as otherwise $H=0$ at $|h|r = |q|$.
Following this line of reasoning on all the individual building blocks of the
metrical factor in Eq.~(\ref{eq:MetFacSU33}), and choosing for convenience $h$
and $q$ to be positive, shows that we must take

\begin{equation}
  \lambda \ >\ \mathrm{max}(|\mu_{1}|,|\mu_{3}|) \;\;\;\mbox{and}\;\;\;
  L\ >\ 2 \; ,
\end{equation}

\noindent
which automatically implies that the mass, Eq.~(\ref{eq:SU33mass}), is
positive.

In order to show that this solution corresponds to the description of a black
hole outside its horizon, we must show that the near origin geometry is that of
a Robinson-Bertotti/$AdS_{2}\times S^{2}$ spacetime. As the $\varphi_{i}$ are
regular at $r=0$, the singularities in the Higgs field come from the $1/r$
terms in Eq.~(\ref{eq:SU33kutje}); it is then easy to see that the near-origin
geometry is indeed of the required type and that the resulting black hole
horizon has entropy

\begin{equation}
  \label{eq:SU33entropy}
  S_{bh} \; =\; \sqrt{\ q\ L\ (L^{2}\ -\ 4)  \ }\; .
\end{equation}

Of course, also in this solution the attractor mechanism is at work as one can see
by calculating the values of the scalar fields at $r=0$, {\em i.e.\/}

\begin{equation}
  \label{eq:SU33attract}
  \lim_{r\rightarrow 0} Z^{\mathbf{1}\oplus\mathbf{8}} \; =\; 
     \frac{iq}{2S_{bh}}\; \mathrm{diag}\left(\ L-2\ ,\ L\ ,\ L+2\ \right) \; .
\end{equation}


\section{The null case}
\label{sec-null}

In the null case the two spinors $\epsilon_{1},\epsilon_{2}$ are proportional
and, following the same procedure as in
Refs.~\cite{Meessen:2006tu,Huebscher:2006mr}, we can write\footnote{The
  scalars $\phi_{I}$ carry a -1 charge and the spinor $\epsilon$ a $+1$
  charge, so $\epsilon_{I}$ is neutral. On the other hand, the $\phi_{I}$s
  have zero K\"ahler weight and $\epsilon$ has K\"ahler weight $1/2$.}
$\epsilon_{I}=\phi_{I}\epsilon$ where the $\phi_{I}$s are normalized
$\phi_{I}\phi^{I}=1$ and can be understood as a unit vector selection a
particular direction in $SU(2)$ or, equivalently, in $S^{3}$. It is useful to
project the equations in the $SU(2)$ directions parallel and perpendicular to
$\phi_{I}$. For the fermions supersymmetry transformation rules we obtain the
following four equations:

\begin{eqnarray}
\label{eq:gravisusyruleparallel}
\phi^{I}\delta_{\epsilon}\psi_{I\, \mu} & = & 
\tilde{\mathfrak{D}}_{\mu}\epsilon\, ,
\\
& & \nonumber \\
\label{eq:gaugsusyruleparallel}
\phi_{I}\delta_{\epsilon}\lambda^{Ii} & = & 
i\not\!\!\mathfrak{D} Z^{i}\epsilon^{*}\, , 
\\
& & \nonumber \\
\label{eq:gaugsusyruleperpendicular} 
-\epsilon_{IJ}\phi^{I}\delta_{\epsilon}\lambda^{Ji} & = & 
[\not\!G^{i\, +} +W^{i}]\epsilon\, ,
\\
& & \nonumber \\
\label{eq:gravisusyruleperpendicular}
-\epsilon^{IJ}\phi_{I}\delta_{\epsilon}\psi_{J\, \mu} & = & 
T^{+}{}_{\mu\nu}\gamma^{\nu}\epsilon^{*} 
+\epsilon^{IJ}\phi_{I}\partial_{\mu}\phi_{J}\epsilon\, .
\end{eqnarray}

The first three equations are formally identical to the supersymmetry
variations of the gravitino, chiralini and gaugini in a gauged $N=1,d=4$
supergravity theory with vanishing superpotential that one would get by
projecting out the component $N=2$ gravitini perpendicular to $\phi_{I}$ (last
equation).  This is no coincidence as we could use the Ansatz
$\epsilon_{I}=\phi_{I}\epsilon$ to perform a truncation of the $N=2,d=4$
theory to an $=1,d=4$ theory\footnote{The Ansatz of
  Refs.~\cite{Andrianopoli:2001zh,Andrianopoli:2001gm} is recovered for the
  particular choice $\phi_{I}=\delta_{I}{}^{1}$.}. Thus, the $N=2$ null case
reduces to an equivalent $N=1$ case modulo some details (the presence of the
fourth equation and the covariant derivative $\tilde{\mathfrak{D}}$) that 
will be discussed later. We shall benefit from this fact by using the results of
Refs.~\cite{Ortin:2008wj,Gran:2008vx} in our analysis. We can also predict the
absence of domain-wall solutions in this case, since they only occur in
$N=1,d=4$ supergravity for non-vanishing superpotential.

Before proceeding, observe that the covariant derivative acting on the
supersymmetry parameter $\epsilon$ in $\phi^{I}\delta_{\epsilon}\psi_{I\,
  \mu}$ is defined by 

\begin{equation}
\tilde{\mathfrak{D}}_{\mu}\epsilon 
\equiv 
\{\nabla_{\mu}+
{\textstyle\frac{i}{2}}\tilde{\mathcal{Q}}_{\mu}\}\epsilon\, ,
\hspace{1cm}
\tilde{\mathcal{Q}}_{\mu} \equiv 
\hat{\mathcal{Q}}_{\mu}+\zeta_{\mu}\, ,
\end{equation}

\noindent
where 

\begin{equation}
\zeta_{\mu}\equiv -2i\phi^{I} \partial_{\mu}\phi_{I}\, ,
\end{equation}

\noindent
is a real $U(1)$ connection associated to the remaining local $U(1)$
freedom that is unfixed by our normalization of $\phi_{I}$.
It can be shown, by
comparing the integrability equations of the above KSEs with the KSIs as in
Refs.~(\cite{Tod:1995jf,Meessen:2006tu,Huebscher:2006mr}), that this connection
is flat\footnote{This can be understood as follows: except for $\zeta_{\mu}$, 
  all the objects that appear
  in the KSEs are related to supergravity fields and,
  when working out the integrability conditions, they end up being related to
  the different terms of the different equations of motion. The terms derived
  from $\zeta_{\mu}$ (components of its curvature) are unrelated to any fields
  and one quickly concludes that they must vanish.} and can be eliminated by
choosing the phase of $\epsilon$ appropriately. We will assume that this has
been done and will ignore it from now on.

The KSEs in the null case are therefore
Eqs.~(\ref{eq:gravisusyruleparallel})-(\ref{eq:gravisusyruleperpendicular})
equalled to zero. To analyze them we add to the system an auxiliary spinor
$\eta$, with the same chirality as $\epsilon$ but with opposite $U(1)$ charges
and normalized as

\begin{equation}
\bar{\epsilon}\eta =
-\bar{\eta}\epsilon = {\textstyle\frac{1}{2}}\, .  
\end{equation}

\noindent
This normalization condition will be preserved {\em iff} $\eta$
satisfies

\begin{equation}
\label{eq:Deta}
\mathfrak{D}_{\mu}\eta +a_{\mu}\epsilon=0\, ,  
\end{equation}

\noindent
for some $a_{\mu}$ with $U(1)$ charges $-2$ times those of $\epsilon$, {\em
  i.e.}

\begin{equation}
\mathfrak{D}_{\mu}a_{\nu} =
(\nabla_{\mu} -i\hat{\mathcal{Q}}_{\mu})a_{\nu}\, ,
\end{equation}

\noindent
to be determined by the requirement that the integrability conditions
of this differential equation be compatible with those of the
differential equation for $\epsilon$.

The introduction of $\eta$ allows for the construction of a null tetrad

\begin{equation}
\label{eq:nulltetraddef}
l_{\mu}=i\sqrt{2}\bar{\epsilon^{*}}\gamma_{\mu}\epsilon\, ,
\hspace{.5cm}
n_{\mu}=i\sqrt{2}\bar{\eta^{*}}\gamma_{\mu}\eta\, ,
\hspace{.5cm}
m_{\mu}=i\sqrt{2}\bar{\epsilon^{*}}\gamma_{\mu}\eta\, ,
\hspace{.5cm}
m_{\mu}^{*}=i\sqrt{2}\bar{\epsilon}\gamma_{\mu}\eta^{*}\, .
\end{equation}

\noindent
$l$ and $n$ have vanishing $U(1)$ charges but $m$ ($m^{*}$) has charge $-1$ (+1),
so that the metric constructed using the tetrad

\begin{equation}
\label{eq:nullcasemetric}
ds^{2}= 2\hat{l}\otimes \hat{n} -2\hat{m}\otimes \hat{m}^{*}\, ,  
\end{equation}

\noindent
is invariant.

The orientation of the null tetrad is important: we choose the complex
null tetrad
$\{e^{u},e^{v},e^{z},e^{z^{*}}\}=\{\hat{l},\hat{n},\hat{m},\hat{m}^{*}\}$ such
that

\begin{equation}
\epsilon^{uvzz^{*}}= \epsilon_{uvzz^{*}} =+i\, ,
\hspace{1cm}
\gamma_{5}\equiv -i \gamma^{0}\gamma^{1}\gamma^{2}\gamma^{3} =
-\gamma^{uv}\gamma^{zz^{*}}\, .  
\end{equation}

We can also construct three independent selfdual 2-forms\footnote{The
  expression of these 2-forms in terms of the vectors are found by studying
  the contractions between the 2-forms and vectors using the Fierz
  identities.}:

\begin{eqnarray}
\Phi^{(1)}{}_{\mu\nu} & = & 
\bar{\epsilon}\gamma_{\mu\nu}\epsilon =
2 l_{[\mu}m^{*}_{\nu]}\, , \\
& & \nonumber \\
\Phi^{(2)}{}_{\mu\nu} & = & \bar{\eta}\gamma_{\mu\nu}\epsilon =
[ l_{[\mu}n_{\nu]} + m_{[\mu}m^{*}_{\nu]}]\, ,\\
& & \nonumber \\
\Phi^{(3)}{}_{\mu\nu} & = & 
\bar{\eta}\gamma_{\mu\nu}\eta =
-2 n_{[\mu}m_{\nu]}\, , 
\end{eqnarray}

\noindent
or, in form language

\begin{eqnarray}
\hat{\Phi}^{(1)} & = & \hat{l}\wedge \hat{m}^{*}\, ,\\
& & \nonumber \\  
\hat{\Phi}^{(2)} & = &
{\textstyle\frac{1}{2}}
[\hat{l}\wedge \hat{n} +\hat{m}\wedge \hat{m}^{*}]\, ,\\
& & \nonumber \\  
\hat{\Phi}^{(3)} & = & 
-\hat{n}\wedge \hat{m}\, .
\end{eqnarray}


\subsection{Killing equations for the vector bilinears and first consequences}

Let us first consider the algebraic KSEs
Eqs.~(\ref{eq:gaugsusyruleparallel}--\ref{eq:gravisusyruleperpendicular})
from them one can immediately obtain

\begin{eqnarray}
\label{eq:dZ}
\mathfrak{D} Z^{i}
& = & 
= A^{i}\hat{l}+B^{i}\hat{m}\, ,\\  
& & \nonumber \\
\label{eq:T+}
T^{+} 
& = & 
{\textstyle\frac{1}{2}}\phi\, \hat{\Phi}^{(1)} \, ,\\
& & \nonumber \\
\label{eq:Gi+}
G^{i\, +} 
& = & 
{\textstyle\frac{1}{2}}\phi^{i}\,  \hat{\Phi}^{(1)}
-{\textstyle\frac{1}{2}} W^{i}  \hat{\Phi}^{(2)}\, ,\\
& & \nonumber \\
\label{eq:dfi}
\epsilon^{IJ}\phi_{I}d\phi_{J} 
& = & 
{\textstyle\frac{i}{\sqrt{2}}}\phi \hat{l}\, ,
\end{eqnarray}

\noindent
where $\phi$, $\phi^{i}$, $A^{i}$ and $B^{i}$ are complex functions to be
determined. 

The last equation combined with the vanishing of $\zeta_{\mu}$ imply that

\begin{equation}
\label{eq:dfisl}
d\phi_{I}\sim \hat{l}\, ,
\hspace{1cm}
d\phi \sim \hat{l}\, .
\end{equation}

The resulting vector field strengths $F^{\Lambda\, +}$ are of the form

\begin{equation}
\label{eq:FL}
F^{\Lambda\, +}= {\textstyle\frac{1}{2}}\phi^{\Lambda} 
\hat{\Phi}^{(1)}
-{\textstyle\frac{i}{2}}\mathcal{D}^{\Lambda}\hat{\Phi}^{(2)}\, ,   
\end{equation}

\noindent
where the $\phi^{\Lambda}$ are complex functions related to $\phi$ and $\phi^{i}$ by

\begin{equation}
\label{eq:fis}
\phi^{\Lambda} =i\mathcal{L}^{*\Lambda}\phi +2f^{\Lambda}{}_{i}\phi^{i}\, ,
\end{equation}

\noindent
and we have defined

\begin{equation}
\mathcal{D}^{\Lambda} \; \equiv\; -2if^{\Lambda}{}_{i}W^{i}\, .
\end{equation}

\noindent
Observe that as

\begin{equation}
\label{eq:RfW}
\mathcal{D}^{\Lambda} = 
-igf_{\Sigma\Omega}{}^{\Lambda}\mathcal{L}^{\Omega}
\mathcal{L}^{*\Sigma} =  
{\textstyle\frac{1}{2}}g\Im{\rm m}\mathcal{N}^{-1|\Lambda\Sigma}\mathcal{P}_{\Sigma}\, ,
\end{equation}

\noindent
is real, we find that the field strengths are given by

\begin{equation}
\label{eq:Fnull}
F^{\Lambda} =  -{\textstyle\frac{1}{2}}
(\phi^{*\Lambda}  \hat{m}+\phi^{\Lambda}  \hat{m}^{*})\wedge \hat{l}
-{\textstyle\frac{i}{2}}\mathcal{D}^{\Lambda}\hat{m} \wedge \hat{m}^{*}\, . 
\end{equation}

Let us consider the differential KSE $\mathfrak{D}_{\mu}\epsilon
=0$ and the auxiliar KSE Eq.~(\ref{eq:Deta}): a straightforward
calculation results in

\begin{eqnarray}
\label{eq:dtetrad1}
\mathfrak{D}_{\mu} l_{\nu} 
& = &
\nabla_{\mu} l_{\nu}  =0\, ,\\
& & \nonumber \\
\label{eq:dtetrad2}
\mathfrak{D}_{\mu} n_{\nu} 
& = & 
\nabla_{\mu}n_{\nu} = -a^{*}_{\mu}m_{\nu} -a_{\mu}m^{*}_{\nu}\, ,\\
& & \nonumber \\
\label{eq:dtetrad3}
  \mathfrak{D}_{\mu} m_{\nu} 
& = & 
(\nabla_{\mu}-i\hat{\mathcal{Q}}_{\mu})m_{\nu} 
=-a_{\mu}l_{\nu}\, .
\end{eqnarray}

The first of these equations implies that $l^{\mu}$ is a covariantly constant
null Killing vector, Eq.~(\ref{eq:dtetrad1}), which tells us that the
spacetime is a Brinkmann $pp$-wave \cite{kn:Br1}.  Since $l^{\mu}$ is a
Killing vector and $d\hat{l}=0$ we can introduce the coordinates $u$ and $v$
such that

\begin{eqnarray}
\hat{l}= l_{\mu}dx^{\mu} & \equiv & du\, , \label{u}\\
& & \nonumber \\
l^{\mu}\partial_{\mu}  & \equiv & \frac{\partial}{\partial v}\, . \label{v}
\end{eqnarray}

\noindent
We can also define a complex coordinate $z$ by 

\begin{equation}
\label{eq:z}
\hat{m} = e^{U}dz\, ,  
\end{equation}

\noindent
where $U$ may depend on $z,z^{*}$ and $u$ but not on $v$. Given the chosen coordinates, 
the most general form of $\hat{n}$ is

\begin{equation}
\hat{n}= dv + H du +\hat{\omega}\, ,   
\hspace{1cm}
\hat{\omega}=\omega_{\underline{z}}dz +\omega_{\underline{z}^{*}}dz^{*}\, ,
\end{equation}

\noindent
where all the functions in the metric are independent of $v$.
Either $H$ or the 1-form $\hat{\omega}$ could, in principle, be removed by a
coordinate transformation, but we have to check that the tetrad
integrability equations (\ref{eq:dtetrad1})-(\ref{eq:dtetrad3}) are
satisfied by our choices of $e^{U},H$ and $\hat{\omega}$.

With above choice of coordinates, Eq.~(\ref{eq:nullcasemetric}) leads to the
metric

\begin{equation}
\label{eq:Brinkmetric}
ds^{2} = 2 du (dv + H du +\hat{\omega})
-2e^{2U}dzdz^{*}\, .
\end{equation}

Let us then consider the tetrad integrability equations
(\ref{eq:dtetrad1})-(\ref{eq:dtetrad3}): the first equation is solved because
the metric does not depend on $v$. The third equation, with
the choice (\ref{eq:z}) for the coordinate $z$
implies

\begin{eqnarray}
\hat{a} & = & n^{\mu}
[\partial_{\mu}U -i\hat{\mathcal{Q}}_{\mu}]\hat{m}+D\hat{l}\, ,\\
& & \nonumber \\
\label{eq:mDU}
0 & = & m^{\mu}[\partial_{\mu}U-i \hat{\mathcal{Q}}_{\mu}] \, ,\\
& & \nonumber \\
0 & = &  l^{\mu}A^{\Lambda}{}_{\mu}\Im {\rm m}\, \lambda_{\Lambda}\, ,
\end{eqnarray}

\noindent
where $D$ is a function to be determined. The last equation can be solved by
the gauge choice

\begin{equation}
\label{eq:gauge1}
l^{\mu}A^{\Lambda}{}_{\mu}=0\, .  
\end{equation}

\noindent
In this gauge the complex scalars $Z^{i}$ are $v$-independent. The remaining
components of the gauge field $A^{\Lambda}{}_{\mu}$ are also $v$-independent
as is indicated by the absence of a $\hat{l}\wedge \hat{n}$, $\hat{m}\wedge \hat{n}$ or a
$\hat{m}^{*}\wedge \hat{n}$ term in the vector field strength.
This in its turn, implies the $v$-independence of all the components of the
vector field strengths, of the functions $\phi^{i}$ and, finally, of $A^{i}$
and $B^{i}$.

The above condition does not completely fix the gauge freedom of the
system, since $v$-independent gauge transformations preserve it. We can use
this residual gauge freedom to remove the
$A^{\Lambda}{}_{\underline{u}}$ component of the gauge potential by means of a
$v$-independent gauge transformation. This leaves us with only one complex
independent component $A^{\Lambda}{}_{\underline{z}}(z,z^{*},u)=
(A^{\Lambda}{}_{\underline{z}^{*}})^{*}$ and 

\begin{eqnarray}
\label{eq:Fuz}
F^{\Lambda}{}_{\underline{u}\underline{z}} 
& = & 
\partial_{\underline{u}}A^{\Lambda}{}_{\underline{z}} 
= {\textstyle\frac{1}{2}}e^{U}\phi^{\Lambda}\, ,\\
& & \nonumber \\
F^{\Lambda}{}_{\underline{z}\underline{z}^{*}} 
& = & \partial_{\underline{z}}A^{\Lambda}{}_{\underline{z}^{*}} 
+{\textstyle\frac{1}{2}}g 
f_{\Sigma\Omega}{}^{\Lambda}A^{\Sigma}{}_{\underline{z}} 
A^{\Omega}{}_{\underline{z}^{*}} -\mathrm{c.c.} 
=
-{\textstyle\frac{i}{2}}e^{2U}\mathcal{D}^{\Lambda}\, .
\end{eqnarray}



We can then treat $F^{\Lambda}{}_{\underline{z}\underline{z}^{*}} dz\wedge
dz^{*}$ as a 2-dimensional YM field strength on the 2-dimensional space with
Hermitean metric $2e^{2U}dzdz^{*}$, both of them depending on the parameter
$u$. This implies that we can always write

\begin{equation}
F^{\Lambda}{}_{\underline{z}\underline{z}^{*}} = 
2i\partial_{\underline{z}}\partial_{\underline{z}^{*}} Y^{\Lambda}\, ,  
\end{equation}

\noindent
for some real $Y^{\Lambda}(z,z^{*},u)$. In the Abelian, {\em i.e.\/} ungauged, case

\begin{equation}
\label{eq:abelian}
A^{\Lambda}{}_{\underline{z}}=-i\partial_{\underline{z}}Y^{\Lambda}\, .  
\end{equation}

Using Eq~(\ref{eq:Kconservation}) we can express the second of the tetrad conditions, Eq.~(\ref{eq:mDU}), as

\begin{equation}
\label{eq:mDU2}
\partial_{\underline{z}^{*}}(U+\mathcal{K}/2) = 
-gA^{\Lambda}{}_{\underline{z}^{*}} \lambda_{\Lambda}\, . 
\end{equation}

\noindent
In the ungauged case this equation (and its complex conjugate) can be
immediately integrated to give $U=-\mathcal{K}/2+h(u)$. The function $h(u)$ can be
eliminated by a coordinate redefinition that does not change the form
of the Brinkmann metric.  

In the Abelian case of the pure $N=1,d=4$ theory, it is possible to have
constant momentum maps (D-terms), as considered in
Ref.~\cite{Gutowski:2001pd}, and $\lambda_{\Lambda}=-i\mathcal{P}_{\Lambda}$
and Eq.~(\ref{eq:abelian}) would lead to

\begin{equation}
\label{eq:mDU3}
\partial_{\underline{z}^{*}}(U+\mathcal{K}/2 
+gY^{\Lambda}\mathcal{P}_{\Lambda}) = 0\, ,
\end{equation}
 
\noindent 
which is solved by $U=-\mathcal{K}/2 -gY^{\Lambda}\mathcal{P}_{\Lambda}+h(u)$;
$h(u)$ can still be eliminated by a coordinate transformation. In the $N=2,d=4$ theory,
however, it is not possible to use constant momentum maps to gauge an Abelian
symmetry and the situation is slightly more complicated. The integrability condition of
Eq.~(\ref{eq:mDU2}) and its complex conjugate is solved by

\begin{equation}
\label{eq:integrabilitymDU}
A^{\Lambda}{}_{\underline{z}^{*}}\lambda_{\Lambda}
=\partial_{\underline{z}^{*}}
[R(z,z^{*},u) +S^{*}(z^{*},u)]\, ,  
\end{equation}

\noindent
where $R$ is a real function and $S(z,u)$ a holomorphic function of $z$,
which then implies

\begin{equation}
U=-\mathcal{K}/2 -g(R+S+S^{*})\, .  
\end{equation}

{}Finally, the second tetrad integrability equation (\ref{eq:dtetrad2}) implies

\begin{eqnarray}
D & = & e^{-U}(\partial_{\underline{z}^{*}}H 
-\dot{\omega}_{\underline{z}^{*}})\, ,\\
& & \nonumber \\
\label{eq:doQ}
(d\omega)_{\underline{z}\underline{z}^{*}} & = & 
2ie^{2U}n^{\mu}\hat{\mathcal{Q}}_{\mu}\, ,
\end{eqnarray}

\noindent
whence $\hat{a}$ is given by

\begin{equation}
\label{eq:exprehata}
\hat{a} =   [\dot{U} -{\textstyle\frac{1}{2}}e^{-2U}
(d\omega)_{\underline{z}\underline{z}^{*}}]\hat{m}
+e^{-U}(\partial_{\underline{z}^{*}}H 
-\dot{\omega}_{\underline{z}^{*}})\hat{l}\, .
\end{equation}


\subsection{Killing spinor equations}
\label{sec-killingspinorequations}

In the previous sections we have shown that supersymmetric configurations
belonging to the null case
must necessarily have a metric of the form Eq.~(\ref{eq:Brinkmetric}), vector
field strengths of the form Eq.~(\ref{eq:Fnull}), and scalar field strengths of
the form Eq.~(\ref{eq:dZ}); they must further satisfy Eqs.~(\ref{eq:dfi},\ref{eq:mDU}) and
(\ref{eq:doQ}) for some $SU(2)$ vector $\phi_{I}$. We now want to show that
these conditions are sufficient for a field configuration
$\{g_{\mu\nu},A^{\Lambda},F^{\Lambda},\mathfrak{D}Z^{i}\}$ to be
supersymmetric.

It takes little to no time to see that all the components of the KSEs are satisfied for
constant Killing spinors (in the chosen gauge, frame, etc.) that
obey the condition 

\begin{equation}
\gamma^{u}\epsilon^{I}=0\, .  
\end{equation}

\noindent
This constraint, which is equivalent to $\gamma^{z}\epsilon^{I}=0$,
together with chirality, imply that the Killing spinors live in a complex
1-dimensional space, whence we can write
$\epsilon^{I}=\xi^{I}\epsilon=0$. Up to normalization, solving the KSEs
requires that $\xi^{I}=\phi^{I}$, where the functions $\phi^{I}$ are given as
part of the definition of the supersymmetric field configuration.  As a
result, the supersymmetric configurations of this theory preserve,
generically, $1/2$ of the 8 supercharges.

Observe that in order to prove the existence of Killing spinors it has not
been necessary to impose the integrability conditions of the field strengths,
i.e.~the Bianchi identities of the vector field strengths etc., {\em nor} the
integrability constraints of Eqs.~(\ref{eq:dfi},\ref{eq:mDU}) and
(\ref{eq:doQ}).
We are however forced to do so in order to have well-defined field
configurations in terms of the fundamental fields
$\{g_{\mu\nu},A^{\Lambda},Z^{i}\}$. We will deal with these integrability
conditions and the equations of motion in the next section.


\subsection{Supersymmetric null solutions}
\label{sec-nullsolutions}

Let us start by computing the Bianchi identities and Maxwell equations
taking the expression for $F^{\Lambda\, +}$ in (\ref{eq:FL}) as our starting point.
We find

\begin{equation}
\begin{array}{rcl}
\mathfrak{D}F^{\Lambda\, +} 
& = & 
\left\{    
{\textstyle\frac{1}{2}}m^{*\, \mu}\mathfrak{D}_{\mu}\phi^{\Lambda} 
-{\textstyle\frac{i}{4}}n^{\mu}\mathfrak{D}_{\mu}\mathcal{D}^{\Lambda}
-{\textstyle\frac{i}{2}}\mathcal{D}^{\Lambda} 
n^{\mu}[\partial_{\mu}U-i\hat{\mathcal{Q}}_{\mu}]
\right\} \hat{l}\wedge \hat{m} \wedge \hat{m}^{*}
\\
& & \\
& & 
+{\textstyle\frac{i}{4}}
\left\{
m^{*\, \mu}\mathfrak{D}_{\mu}\mathcal{D}^{\Lambda}
\hat{l}\wedge \hat{n} \wedge \hat{m}
+\mathrm{c.c.}
\right\}\, .
\end{array}
\end{equation}

\noindent
Observe that the terms in the second line are purely imaginary, so that

\begin{equation}
  \begin{array}{rcl}
\star \mathcal{B}^{\Lambda}
& = &
-2\Re{\rm e}\, \mathfrak{D}F^{\Lambda\, +} \\
& & \\
& = & 
-i\left\{
\Im {\rm m}(
m^{*\, \mu}\mathfrak{D}_{\mu}\phi^{\Lambda} )
-{\textstyle\frac{1}{2}}n^{\mu}\mathfrak{D}_{\mu}\mathcal{D}^{\Lambda}
-\mathcal{D}^{\Lambda} n^{\mu}\partial_{\mu}U
\right\} \hat{l}\wedge \hat{m} \wedge \hat{m}^{*}\, .
\end{array}
\end{equation}

A similar calculation for $F_{\Lambda}$ leads to

\begin{equation}
  \begin{array}{rcl}
-\mathfrak{D}F_{\Lambda}
& = & 
-2\Re{\rm e}\, \mathfrak{D}(\mathcal{N}^{*}_{\Lambda\Sigma}F^{\Sigma\, +})
\\
& & \\
& = & 
-i\left\{
\Im{\rm m}\, (m^{*\, \mu}\mathfrak{D}_{\mu}\phi_{\Lambda} )
-{\textstyle\frac{1}{2}}n^{\mu}\mathfrak{D}_{\mu}\Re{\rm e}\, \mathcal{D}_{\Lambda}
-\Re{\rm e}\, \mathcal{D}_{\Lambda} 
n^{\mu}\partial_{\mu}U 
-\Im{\rm m}\, \mathcal{D}_{\Lambda}n^{\mu}\hat{Q}_{\mu}
\right\} \hat{l}\wedge \hat{m} \wedge \hat{m}^{*}
\\
& & \\
& & 
+
\Re {\rm e}
\left[
m^{*\, \mu}\mathfrak{D}_{\mu}\Im{\rm m}\, \mathcal{D}_{\Lambda} 
\hat{l}\wedge \hat{n} \wedge \hat{m}
\right]
\, ,
\end{array}
\end{equation}

\noindent
where

\begin{equation}
\phi_{\Lambda} \equiv \mathcal{N}^{*}_{\Lambda\Sigma} \phi^{\Sigma}\, ,
\hspace{1cm}
\mathcal{D}_{\Lambda} \equiv \mathcal{N}^{*}_{\Lambda\Sigma}
\mathcal{D}^{\Sigma}\, , \,\,\,
\Rightarrow 
\Im {\rm m}\, \mathcal{D}_{\Lambda} =
-{\textstyle\frac{1}{2}}g \mathcal{P}_{\Lambda}\, .
\end{equation}

\noindent
Of course we can also calculate

\begin{equation}
{\textstyle\frac{1}{2}}g
\star \Re{\rm e}\, (k^{*}_{\Lambda\, i}\mathfrak{D}Z^{i})  
= 
{\textstyle\frac{i}{2}}g
\Im{\rm m}\, (n^{\mu}\mathfrak{D}_{\mu}Z^{i}\partial_{i}\mathcal{P}_{\Lambda})  
 \hat{l}\wedge \hat{m} \wedge \hat{m}^{*}
+
{\textstyle\frac{1}{2}}g
\Re{\rm e}\, [m^{*\mu}\mathfrak{D}_{\mu}Z^{i}\partial_{i}\mathcal{P}_{\Lambda} 
\hat{l}\wedge \hat{n} \wedge \hat{m}
]\, ,
\end{equation}

\noindent
which means that the Maxwell equation can be expressed as

\begin{equation}
  \begin{array}{rcl}
\star \mathcal{E}_{\Lambda} 
& = &    
-\mathfrak{D}F_{\Lambda}
+
{\textstyle\frac{1}{2}}g
\star \Re{\rm e}\, (k^{*}_{\Lambda\, i}\mathfrak{D}Z^{i})  
\\
& & \\
& = &
-i\biggl\{
\Im{\rm m}(
m^{*\, \mu}\mathfrak{D}_{\mu}\phi_{\Lambda})
-{\textstyle\frac{1}{2}}n^{\mu}\mathfrak{D}_{\mu}\Re{\rm e}\, \mathcal{D}_{\Lambda}
-\Re{\rm e}\, \mathcal{D}_{\Lambda} 
n^{\mu}\partial_{\mu}U 
\\
& & \\
& & 
\left.
-\Im{\rm m}\, \mathcal{D}_{\Lambda}n^{\mu}\hat{Q}_{\mu}
-{\textstyle\frac{1}{2}}g
\Im{\rm m}\, (n^{\mu}\mathfrak{D}_{\mu}Z^{i}\partial_{i}\mathcal{P}_{\Lambda})  
\right\} \hat{l}\wedge \hat{m} \wedge \hat{m}^{*}
\\
\end{array}
\end{equation}

In concordance with the KSIs, the Maxwell equations and Bianchi identities have only one non-trivial
component, wherefore all the KSIs that involve them are automatically
satisfied.










Finally, the only non-automatically satisfied component of the Einstein equations
is

\begin{equation}
\mathcal{E}_{\underline{u}\underline{u}} = R_{\underline{u}\underline{u}} 
+2 \mathcal{G}_{ij^{*}}A^{i}A^{*\, j^{*}}
-2\Im{\rm m}\mathcal{N}_{\Lambda\Sigma} \phi^{\Lambda}\phi^{*\Sigma} =0\, .
\end{equation}

Using our coordinate and gauge choices
$l^{\mu}A^{\Lambda}{}_{\mu}=A^{\Lambda}{}_{\underline{v}}=0$ and
$n^{\mu}A^{\Lambda}{}_{\mu}=A^{\Lambda}{}_{\underline{u}}=0$, we can rewrite
the above Bianchi identities, Maxwell equations and Einstein equation as

\begin{eqnarray}
\label{eq:eom1}
\Im {\rm m}\, \mathfrak{D}_{\underline{z}} (e^{U}\phi^{\Lambda}) & = &
-{\textstyle\frac{1}{2}}
\partial_{\underline{u}}(e^{2U}\mathcal{D}^{\Lambda})\, ,
\\
& & \nonumber \\
\label{eq:eom2}
\Im {\rm m}\, \mathfrak{D}_{\underline{z}} (e^{U}\phi_{\Lambda}) & = &  
-{\textstyle\frac{1}{2}}\partial_{\underline{u}}(e^{2U}\Re{\rm e}\,
\mathcal{D}_{\Lambda})
-{\textstyle\frac{1}{2}}g\Im{\rm m}\, 
[\partial_{\underline{u}}Z^{i}e^{\mathcal{K}}
\partial_{i}(e^{-\mathcal{K}}\mathcal{P}_{\Lambda})]\, ,
\\
& & \nonumber \\
\partial_{\underline{z}}\partial_{\underline{z}^{*}}H 
& = & 
\partial_{\underline{z}}\dot{\omega}_{\underline{z}^{*}}
+e^{2U}\{\partial_{\underline{u}} +[\dot{U} -{\textstyle\frac{1}{2}}e^{-2U}
(d\omega)_{\underline{z}\underline{z}^{*}}] \} 
[\dot{U} -{\textstyle\frac{1}{2}}e^{-2U}
(d\omega)_{\underline{z}\underline{z}^{*}}] 
\nonumber \\
& & \nonumber \\
& & 
+e^{2U}\mathcal{G}_{ij^{*}}(A^{i}A^{*\, j^{*}}+2\phi^{i}\phi^{*j^{*}}) 
+{\textstyle\frac{1}{2}} e^{2U} |\phi|^{2}\, . 
\label{eq:eom3}
\end{eqnarray}

\noindent
where we made used of

\begin{eqnarray}
\mathfrak{D}_{\underline{z}^{*}} (e^{U}\phi^{\Lambda})
 & \equiv&
\partial_{\underline{z}^{*}} (e^{U}\phi^{\Lambda})
+gf_{\Sigma\Omega}{}^{\Lambda}A^{\Sigma}{}_{\underline{z}^{*}}
e^{U}\phi^{\Omega}\, , \\
\mathfrak{D}_{\underline{z}^{*}} (e^{U}\phi_{\Lambda})
& \equiv&
\partial_{\underline{z}^{*}} (e^{U}\phi_{\Lambda})
+gf_{\Lambda\Sigma}{}^{\Omega}A^{\Sigma}{}_{\underline{z}^{*}}
e^{U}\phi^{\Omega}\, .
\end{eqnarray}

To summarize our results, supersymmetric configurations have vector and scalar
field strengths and metric given by Eqs.~(\ref{eq:Fnull},\ref{eq:dZ}) and (\ref{eq:Brinkmetric}) 
and must satisfy the first-order
differential Eqs.~(\ref{eq:doQ}) and (\ref{eq:mDU2}). We must also find
$\phi_{I}$ and $\phi$ such that

\begin{equation}
\epsilon^{IJ}\phi_{I}\partial_{\underline{u}}\phi_{J} =
{\textstyle\frac{i}{\sqrt{2}}}\phi\, .  
\end{equation}

\noindent
If a supersymmetric configuration satisfies the second-order differential
Eqs.~(\ref{eq:eom1}-\ref{eq:eom3}) then it satisfies all the classical
equations of motion and is supersymmetric solutions.


\subsubsection{$u$-independent supersymmetric null solutions}
\label{sec-uindependentnullsolutions}

In the $u$-independent case the equations that we have to solve simplify
considerably. First of all, since the complex scalars $Z^{i}$ are
$u$-independent, we have $A^{i}=0$ and $(d\omega)_{zz^{*}}=0$, whence we can take
$\hat{\omega}=0$. Furthermore, $\phi^{\Lambda}=0$ (see Eq.~(\ref{eq:Fuz})), which
implies $\phi=\phi^{i}=0$ (see Eq.~(\ref{eq:fis})) and the constancy of
$\phi_{I}$, which is otherwise arbitrary. We need to solve
Eq.~(\ref{eq:mDU2}), which is only possible if its integrability condition
Eq.~(\ref{eq:integrabilitymDU}), which we repeat here for clarity,

\begin{equation}
A^{\Lambda}{}_{\underline{z}^{*}}\lambda_{\Lambda}
=\partial_{\underline{z}^{*}}
[R(z,z^{*},u) +S^{*}(z^{*},u)]\, ,  
\end{equation}

\noindent
is satisfied. Then, the solution is

\begin{equation}
U=-\mathcal{K}/2 -g(R+S+S^{*})\, .  
\end{equation}

We also need to find covariantly-holomorphic functions $Z^{i}(z,z^{*})$ by solving

\begin{equation}
\partial_{\underline{z}^{*}}Z^{i}
+gA^{\Lambda}{}_{\underline{z}^{*}}k_{\Lambda}{}^{i}=0\, ,
\end{equation}

\noindent
which depends strongly on the model.

Finally, the only e.o.m.~need to solve is the Einstein equation Eq.~(\ref{eq:eom3}):
in this case it reduces to the 2-dimensional Laplace equation and is solved by
real harmonic functions $H$ on $\mathbb{R}^{2}$.

In spite of the apparent simplicity of this system, we have not been able to
find solutions different from those of the ungauged theory.

\section{Conclusions and outlook}
\label{sec:concl}
In this paper we have analyzed the conditions that fields have to satisfy in
hyperless $N=2,d=4$ gauged supergravity ($N=2,d=4$ super-Einstein-Yang-Mills
theory) in order to give rise to a supersymmetric solution.

We have presented and analyzed some spherically-symmetric solutions in the
timelike class, which describe monopoles and hairy black holes. As the
monopole solutions to the Bogomol'nyi equations are regular on
$\mathbb{R}^{3}$, we investigated the question of whether this regularity can
be extended to the full supergravity solution, which we called global
regularity. This is a tricky question whose answer, perhaps disappointingly,
is that it depends on the model. As should be clear from the results of
Section~\ref{sec:EYMsols}, the biggest obstruction to generating
globally-regular supergravity solutions out of spherically-symmetric monopoles
can also be one of its virtues, namely that at the origin the Higgs field
vanishes; as long as the model we are using has extra Abelian fields, this
`problem' can be obviated, but otherwise, such as happens in the $SO^{*}(12)$
model, it is a real showstopper.

The hairy black holes were generated by the introduction of a parameter $s>0$
called the \textit{Protogenov hair}. The introduction of this parameter in the
solutions is straightforward and basically consists of doing a coordinate
shift in the exponential parts of the explicit expressions for the gauge
connection and the Higgs field. The effect of this coordinate shift w.r.t.the
monopole solution is to leave unchanged the asymptotic behaviour of the
solution, but to change the behaviour of the solution at the origin. In fact,
due to the positivity of $s$, the singularity is of Coulomb type and opens up
the possibility of creating black holes similar to the ones occurring in
Abelian theories.  The solutions we studied show that the asymptotic data
needed to specify an $N=2$ $d=4$ sugra black hole ({\em i.e.\/} the asymptotic
mass, the moduli and the asymptotic charges) are independent of the parameter
$s$ which is, however, needed in order to specify the black hole fully and
demonstrates the failure of the no-hair theorem for gravity coupled to
YM fields in an explicit and analytic manner.\footnote{There can of course be
  more hairy parameters than just the Protogenov hair. In fact, the
  \textit{cloud} parameter $a$ in Eqs.~(\ref{eq:3Wmon}) and (\ref{eq:WH1a})
  should also be considered as hair.} More surprisingly, the hair parameters
don't show up in other relevant quantities such as the entropy of the black
hole or the attractor values for the scalars at the horizon: a general
understanding of why this happens is lacking but needed. 

The attractor mechanism that holds for the scalars of the Abelian black holes
still works, but in a generalized way: the Higgs field is not gauge-invariant
and one can only expect ``attraction'' up to gauge
transformations. Gauge-invariant combinations of the scalar fields do have
fixed points on the horizon

The question about the multi-monopoles and the multi-non-Abelian black holes
comes quite naturally, not only as their embedding into sugra could defy
Israel's theorem: even though there is a humongous literature on the subject
of multi-monopoles, most of the solutions are not known in explicit form. The
general 2-monopole solution to the $SO(3)$ Bogomol'nyi equation was, after
considerable effort, generated by Panagopoulos \cite{Panagopoulos:1983yx}, who
however did not publish the explicit solution.  The limiting case of the 2
constituents coinciding corresponds to Ward's axisymmetric 2-monopole solution
\cite{Ward:1981jb}, who gives explicit formulae for the Higgs field on the
symmetry axis, taken to coincide with the $z$-axis, and on the $z=0$ plane.
These expressions satisfy the bounds for the regularity of the embedding, but
hardly constitute a definite answer. Work in this direction is in progress.


Recently the magic supergravities were obtained from superstring theory by
means of an asymmetric orbifold construction in Refs.~\cite{Dolivet:2007sz}
and \cite{Bianchi:2007va}.  It would be interesting if these constructions
were to be generalized to the gauged models, which would shed more light on
the stringy properties of the hairy black holes.

On the other hand, the gauged $N=2,d=4$ supergravities that we have considered
here are certainly not the most general ones. One could gauge R-symmetry and
the isometries of the hyperscalar manifold, should there be one. The gauging
of R-symmetry in absence of hyperscalars has been recently studied in
Ref.~\cite{Cacciatori:2008ek} and the timelike case has been completely
solved. The next step would be to include hypermultiplets and the most general
gauging of the hyperscalar manifold (which includes, in a certain limit, the
gauging of R-symmetry) combined with the gaugings considered in this paper.
The null case of the $N=2,d=4$ SEYM theories considered in this paper was
related to gauged $N=1,d=4$ supergravity without a superpotential (but with a
kinetic matrix equal to the complex conjugate of the $N=2$ period matrix).  In
the null case of the most general theory $N=2,d=4$ that we can consider one
should recover gauged $N=1,d=4$ supergravity with both non-trivial kinetic
matrix and superpotential, opening th epossibility of having supersymmetric
domain-wall solutions in this sector.  We hope to present new results in this
direction soon \cite{kn:superpaper}.


\section*{Acknowledgements}

This work has been supported in part by the Spanish Ministry of Science and
Education grants FPU AP2004-2574 (MH), FPA2006-00783 and PR2007-0073 (TO), the
Comunidad de Madrid grant HEPHACOS P-ESP-00346, by the EU Research Training
Network \textit{Constituents, Fundamental Forces and Symmetries of the
  Universe} MRTN-CT-2004-005104, the Spanish Consolider-Ingenio 2010 program
CPAN CSD2007-00042 and by the {\em Fondo Social Europeo} through an
I3P-doctores scholarship (PM). The authors wish to thank
M.~Garc\'{\i}a-P\'erez, R.~Hern\'andez,
D.~Klemm, K.~Landsteiner, E. L\'opez, D.~Mansi, C. Pena and E.~Zorzan for
discussions.
Further, TO wishes to express his gratitude to M.M. Fern\'andez for her
unwavering support.

\appendix

\section{Gauging holomorphic isometries of special K\"ahler manifolds}
\label{sec-gauging}

In this appendix we will review some basics of the gauging of holomorphic
isometries of the special K\"ahler manifold in $N=2,d=4$ supergravities
coupled to vector supermultiplets with the aim of fixing our conventions.

We start by assuming that the Hermitean metric $\mathcal{G}_{ij^{*}}$ admits a
set of Killing vectors\footnote{The index $\Lambda$ always takes values from
  $1$ to $\bar{n}$, but some (or all) the Killing vectors may be zero.}
$\{K_{\Lambda}= k_{\Lambda}{}^{i}\partial_{i}
+k^{*}_{\Lambda}{}^{i^{*}}\partial_{i^{*}}\}$
satisfying the Lie algebra 

\begin{equation}
\label{eq:Liealgebra}
[K_{\Lambda},K_{\Sigma}]= -f_{\Lambda\Sigma}{}^{\Omega} K_{\Omega}\, ,
\end{equation}

\noindent
of the group $G_{V}$ that we want to gauge.

Hermiticity and the $ij$ and $i^{*}j^{*}$ components of the Killing equation
imply that the components $k_{\Lambda}{}^{i}$ and $k^{*}_{\Lambda}{}^{i^{*}}$
of the Killing vectors are, respectively, holomorphic and anti-holomorphic and
satisfy, separately, the above Lie algebra. Once (anti-) holomorphicity is
taken into account, the only non-trivial components of the Killing equation
are

\begin{equation}
\label{eq:Killingeq}
{\textstyle\frac{1}{2}}\pounds_{\Lambda}\mathcal{G}_{ij^{*}}= 
\nabla_{i^{*}}k^{*}_{\Lambda\, j}+ \nabla_{j}k_{\Lambda\, i^{*}} = 0\, ,
\end{equation}

\noindent
where $\pounds_{\Lambda}$ stands for the Lie derivative
w.r.t.~$K_{\Lambda}$. 

The standard $\sigma$-model kinetic term
$\mathcal{G}_{ij^{*}}\partial_{\mu}Z^{i} \partial^{\mu}Z^{*j^{*}}$ is
automatically invariant under infinitesimal reparametrizations of the form

\begin{equation}
\label{eq:deltazio}
\delta_{\alpha} Z^{i} = \alpha^{\Lambda} k_{\Lambda}{}^{i}\, ,  
\end{equation}

\noindent
if the $\alpha^{\Lambda}$s are constants.  If they are arbitrary functions of
the spacetime coordinates $\alpha^{\Lambda}(x)$ we need to introduce a
covariant derivative using as connection the vector fields present in the
theory. The covariant derivative is

\begin{equation}
\label{eq:nablazio}
\mathfrak{D}_{\mu} Z^{i} = \partial_{\mu} Z^{i}+gA^{\Lambda}{}_{\mu} 
k_{\Lambda}{}^{i}\, , 
\end{equation}

\noindent
and transforms as

\begin{equation}
\label{eq:igaugetrans}
\delta_{\alpha}\mathfrak{D}_{\mu} Z^{i}=
\alpha^{\Lambda}(x) \partial_{j} k_{\Lambda}{}^{i}\mathfrak{D}_{\mu} Z^{j}
=-\alpha^{\Lambda}(x)(\pounds_{\Lambda}-K_{\Lambda})
\mathfrak{D}_{\mu} Z^{j} \, ,
\end{equation}

\noindent
provided that the gauge potentials transform as

\begin{equation}
\delta_{\alpha} A^{\Lambda}{}_{\mu} = 
-g^{-1}\mathfrak{D}_{\mu}\alpha^{\Lambda} 
\equiv
-g^{-1}(\partial_{\mu}\alpha^{\Lambda}
+gf_{\Sigma\Omega}{}^{\Lambda}
A^{\Sigma}{}_{\mu} \alpha^{\Omega})\, .
\end{equation}

For any tensor\footnote{Spacetime and target space tensor indices are not
  explicitly shown.} $\Phi$ transforming covariantly under gauge
transformations, i.e.~tranforming as

\begin{equation}
\delta_{\alpha}\Phi = -\alpha^{\Lambda}(x)(\pounds_{\Lambda}
-K_{\Lambda}) \Phi\, ,  
\end{equation}

\noindent
the gauge covariant derivative is given by 

\begin{equation}
\mathfrak{D}_{\mu}\Phi = 
\{\nabla_{\mu} +\mathfrak{D}_{\mu}Z^{i}\Gamma_{i}  
+\mathfrak{D}_{\mu}Z^{*i^{*}}\Gamma_{i^{*}}  
-gA^{\Lambda}{}_{\mu}(\pounds_{\Lambda} 
-K_{\Lambda})\}\Phi\, .
\end{equation}

\noindent
In particular, on $\mathfrak{D}_{\mu}Z^{i}$

\begin{eqnarray}
\mathfrak{D}_{\mu}  \mathfrak{D}_{\nu} Z^{i} & = & 
\nabla_{\mu}\mathfrak{D}_{\nu} Z^{i}+\Gamma_{jk}{}^{i}\mathfrak{D}_{\mu} Z^{j}
\mathfrak{D}_{\nu} Z^{k} +gA^{\Lambda}{}_{\mu}\partial_{j}k_{\Lambda}{}^{i}
\mathfrak{D}_{\nu} Z^{j}\, ,  \\
& & \nonumber \\
\left[ \mathfrak{D}_{\mu},\mathfrak{D}_{\nu} \right]Z^{i}
& = & gF^{\Lambda}{}_{\mu\nu} k_{\Lambda}{}^{i}\, ,
\end{eqnarray}

\noindent
where 

\begin{equation}
F^{\Lambda}{}_{\mu\nu} = 2\partial_{[\mu}A^{\Lambda}{}_{\nu]}
+gf_{\Sigma\Omega}{}^{\Lambda}A^{\Sigma}{}_{[\mu}A^{\Omega}{}_{\nu]}\, , 
\end{equation}

\noindent
is the gauge field strength and transforms under gauge transformations as

\begin{equation}
\delta_{\alpha} F^{\Lambda}{}_{\mu\nu} = 
-\alpha^{\Sigma}(x)f_{\Sigma\Omega}{}^{\Lambda} F^{\Omega}{}_{\mu\nu}\, .
\end{equation}

An important case is that of tensors which only depend on the spacetime
coordinates through the complex scalars $Z^{i}$ and their complex conjugates
so that $\nabla_{\mu} \Phi = \partial_{\mu} \Phi = \partial_{\mu} Z^{i}
\partial_{i} \Phi +\partial_{\mu} Z^{*i^{*}} \partial_{i^{*}} \Phi$. This can
only be true irrespectively of gauge transformations if the tensor $\Phi$ is
invariant, that is

\begin{equation}
\pounds_{\Lambda} \Phi = 0\, .  
\end{equation}

\noindent
The gauge covariant derivative of invariant tensors is always the covariant
pullback of the target covariant derivative:

\begin{equation}
\mathfrak{D}_{\mu}\Phi =
\mathfrak{D}_{\mu}Z^{i}\nabla_{i} \Phi  
+\mathfrak{D}_{\mu}Z^{*i^{*}}\nabla_{i^{*}} \Phi\, .  
\end{equation}



Now, to make the $\sigma$-model kinetic gauge invariant it is enough to
replace the partial derivatives by covariant derivatives.

In $N=2,d=4$ supergravity, however, the scalar manifold is not just Hermitean,
but special K\"ahler, and simple isometries of the metric are not necessarily
symmetries of the theory: they must respect the special K\"ahler structure.
Let us first study how the K\"ahler structure is preserved.

The transformations generated by the Killing vectors will preserve the
K\"ahler structure if they leave the K\"ahler potential invariant up to
K\"ahler transformations, i.e., for each Killing vector $K_{\Lambda}$

\begin{equation}
\label{eq:Kconservation}
\pounds_{\Lambda}\mathcal{K}\equiv 
k_{\Lambda}{}^{i}\partial_{i}\mathcal{K}  
+
k^{*}_{\Lambda}{}^{i^{*}}\partial_{i^{*}}\mathcal{K} 
=
\lambda_{\Lambda}(Z)+ \lambda^{*}_{\Lambda}(Z^{*})\, .
\end{equation}

\noindent
{}From this condition it follows that 

\begin{equation}
\label{eq:lambdaalgebra}
\pounds_{\Lambda}\lambda_{\Sigma} 
-\pounds_{\Sigma}\lambda_{\Lambda} 
=
-f_{\Lambda\Sigma}{}^{\Omega}\lambda_{\Omega}\, .
\end{equation}

On the other hand, the preservation of the K\"ahler structure implies the
conservation of the K\"ahler 2-form $\mathcal{J}$ 

\begin{equation}
\label{eq:Jconservation}
\pounds_{\Lambda}\mathcal{J}=0\, .
\end{equation}

\noindent
The closedness of $\mathcal{J}$ implies that $\pounds_{\Lambda}\mathcal{J}
= d(i_{k_{\Lambda}}\mathcal{J})$ and therefore the preservation of the
K\"ahler structure implies the existence of a set of real 0-forms
$\mathcal{P}_{\Lambda}$ known as \textit{momentum map} such that

\begin{equation}
i_{k_{\Lambda}}\mathcal{J}= \mathcal{P}_{\Lambda}\, .
\end{equation}

A local solution for this equation is provided by

\begin{equation}
i\mathcal{P}_{\Lambda}
=
k_{\Lambda}{}^{i}\partial_{i}\mathcal{K} -\lambda_{\Lambda}\, ,   
\end{equation}

\noindent
which, on account of Eq.~(\ref{eq:Kconservation}) is equivalent to

\begin{equation}
i\mathcal{P}_{\Lambda}
=
-(k^{*}_{\Lambda}{}^{i^{*}}\partial_{i^{*}}\mathcal{K} 
-\lambda^{*}_{\Lambda})\, ,
\end{equation}

\noindent
or

\begin{equation}
\mathcal{P}_{\Lambda}
=
i_{k_{\Lambda}}\mathcal{Q}
-{\textstyle\frac{1}{2i}}(\lambda_{\Lambda}-\lambda^{*}_{\Lambda})\, .
\end{equation}

The momentum map can be used as a prepotential from which the Killing vectors
can be derived:

\begin{equation}
\label{eq:prepo}
k_{\Lambda\, i^{*}} =i\partial_{i^{*}}\mathcal{P}_{\Lambda}\, .  
\end{equation}

Using Eqs.~(\ref{eq:Liealgebra}),(\ref{eq:Kconservation}) and
(\ref{eq:lambdaalgebra}) one finds

\begin{equation}
\label{eq:momentummaptransformationrule}
\pounds_{\Lambda}\mathcal{P}_{\Sigma} = 
2i k_{[\Lambda}{}^{i}k^{*}_{\Sigma]}{}^{j^{*}}\mathcal{G}_{ij^{*}} =
-f_{\Lambda\Sigma}{}^{\Omega} \mathcal{P}_{\Omega}\, .  
\end{equation}

The gauge transformation rule a symplectic section $\Phi$ of K\"ahler weight
$(p,q)$ is\footnote{Again, spacetime and target space tensor indices are not
  explicitly shown. Symplectic indices are not shown, either.}

\begin{equation}
\delta_{\alpha}\Phi = -\alpha^{\Lambda}(x)(\mathbb{L}_{\Lambda}
-K_{\Lambda}) \Phi\, ,  
\end{equation}

\noindent
where $\mathbb{L}_{\Lambda}$ stands for the symplectic and K\"ahler-covariant
Lie derivative w.r.t.~$K_{\Lambda}$ and is given by

\begin{equation}
\mathbb{L}_{\Lambda} \Phi = \{\pounds_{\Lambda}
-[\mathcal{S}_{\Lambda}
-{\textstyle\frac{1}{2}}(p\lambda_{\Lambda}+q\lambda^{*}_{\Lambda})]\}
\Phi\, ,  
\end{equation}

\noindent
where the $\mathcal{S}_{\Lambda}$ are $\mathfrak{sp}(2\bar{n})$ matrices that
provide a representation of the Lie algebra of the gauge group $G_{V}$:

\begin{equation}
\label{eq:SLiealgebra}
[\mathcal{S}_{\Lambda},\mathcal{S}_{\Sigma}]= 
+f_{\Lambda\Sigma}{}^{\Omega} \mathcal{S}_{\Omega}\, .
\end{equation}

\noindent
The gauge covariant derivative acting on these sections is given by 

\begin{equation}
  \begin{array}{rcl}
\mathfrak{D}_{\mu}\Phi & = &  
\{\nabla_{\mu} +\mathfrak{D}_{\mu}Z^{i}\Gamma_{i}  
+\mathfrak{D}_{\mu}Z^{*i^{*}}\Gamma_{i^{*}}  
+{\textstyle\frac{1}{2}}(pk_{\Lambda}{}^{i}\partial_{i}\mathcal{K}+
qk^{*}_{\Lambda}{}^{i^{*}}\partial_{i^{*}}\mathcal{K})  \\
& & \\
& & 
+gA^{\Lambda}{}_{\mu}[\mathcal{S}_{\Lambda} 
+{\textstyle\frac{i}{2}}(p-q)\mathcal{P}_{\Lambda} -(\pounds_{\Lambda} 
-K_{\Lambda})]\}\Phi\, .\\
\end{array}
\end{equation}

Invariant sections are those for which 

\begin{equation}
\mathbb{L}_{\Lambda} \Phi =0\, ,\,\,\,\, \Rightarrow\,\,\,\,  
\pounds_{\Lambda}\Phi = 
[\mathcal{S}_{\Lambda}
-{\textstyle\frac{1}{2}}(p\lambda_{\Lambda}+q\lambda^{*}_{\Lambda})]
\Phi\, ,
\end{equation}

\noindent
and their gauge covariant derivatives are, again, the covariant pullbacks of
the K\"ahler-covariant derivatives:

\begin{equation}
\mathfrak{D}_{\mu}\Phi = \mathfrak{D}_{\mu}Z^{i}\mathcal{D}_{i}\Phi  
+\mathfrak{D}_{\mu}Z^{*i^{*}}\mathcal{D}_{i^{*}}\Phi\, .  
\end{equation}

By hypothesis (preservation of the special K\"ahler structure), the canonical
weight $(1,-1)$ section $\mathcal{V}$ is an invariant section

\begin{equation}
\label{eq:KLV}
K_{\Lambda}\mathcal{V} = 
[\mathcal{S}_{\Lambda}
-{\textstyle\frac{1}{2}}(\lambda_{\Lambda}-\lambda^{*}_{\Lambda})]
\mathcal{V}\, ,
\end{equation}

\noindent
and its gauge covariant derivative is given by

\begin{equation}
\mathfrak{D}_{\mu}\mathcal{V} = 
\mathfrak{D}_{\mu}Z^{i}\mathcal{D}_{i}\mathcal{V}= 
\mathfrak{D}_{\mu}Z^{i}\mathcal{U}_{i}\, .
\end{equation}

\noindent
Using the covariant holomorphicity of $\mathcal{V}$ one can write 

\begin{equation}
K_{\Lambda}\mathcal{V} = k_{\Lambda}{}^{i}\mathcal{U}_{i}
-i\mathcal{P}_{\Lambda}\mathcal{V} -{\textstyle\frac{1}{2}}
(\lambda_{\Lambda}-\lambda^{*}_{\Lambda})\mathcal{V}\, ,
\end{equation}

\noindent
and, comparing with Eq.~(\ref{eq:KLV}) and taking the symplectic product with
$\mathcal{V}^{*}$, we find another expression for the momentum map

\begin{equation}
\label{eq:altmomentum}
\mathcal{P}_{\Lambda} = \langle\, \mathcal{V}^{*} \mid
\mathcal{S}_{\Lambda}\mathcal{V}\, \rangle\, ,  
\end{equation}

\noindent
which leads, via Eq.~(\ref{eq:prepo}) to another expression for the Killing
vectors

\begin{equation}
\label{eq:altkilling}
k_{\Lambda}{}^{i}=  i\partial^{i}\mathcal{P}_{\Lambda}=
i\langle\, \mathcal{V} \mid
\mathcal{S}_{\Lambda}\mathcal{U}^{*i}\, \rangle\, .
\end{equation}

\noindent
If we take the symplectic product with $\mathcal{V}$ instead, we get the
following condition 

\begin{equation}
\label{eq:VSLV}
\langle\, \mathcal{V} \mid
\mathcal{S}_{\Lambda}\mathcal{V}\, \rangle  =0\, .
\end{equation}

\noindent
Using the same identity and $\mathcal{G}_{ij^{*}}=
-i \langle\, \mathcal{U}_{i} \mid \mathcal{U}^{*}_{j^{*}}\, \rangle$ 
one can also show that

\begin{equation}
\label{eq:klks}
k_{\Lambda}{}^{i}k^{*}_{\Sigma}{}^{j^{*}}\mathcal{G}_{ij^{*}}=
\mathcal{P}_{\Lambda}  \mathcal{P}_{\Sigma} -i
\langle\, \mathcal{S}_{\Lambda}\mathcal{V}\mid
\mathcal{S}_{\Sigma}\mathcal{V}^{*}\, \rangle\, .
\end{equation}

It follows that
\begin{equation}
\langle\, \mathcal{S}_{[\Lambda}\mathcal{V} \mid
\mathcal{S}_{\Sigma]}\mathcal{V}^{*}\, \rangle\,=
-{\textstyle\frac{1}{2}}f_{\Lambda\Sigma}{}^\Omega\mathcal P_\Omega.
\end{equation}

The gauge covariant derivative of $\mathcal{U}_{i}$ is

\begin{equation}
\mathfrak{D}_{\mu}\mathcal{U}_{i} = 
\mathfrak{D}_{\mu}Z^{j}\mathcal{D}_{j}\mathcal{U}_{i} 
+\mathfrak{D}_{\mu}Z^{*j^{*}}\mathcal{D}_{j^{*}}\mathcal{U}_{i}
= 
i\mathcal{C}_{ijk}\mathcal{U}^{*j}
\mathfrak{D}_{\mu}Z^{k}
+\mathcal{G}_{ij^{*}}\mathcal{V}\mathfrak{D}_{\mu}Z^{*j^{*}}\, .  
\end{equation}

On the supersymmetry parameters $\epsilon_{I}$, which have $(1/2,-1/2)$ weight

\begin{equation}
\label{eq:dei}
\mathfrak{D}_{\mu}\epsilon_{I} =
\left\{
\nabla_{\mu}+{\textstyle\frac{i}{2}}\hat{\mathcal{Q}}_{\mu}
\right\}\epsilon_{I}\, ,
\end{equation}

\noindent
where we have defined

\begin{equation}
\hat{\mathcal{Q}}_{\mu}
\equiv \mathcal{Q}_{\mu}+gA^{\Lambda}{}_{\mu}\mathcal{P}_{\Lambda}\, .  
\end{equation}

The formalism, so far, applies to any group $G_{V}$ of isometries. However, we
will restrict ourselves to those for which the matrices 

\begin{equation}
\mathcal{S}_{\Lambda} =
\left(
  \begin{array}{cc}
a_{\Lambda}{}^{\Omega}{}_{\Sigma} & b_{\Lambda}{}^{\Omega\Sigma} \\
& \\
c_{\Lambda\Omega\Sigma} & d_{\Lambda\Omega}{}^{\Sigma} \\
\end{array}
\right)\, , 
\end{equation}

\noindent
have $b=c=0$. The symplectic transformations with $b\neq 0$ are not symmetries
of the action and the gauging of symmetries with $c\neq 0$ leads to the
presence of complicated Chern-Simons terms in the action. The matrices $a$ and
$d$ are 

\begin{equation}\label{eq:GGchoice}
a_{\Lambda}{}^{\Omega}{}_{\Sigma} =f_{\Lambda\Sigma}{}^{\Omega}\, ,
\hspace{1cm}
d_{\Lambda\Omega}{}^{\Sigma} =  -f_{\Lambda\Omega}{}^{\Sigma}\, .
\end{equation}

\noindent
These restrictions lead to additional identities. First, observe that the
condition Eq.~(\ref{eq:VSLV}) takes the form

\begin{equation}
\label{eq:VSLV2}
f_{\Lambda\Sigma}{}^{\Omega}\mathcal{L}^{\Sigma}\mathcal{M}_{\Omega}=0\, ,
\end{equation}

\noindent
and the covariant derivative of Eq.~(\ref{eq:VSLV}) $\langle\, \mathcal{V}
\mid \mathcal{S}_{\Lambda}\mathcal{U}_{i}\, \rangle =0$

\begin{equation}
\label{eq:VSLV3}
f_{\Lambda\Sigma}{}^{\Omega}(f^{\Sigma}{}_{i}\mathcal{M}_{\Omega}
+h_{\Omega\, i}\mathcal{L}^{\Sigma})=0\, .
\end{equation}

\noindent
Then, using Eqs.~(\ref{eq:altmomentum}) and (\ref{eq:altkilling}) and
Eqs.~(\ref{eq:VSLV}),(\ref{eq:VSLV2})  and (\ref{eq:VSLV3}) we find that

\begin{eqnarray}
\label{eq:LP0}
\mathcal{L}^{\Lambda}\mathcal{P}_{\Lambda} & = & 0\, ,\\
& & \nonumber \\
\label{eq:LK0}
\mathcal{L}^{\Lambda}k_{\Lambda}{}^{i} & = & 0\, ,\\  
& & \nonumber \\
\label{eq:LKfP}
\mathcal{L}^{*\Lambda}k_{\Lambda}{}^{i} & = & 
-if^{*\Lambda\, i}\mathcal{P}_{\Lambda}\, .  
\end{eqnarray}

{}From the first two equations it follows that

\begin{equation}
\mathcal{L}^{\Lambda}\lambda_{\Lambda}=0\, .  
\end{equation}

Some further equations that can be derived and are extensively used in the
calculation throughout the text are explicit versions of
Eqs. (\ref{eq:altmomentum}) and (\ref{eq:altkilling}), {\em i.e.\/}

\begin{equation}
\label{eq:1}
\mathcal{P}_{\Lambda} \; =\; 2f_{\Lambda\Sigma}{}^{\Gamma}
\Re{\rm e}\,\left( \mathcal{L}^{\Sigma} \mathcal{M}^{*}_{\Gamma}\right)\, ,
\hspace{1cm}
k_{\Lambda\, i^{*}} \; =\; 
if_{\Lambda\Sigma}{}^{\Gamma}\left( f^{*\Sigma}_{i^{*}}M_{\Gamma} 
+ \mathcal{L}^{\Sigma}h^{*}_{\Gamma i^{*}}
 \right) \; .
\end{equation}

Finally, notice the identity

\begin{equation}
\label{eq:laleche}
k_{\Lambda\, i^{*}}\mathfrak{D}Z^{*i^{*}}  
-k^{*}_{\Lambda i}\mathfrak{D}Z^{i} =i\mathfrak{D}\mathcal{P}_{\Lambda}
= i(d\mathcal{P}_{\Lambda}
+f_{\Lambda\Sigma}{}^{\Omega}A^{\Sigma}\mathcal{P}_{\Omega}) \, . 
\end{equation}

The absolutely last comment in this appendix is the following: if we start
from the existence of a prepotential $\mathcal{F}(\mathcal{X})$, then
Eq.~(\ref{eq:VSLV}) implies

\begin{equation}
\label{eq:2PreP}
0 \; =\; f_{\Lambda\Sigma}{}^{\Gamma}\ \mathcal{X}^{\Sigma}\partial_{\Gamma}\ 
\mathcal{F} \; , 
\end{equation}

\noindent
the meaning of which is that one can gauge only the invariances of the
prepotential.  To put it differently: if you want to construct a model having
$\mathfrak{g}$ as the gauge algebra, you need to pick a prepotential that is
$\mathfrak{g}$-invariant.


\section{The $\mathcal{ST}[2,n]$ models}
\label{sec:ST2n}

The $\mathcal{ST}[2,n]$ models have as their K\"ahler geometry the homogeneous
space $\scriptstyle{\frac{SU(1,1)}{U(1)}\times\frac{SO(2,n)}{SO(2)\otimes
    SO(n)}}$, which is of complex-dimension $n+1$, and must therefore be
embedded into $Sp(n+1;\mathbb{R})$.  As we are mainly interested in the
solution to the stabilization equations, which for this model were solved in
Ref.~\cite{Kallosh:1996tf}, and also in the gaugeability of the model, it is
convenient to start with the parametrization of the symplectic section for
which no prepotential exists. One advantage of this parametrization is that
the $SO(2,n)$ symmetry is obvious as one can see from

\begin{equation}
\label{eq:ST1}
\mathcal{V}^{T} \ =\ 
\left(\mathcal{L}^{\Lambda}\ ,\ \eta_{\Lambda\Sigma}\ \mathrm{S}\mathcal{L}^{\Sigma}\right)
\;\;\mbox{where}\;\; \eta =\mathrm{diag}\left( [+]^{2},[-]^{n}\right) 
\;\;\mbox{and}\;\;   \mathcal{L}^{T}\eta\mathcal{L}\ =\ 0 \; ,
\end{equation}

\noindent
where the constraint is necessary to ensure the correct number of degrees of
freedom.  Also, and for want of a better place to say so, we take the
symplectic indices to run over $\Lambda = (\underline{1},0,\ldots ,n)$.

In order to declutter the solution to the stabilization equation $\mathcal{I}
=\Im{\rm m}\,\left(\mathcal{V}/X\right)$, we absorb the $X$ into the
$\mathcal{L}$ and introduce the abbreviations
$p^{\Lambda}=\mathcal{I}^{\Lambda}$ and
$q_{\Lambda}=\mathcal{I}_{\Lambda}$. If we then also use $\eta$ to raise and
lower the indices, we can write the stabilization equation as

\begin{equation}
\label{eq:ST2}
2i\ p^{\Lambda} 
\; =\; 
\mathcal{L}^{\Lambda} -{\mathcal{L}}^{* \Lambda} \;\; ,\;\;
2i\ q^{\Lambda} 
\; =\; 
\mathrm{S}\ \mathcal{L}^{\Lambda} -{\mathrm{S}}^{*}\ {\mathcal{L}}^{* \Lambda}
\;\; \longrightarrow\;\;
\mathcal{L}^{\Lambda} 
\; =\; 
\frac{q^{\Lambda}\ -\ 
{\mathrm{S}}^{*}\ p^{\Lambda}}{\Im{\rm m}\,\mathrm{S}} \; .
\end{equation}

\noindent
The function $\mathrm{S}$ is then easily found by solving the constraint 
$\mathcal{L}_{\Lambda}\mathcal{L}^{\Lambda}=0$, and gives

\begin{equation}
\label{eq:ST3}
\mathrm{S} 
\; =\; 
\frac{p\cdot q}{p^{2}} \; -i\; \frac{\sqrt{\ p^{2}q^{2} \ -\ (p\cdot q)^{2}\ }}{p^{2}} \; ,
\end{equation}

\noindent
so that we have the constraint $p^{2}q^{2}>(p\cdot q)^{2}$; the sign of
$\Im{\rm m}\,\mathrm{S}$ is fixed by the positivity of the metrical
function, which with the above sign reads

\begin{equation}
\label{eq:ST4}
\frac{1}{2|X|^{2}} \; =\; 2\sqrt{\ p^{2}q^{2}\ -\ (p\cdot q)^{2}\ } \; .
\end{equation}

We would like to stress that this solution is manifestly $SO(2,n)$
(co/in)variant and automatically solves the constraint
$\mathcal{L}^{T}\eta\mathcal{L}=0$, without any constraints on $p^{\Lambda}$
nor on $q_{\Lambda}$.

For our applications, namely the regularity of the embeddings of monopoles and
the attractor mechanism, it is important to to know the expression of the
moduli in terms of $(n+1)$ unconstrained fields, one of which should be
$\mathrm{S}$ as it corresponds to the axidilaton. This means that we should
have $n$ unconstrained fields $Z^{a}$ ($a=0,1,\ldots ,n-1$) and express them
in terms of $p$'s and $q$'s.

One way of doing this is through the introduction of so-called
Calabi-Visentini coordinates which means that ($a=1,\ldots ,n$)

\begin{equation}
  \label{eq:ST5}
  \mathcal{L}^{\underline{1}} \; =\; \textstyle{1\over 2}\ Y^{0}\ \left(1+\vec{Z}^{2}\right)
  \; ,\;
  \mathcal{L}^{0} \; =\; \textstyle{i\over 2}\ Y^{0}\ \left( \vec{Z}^{2}-1\right)
  \; ,\;
  \mathcal{L}^{a} \; =\; Y^{0}\ Z^{a} \; ,
\end{equation}

\noindent
which after solving for $Y^{0}$ means that the scalar fields are given by 

\begin{equation}
  \label{eq:ST6}
  Z^{a} \ =\ \frac{\ q^{a}\ -\ \mathrm{S}^{*}\ p^{a}\ }
{q^{\underline{1}}+iq^{0}\ -\ \mathrm{S}^{*}\left( p^{\underline{1}}+ip^{0}\right) }\; , 
\end{equation}

\noindent
and $\mathrm{S}$ is given by expression (\ref{eq:ST3}). Observe that in this
parametrization the $SO(n)$ invariance is manifest.

In order to discuss the possible groups that can be gauged in these models,
let us recall that a given compact simple Lie algebra $\mathfrak{g}$ of a
group $G$ is a subalgebra of $\mathfrak{so}(\mathrm{dim}(\mathfrak{g}))$ and
furthermore the latter's vector representation branches into $\mathfrak{g}$'s
adjoint representation. This then implies that in an $\mathcal{ST}[2,n]$-model
one can always gauge a group $G$ as long as $n\geq
\mathrm{dim}(\mathfrak{g})$.

In Section~\ref{sec:EYMsols} the explicit details are given for the
$\overline{\mathbb{CP}}^{n}$ models, but at least as far as the embedding of
the monopoles are concerned, the embedding into the $\mathcal{ST}$-models is
similar. In order to show that this is the case, consider the case of a purely
magnetic solution, so that $q^{a}=0$, and take furthermore
$q_{0}=p^{\underline{1}}=0$ and normalize $q_{\underline{1}}=1$.  Using this
Ansatz in Eq.~(\ref{eq:ST4}) we obtain

\begin{equation}
  \frac{1}{2|X|^{2}} \; =\; 2\sqrt{\ p^{2}\ }
                     \; =\; 2\sqrt{\ (p^{0})^{2} \ -\ (p^{a})^{2}\ } \; ,
\end{equation}

\noindent
which, apart from the $\textstyle{\sqrt{~}}$, is just the same expression as
obtained in the $\overline{\mathbb{CP}}^{n}$-models and leads to the same
conditions for the global regularity of the metric.  Using the same Ansatz in
Eq.~(\ref{eq:ST6}) for the scalars, one finds

\begin{equation}
  Z^{a} \; =\; -i\frac{\sqrt{p^{2}}}{p^{2}\ +\ p^{0}\sqrt{p^{2}}}\ p^{a} \; .
\end{equation}

This then means that as long as $p^{0}>0$ and $p^{2}$ is regular and positive
definite, as is the case for the solutions in section (\ref{sec:EYMsols}), the
embeddings of the monopoles is a globally regular supergravity solution.

\section{The Wilkinson-Bais monopole in $SU(3)$}
\label{sec:bais}

In Ref.~\cite{Wilkinson:1978zh}, Bais and Wilkinson derived the general
spherically symmetric monopoles to the $SU(N)$ Bogomol'nyi equations.
In this case we are going to discuss their monopole for the case of 
$SU(3)$ as it can be embedded into the $\overline{\mathbb{CP}}^{8}$,
$ST[2,8]$ and the $SU(3,3)/S[U(3)\otimes U(3)]$ model.

The derivation is best done using Hermitean generators and in the 
fundamental, which means that we use the definitions

\begin{equation}
  \label{eq:WB1}
  \mathfrak{D}\Phi \; =\; d\Phi -i\left[ A,\Phi\right] \;\; ,\;\;
  F \; =\; dA \ -\ i\ A\wedge A \; ,
\end{equation}

\noindent
where $A$ and $\Phi$ are $\mathfrak{su}(3)$-valued, and we have taken $g=1$.

The maximal form of the fields compatible with spherical symmetry are given by

\begin{eqnarray}
  \label{eq:WB2a}
  \Phi & =& \textstyle{1\over 2}\mathrm{diag}\left[ 
                \phi_{1}(r)\ ;\ \phi_{2}(r)-\phi_{1}(r)\ ;\ -\phi_{2}(r)
             \right] \; , \\
    & & \nonumber\\
  \label{eq:WB2b}
  A  & =& J_{3}\ \cos (\theta )d\varphi 
     \ +\ \textstyle{i\over 2}\left[ C -C^{\dagger}\right]\ d\theta
     \ +\ \textstyle{1\over 2}\left[ C+C^{\dagger}\right]\ \sin (\theta )d\varphi \; ,
\end{eqnarray}

\noindent
where $J_{3} =\mathrm{diag}(1;0;-1)$ and $C$ is the real and upper-triangular matrix

\begin{equation}
  \label{eq:WB3}
  C \ =\ \left(\begin{array}{ccc}
             0 & a_{1}(r) & 0 \\
             0 & 0 & a_{2}(r) \\
             0 & 0 & 0 
         \end{array}\right)\; .
\end{equation}

\noindent
Plugging the above Ans\"atze into the Bogomol'nyi equation $\mathfrak{D}\Phi
=\star F$, leads to the following equations ($i=1,2$)

\begin{equation}
  \label{eq:WB4}
  r^{2}\partial_{r}\phi_{i} \ =\ a_{i}^{2} -2 \;\; ,\;\;
  2\partial_{r}a_{1} \ =\ a_{1}\left( 2\phi_{1} -\phi_{2}\right) \;\; ,\;\;
  2\partial_{r}a_{2} \ =\ a_{2}\left( 2\phi_{2} -\phi_{1}\right) \; .
\end{equation}

\noindent
Following Wilkinson and Bais \cite{Wilkinson:1978zh}, we solve the equations for 
the $a_{i}$ by defining new functions $Q_{i}(r)$ through

\begin{equation}
  \label{eq:WB5}
  \phi_{i} \ =\ -\partial_{r}\log{Q_{i}} +{\textstyle\frac{2}{r}}
  \;\; ,\;\;
  a_{1}\ \equiv\ \frac{r\sqrt{Q_{2}}}{Q_{1}} 
  \;\; ,\;\;
  a_{2}\ \equiv\ \frac{r\sqrt{Q_{1}}}{Q_{2}} \; , 
\end{equation}

\noindent
after which the remaining equations are

\begin{equation}
  \label{eq:WB6}
  Q_{2} \ =\ \partial_{r}Q_{1}\partial_{r}Q_{1} \ -\ Q_{1}\partial_{r}^{2}Q_{1}
  \;\; ,\;\;
  Q_{1} \ =\ \partial_{r}Q_{2}\partial_{r}Q_{2} \ -\ Q_{2}\partial_{r}^{2}Q_{2}
\end{equation}

The solution found by Wilkinson {\&} Bais for $SU(3)$ then given by 

\begin{equation}
  \label{eq:WB7}
  \left.\begin{array}{ccc}
      Q_{1} & =& \sum_{a=1}^{3}\ A_{a}\ e^{\mu_{a} r} \\
        & & \\
      Q_{2} & =& \sum_{a=1}^{3}\ A_{a}\ e^{-\mu_{a} r} 
  \end{array}\right\} \;\longleftarrow\;
  \left\{\begin{array}{ccc} 
      0 & =& \sum_{a=1}^{3}\ \mu_{a} \\
        & & \\
      A_{1} & =& -A_{2}A_{3}\ \left( \mu_{2} -\mu_{3}\right)^{2}\\
      A_{2} & =& -A_{3}A_{1}\ \left( \mu_{3} -\mu_{1}\right)^{2}\\
      A_{3} & =& -A_{1}A_{2}\ \left( \mu_{1} -\mu_{2}\right)^{2}\\ 
      \end{array}\right.\; .
\end{equation}

\noindent
The solution to the above equations is 

\begin{equation}
  \label{eq:WB9}
  A_{a} \; =\; \prod_{b\neq a}\left(\mu_{a}-\mu_{b}\right)^{-1} \; .
\end{equation}

Defining the useful quantity $V_{n} \equiv\sum_{a=1}^{3}\ A_{a}\mu_{a}^{n}$,
we can see by direct inspection that $V_{0}=V_{1}=V_{3}=0$ and that $V_{1}=1$.
Using these quantities one can see that around $r=0$ we see that $Q_{i}\sim
r^{2}/2+\mathcal{O}(r^{3})$, which means that the $\phi_{i}\sim -V_{4}/3!\ r +
\mathcal{O}(r^{2})$, implying that the solution is completely regular on
$\mathbb{R}^{3}$. Furthermore, one can show that the $Q$ are monotonic,
positive semi-definite functions on $\mathbb{R}^{+}$ that vanish only at
$r=0$, at which point also its derivative vanishes. This furthermore implies
that the $\phi_{i}$ are negative semi-definite functions on $\mathbb{R}^{+}$.


The asymptotic behaviour of the Higgs field is easily calculated and, choosing
$\mu_{1}<\mu_{2}<\mu_{3}$, is readily seen to be

\begin{equation}
  \label{eq:WB10}
  \lim_{r\rightarrow\infty}\Phi \; =\; -{\textstyle\frac{1}{2}}\ 
        \mathrm{diag}\left(\ \mu_{3}\ ;\ \mu_{2}\ ;\ \mu_{1}\ \right)
        \; +\; \frac{1}{r}\ J_{3} \; +\ldots
\end{equation}

\noindent
from which the breaking of $SU(3)\rightarrow U(1)^{2}$ is paramount.

The above solution does not admit the possibility of having degenerate
$\mu$'s, but as emphasised by Wilkinson {\&} Bais, such a solution can be
obtained as a limiting solution. For this, define $\mu_{1}=-2$, $\mu_{2} =
1-\delta$ and $\mu_{3}=1+\delta$, for $\delta >0$, and calculate the
solution. This solution admits a non-singular $\delta\rightarrow 0$ limit,
which is

\begin{equation}
  \label{eq:WB11}
  Q_{1} \; =\; \textstyle{1\over 9}\left[ e^{-2r}\ +\ (3r-1)e^{r}\right]
  \;\; ,\;\;
  Q_{2} \; =\; \textstyle{1\over 9}\left[ e^{2r}\ -\ (3r+1)e^{-r}\right]
  \; .
\end{equation}

The symmetry breaking pattern in this degenerate case is $SU(3)\rightarrow U(2)$
as becomes clear from the asymptotic behaviour of the Higgs field, {\em i.e.\/}

\begin{equation}
  \label{eq:WB12}
  \lim_{r\rightarrow\infty}\Phi \; =\; -\mathtt{Y} \ +\ \frac{1}{r}\ \mathtt{Y} 
  \;\; \mbox{where}\;\;
  \mathtt{Y} \ =\ \textstyle{1\over 2}\ \mathrm{diag}\left( 1\ ,\ 1\ ,\ -2\right)\; .
\end{equation}


\subsection{A hairy deformation of the W{\&}B monopole}
\label{sec:WBhairy}

The foregoing derivation of Wilkinson {\&} Bais's monopole was cooked up to
give a regular solution, and we would like to have a hairy version of this
monopole.  This is easily achieved by applying the Protogenov trick, which
calls for adding constants in the exponential parts of the monopole fields; in
this case, we simply extend the Ansatz for the $Q_{i}$'s to

\begin{equation}
  \label{eq:WBh1}
  Q_{1} \; =\; \sum_{a=1}^{3}\ A_{a}\ e^{\mu_{a}r +\beta_{a}} \;\; ,\;\;
  Q_{2} \; =\; \sum_{a=1}^{3}\ A_{a}\ e^{-\mu_{a} r -\beta_{a}} \; ,
\end{equation}
and plug it into Eq.~(\ref{eq:WB6}). Obviously this leads to a solution if
$\sum\ \mu_{a} =\sum\beta_{a} =0$ and $A_{a}$ is once again given by
Eq.~(\ref{eq:WB9}).  Furthermore, it is clear that the asymptotic behaviour
does not change and it is the one in Eq.~(\ref{eq:WB10}); what does change is
the behaviour of the solution at $r=0$, which is singular except when
$\beta_{a}=0$.

Using the above expression we can also create a hairy version of the
degenerate monopole: we have to make the same Ansatz as the one used in the
derivation of Eq.~(\ref{eq:WB11}), and also define $\beta_{2} = s + \delta
\gamma /3$, $\beta_{3} = s -\delta \gamma /3$ and $\beta_{1} = -2s$, which is
the maximal possibility compatible with a regular limit.  Taking then the
limit $\delta\rightarrow 0$ we find

\begin{equation}
  \label{eq:WBh2}
  Q_{1} \; =\; \textstyle{1\over 9}\left[ e^{-2(r+s)}\ +\ (3r+\gamma -1)e^{r+s}\right]
  \;\; ,\;\;
  Q_{2} \; =\; \textstyle{1\over 9}\left[ e^{2(r+s)}\ -\ (3r+\gamma +1)e^{-(r+s)}\right]
  \; .
\end{equation}

\noindent
which leads to $\phi_{i}$'s that are singular at $r=0$ but with the asymptotic
behaviour displayed in Eq.~(\ref{eq:WB12}).



\begin{thebibliography}{99}
\bibitem{Gibbons:1982ih}
G.~W.~Gibbons,
Nucl.\ Phys.\  B {\bf 207} (1982) 337.
R.~Kallosh and T.~Ort\'{\i}n,
Phys.\ Rev.\  D {\bf 48} (1993) 742
[\hepth{9302109}].
E. Bergshoeff, R. Kallosh and T. Ort\'{\i}n,
Nucl.\ Phys.\ B {\bf 478} (1996) 156
[\hepth{9605059}].
E.~Lozano-Tellechea and T.~Ort\'{\i}n,
Nucl.\ Phys.\  B {\bf 569} (2000) 435
[\hepth{9910020}].

\bibitem{Ferrara:1995ih}
S.~Ferrara, R.~Kallosh and A.~Strominger,
Phys.\ Rev.\ D {\bf 52} (1995) 5412
[\hepth{9508072}].

\bibitem{Behrndt:1997ny}
K. Behrndt, D. L\"ust and W.A. Sabra,
Nucl.\ Phys.\ B {\bf 510} (1998) 264
[\hepth{9705169}].
G.~Lopes Cardoso, B.~de Wit, J.~Kappeli and T.~Mohaupt,
JHEP {\bf 0012} (2000) 019
[\hepth{0009234}].

\bibitem{Tod:1983pm}
K.P.~Tod,
Phys.\ Lett.\ B {\bf 121} (1983) 241.

\bibitem{Tod:1995jf}
K.P.~Tod,
Class.\ Quant.\ Grav.\  {\bf 12} (1995) 1801.

\bibitem{Caldarelli:2003pb}
M.M.~Caldarelli and D.~Klemm,
JHEP {\bf 0309} (2003) 019
[\hepth{0307022}].

\bibitem{Cacciatori:2004rt}
S.~L.~Cacciatori, M.~M.~Caldarelli, D.~Klemm and D.~S.~Mansi,
JHEP {\bf 0407} (2004) 061
[\hepth{0406238}].

\bibitem{Cacciatori:2007vn}
S.~L.~Cacciatori, M.~M.~Caldarelli, D.~Klemm, D.~S.~Mansi and D.~Roest,
JHEP {\bf 0707} (2007) 046
[\arxiv{0704.0247} [hep-th]].

\bibitem{Meessen:2006tu}
P.~Meessen and T.~Ort\'{\i}n,
Nucl.\ Phys.\  B {\bf 749} (2006) 291
[\hepth{0603099}].

\bibitem{Cacciatori:2008ek}
S.~L.~Cacciatori, D.~Klemm, D.~S.~Mansi and E.~Zorzan,
JHEP {\bf 0805} (2008) 097
[\arxiv{0804.0009} [hep-th]].

\bibitem{Huebscher:2006mr}
M.~H\"ubscher, P.~Meessen and T.~Ort\'{\i}n,
Nucl.\ Phys.\ B {\bf 759} (2006) 228
[\hepth{0606281}].

\bibitem{Bellorin:2005zc}
J. Bellor\'{\i}n and T. Ort\'{\i}n,
Nucl.\ Phys.\ B {\bf 726} (2005) 171
[\hepth{0506056}].

\bibitem{Ortin:2008wj}
T.~Ort\'{\i}n,
JHEP {\bf 0805} (2008) 034
[\arxiv{0802.1799} [hep-th]].

\bibitem{Gran:2008vx}
U.~Gran, J.~Gutowski and G.~Papadopoulos,
``{\em Geometry of all supersymmetric four-dimensional ${\cal N}=1$ supergravity
       backgrounds}'',
\arxiv{0802.1779} [hep-th].


\bibitem{Chamseddine:1997nm}
A.H.~Chamseddine and M.S.~Volkov,
Phys.\ Rev.\ Lett.\  {\bf 79} (1997) 3343
\hepth{9707176}.
A.H.~Chamseddine and M.S.~Volkov,
Phys.\ Rev.\  D {\bf 57} (1998) 6242
\hepth{9711181}.

\bibitem{Huebscher:2007hj}
M.~H\"ubscher, P.~Meessen, T.~Ort\'{\i}n and S.~Vaul\`a,
 ``{\em Supersymmetric N=2 Einstein-Yang-Mills monopoles and covariant
         attractors}'',
\arxiv{0712.1530} [hep-th].

\bibitem{Meessen:2008kb}
P.~Meessen,
\arxiv{0803.0684} [hep-th]. To be oublished in 
\textit{Physics Letters}~\textbf{B}.

\bibitem{kn:Bog} E.~Bogomol'nyi,
                 {\it Sov.~J.~Nucl.~Phys.}~\textbf{24} (1976) 449.

\bibitem{Harvey:1991jr}
J.~A.~Harvey and J.~Liu,
Phys.\ Lett.\  B {\bf 268} (1991) 40.

\bibitem{Galtsov:1989ip}
D.V.~Gal'tsov and A.A.~Ershov,
Phys.\ Lett.\  A {\bf 138} (1989) 160.
A.A.~Ershov and D.V.~Gal'tsov,
Phys.\ Lett.\  A {\bf 150} (1990) 159.
P.~Bizon and O.T.~Popp,
Class.\ Quant.\ Grav.\  {\bf 9} (1992) 193.

\bibitem{Volkov:1998cc}
M.S.~Volkov and D.V.~Gal'tsov,
Phys.\ Rept.\  {\bf 319} (1999) 1
[\hepth{9810070}].

\bibitem{Bartnik:1988am}
R.~Bartnik and J.~Mckinnon,
Phys.\ Rev.\ Lett.\  {\bf 61} (1988) 141.

\bibitem{Bizon:1990sr}
  P.~Bizon:
  Phys. Rev. Lett. {\bf 64}(1990), 2844;
  H.P.~K\"unzle, A.K.M. Masood-ul-Alam:
  J.\ Math.\ Phys.\  {\bf 31} (1990) 928;
  M.S.~Volkov, D.V.~Gal'tsov:
  Sov.\ J.\ Nucl.\ Phys.\  {\bf 51} (1990) 747
%
\bibitem{Strominger:1996kf}
A.~Strominger,
Phys.\ Lett.\ B {\bf 383} (1996) 39
[\hepth{9602111}].
S. Ferrara and R. Kallosh,
Phys.\ Rev.\ D {\bf 54} (1996) 1514
[\hepth{9602136}].
S. Ferrara and R. Kallosh,
Phys.\ Rev.\ D {\bf 54} (1996) 1525
[\hepth{9603090}].
%
\bibitem{Andrianopoli:1996cm}
L. Andrianopoli, M. Bertolini, A. Ceresole, R. D'Auria, S. Ferrara, P. Fr\'e and T. Magri,
J.\ Geom.\ Phys.\  {\bf 23} (1997) 111
[\hepth{9605032}].
%
\bibitem{kn:toinereview}
A. Van Proeyen,
``{\em $N=2$ supergravity in $d=4$, $5$, $6$ and its matter couplings}'',
lectures given at the Institute Henri Poincar\'e, Paris, November 2000.
\href{http://itf.fys.kuleuven.ac.be/~toine/LectParis.pdf}{\tt http://itf.fys.kuleuven.ac.be/\~{}toine/LectParis.pdf}
%
\bibitem{deWit:1984pk}
 B.~de Wit and A.~Van Proeyen,
Nucl.\ Phys.\ B {\bf 245} (1984) 89.
%
\bibitem{deWit:1984px}
B. de Wit, P.G. Lauwers and A. Van Proeyen,
Nucl.\ Phys.\ B {\bf 255} (1985) 569.

\bibitem{Kallosh:1993wx}
R. Kallosh and T. Ort\'{\i}n,
``{\em Killing spinor identities}'',
\hepth{9306085}.

\bibitem{Bellorin:2005hy}
J. Bellor\'{\i}n and T. Ort\'{\i}n,
Phys.\ Lett.\ B {\bf 616} (2005) 118
[\hepth{0501246}].

\bibitem{Bellorin:2006xr}
J.~Bellor\'{\i}n, P.~Meessen and T.~Ort\'{\i}n,
Nucl.\ Phys.\  B {\bf 762} (2007) 229
[\hepth{0606201}].
%
\bibitem{Denef:2000nb}
F.~Denef,
JHEP {\bf 0008} (2000) 050
[\hepth{0005049}].
%
\bibitem{Bates:2003vx}
B.~Bates and F.~Denef,
``{\em Exact solutions for supersymmetric stationary black hole composites}'',
\hepth{0304094}.
%
\bibitem{Andrianopoli:2002vq}
  L.~Andrianopoli, R.~D'Auria, S.~Ferrara and M.A.~Lled\'o,
  JHEP {\bf 0301} (2003) 045
  [arXiv:hep-th/0212236];
  Nucl.\ Phys.\  B {\bf 640} (2002) 46
  [arXiv:hep-th/0202116];
  Nucl.\ Phys.\  B {\bf 640} (2002) 63
  [arXiv:hep-th/0204145];
  Mod.\ Phys.\ Lett.\  A {\bf 18} (2003) 1001
  [arXiv:hep-th/0212141].
\bibitem{Weinberg:1982jh}
  E.J.~Weinberg,
  Phys.\ Lett.\  B {\bf 119} (1982) 151.
%
\bibitem{Goddard:1977da}
  P.~Goddard and D.~I.~Olive,
  Rept.\ Prog.\ Phys.\  {\bf 41} (1978) 1357.
%
\bibitem{Protogenov:1977tq}
  A.P.~Protogenov,
  Phys.\ Lett.\  B {\bf 67} (1977) 62.
%
\bibitem{Guo:2008ks}
  H.~Guo and E.~J.~Weinberg,
  ``{\em Instabilities of chromodyons in $SO(5)$ gauge theory}'',
  arXiv:0803.0736 [hep-th].
%
\bibitem{Slansky:1981yr}
R.~Slansky,
Phys.\ Rept.\  {\bf 79} (1981) 1.
%
%
\bibitem{Ferrara:2006yb}
  S.~Ferrara, E.G.~Gimon and R.~Kallosh,
  Phys.\ Rev.\  D {\bf 74} (2006) 125018
  [arXiv:hep-th/0606211].
%
\bibitem{Wilkinson:1978zh}
D.~Wilkinson and F.A.~Bais,
Phys.\ Rev.\  D {\bf 19} (1979) 2410.

\bibitem{Andrianopoli:2001zh}
L.~Andrianopoli, R.~D'Auria and S.~Ferrara,
JHEP {\bf 0203} (2002) 025
[\hepth{0110277}].


\bibitem{Andrianopoli:2001gm}
L.~Andrianopoli, R.~D'Auria and S.~Ferrara,
Nucl.\ Phys.\  B {\bf 628} (2002) 387
[\hepth{0112192}].

\bibitem{kn:Br1} H.W. Brinkmann,
                 {\it Proc.~Natl.~Acad.~Sci.~U.S.}~\textbf{9} (1923) 1;

\bibitem{Gutowski:2001pd}
J.~Gutowski and G.~Papadopoulos,
Phys.\ Lett.\  B {\bf 514} (2001) 371
\hepth{0102165}


\bibitem{Panagopoulos:1983yx}
  H.~Panagopoulos,
  Phys.\ Rev.\  D {\bf 28} (1983) 380.
%

\bibitem{Ward:1981jb}
  R.S.~Ward,
  Commun.\ Math.\ Phys.\  {\bf 79} (1981) 317.

\bibitem{Dolivet:2007sz}
Y.~Dolivet, B.~Julia and C.~Kounnas,
JHEP {\bf 0802} (2008) 097
[\arxiv{0712.2867} [hep-th]].

\bibitem{Bianchi:2007va}
M.~Bianchi and S.~Ferrara,
JHEP {\bf 0802} (2008) 054
[\arxiv{0712.2976} [hep-th]].
%

\bibitem{kn:superpaper} 
M.~H\"ubscher, P.~Meessen, T.~Ort\'{\i}n and S.~Vaul\`a,
in preparation.


\bibitem{Kallosh:1996tf}
R.~Kallosh, M.~Shmakova and W.~K.~Wong,
Phys.\ Rev.\  D {\bf 54} (1996) 6284
\hepth{9607077}.


\end{thebibliography}
\end{document}